\documentclass[12pt]{article}
\usepackage{graphicx,psfrag,color}

\def\bm#1{\mbox{\boldmath$#1$\unboldmath}}
\newcommand{\bffmath}[1]{\mbox{\boldmath ${#1}$}}
\newcommand{\bff}[1]{{\bf {#1}}}

\usepackage{amsmath,amssymb}
\numberwithin{equation}{section}
\allowdisplaybreaks

\newcommand{\blue}[1]{\textcolor{black}{#1}}

\catcode`@=11
\newcount\@tempcntc
\def\@citex[#1]#2{\if@filesw\immediate\write\@auxout{\string\citation{#2}}\fi
  \@tempcnta\z@\@tempcntb\m@ne\def\@citea{}\@cite{\@for\@citeb:=#2\do
    {\@ifundefined
       {b@\@citeb}{\@citeo\@tempcntb\m@ne\@citea\def\@citea{,}{\bf ?}\@warning
       {Citation `\@citeb' on page \thepage \space undefined}}%
    {\setbox\z@\hbox{\global\@tempcntc0\csname b@\@citeb\endcsname\relax}%
     \ifnum\@tempcntc=\z@ \@citeo\@tempcntb\m@ne
       \@citea\def\@citea{,}\hbox{\csname b@\@citeb\endcsname}%
     \else
      \advance\@tempcntb\@ne
      \ifnum\@tempcntb=\@tempcntc
      \else\advance\@tempcntb\m@ne\@citeo
      \@tempcnta\@tempcntc\@tempcntb\@tempcntc\fi\fi}}\@citeo}{#1}}
\def\@citeo{\ifnum\@tempcnta>\@tempcntb\else\@citea\def\@citea{,}%
  \ifnum\@tempcnta=\@tempcntb\the\@tempcnta\else
   {\advance\@tempcnta\@ne\ifnum\@tempcnta=\@tempcntb \else \def\@citea{--}\fi
    \advance\@tempcnta\m@ne\the\@tempcnta\@citea\the\@tempcntb}\fi\fi}
\catcode`@=12

\begin{document}

\begin{titlepage}

\begin{flushright}
TUM-HEP-919/13\\
TTK-13-26\\
SFB/CPP-13-110 \\
arXiv:1312.4791 [hep-ph]\\
September 07, 2024
\end{flushright}

\vskip1.2cm
\begin{center}
\Large\bf\boldmath Third-order correction to top-quark
pair production near threshold I. Effective theory set-up 
and matching coefficients 
\end{center}

\vspace{1cm}
\begin{center}
{\sc M.~Beneke}$^{a,b}$, {\sc Y. Kiyo}$^c$,
{\sc K. Schuller}$^b$\\[5mm]
  {\it $^a$ Physik Department T31, James-Franck-Stra\ss e~1,}\\
  {\it Technische Universit\"at M\"unchen, D--85748 Garching, Germany}\\[0.1cm]
  {\it $^b$ Institut f{\"u}r Theoretische Teilchenphysik und Kosmologie,}\\
  {\it RWTH Aachen University, D--52056 Aachen, Germany}\\[0.1cm]
  {\it $^c$ Department of Physics, Juntendo University,}\\
  {\it Inzai, Chiba 270-1695, Japan}\\[0.3cm]
\end{center}

\vspace{2cm}
\begin{abstract}
\noindent 
This is the first in a series of two papers, in which 
we compute the third-order QCD corrections to top-antitop 
production near threshold in $e^+ e^-$ collisions. 
The present paper provides a detailed outline of 
the strategy of computation in the framework of 
non-relativistic effective theory and the threshold 
expansion, applicable more generally to heavy-quark pair production 
near threshold.  It summarizes matching coefficients and 
potentials relevant to the next-to-next-to-next-to-leading
order and ends with the master formula for the 
computation of the third-order Green function. The master formula  
is evaluated in part II of the series.  
\end{abstract}
\end{titlepage}

\pagenumbering{roman}
\tableofcontents
\newpage
\pagenumbering{arabic}


\section{Introduction}

Many of the most accurate heavy-quark mass determinations are 
related to the spectral functions of the heavy-quark vector current, 
which can be measured in $e^+ e^-$ collisions. The energy region 
near the pair production threshold is particularly sensitive to 
the mass value. The bottom and charm masses are usually inferred 
from suitable averages of the pair production cross section. 
Looking to the future, the measurement of the top quark pair 
production cross section in the threshold region at 
\blue{an electron-positron collider with centre-of-mass energy above 
350~GeV} would lead to a very precise 
knowledge of the top mass directly from the energy dependence 
of the cross section, even though the toponium resonances are smeared 
out due to the large top quark width~\cite{Bigi:1986jk}. To 
put this into perspective, the top mass value 
from direct production at the Fermilab Tevatron is 
\blue{$m_t=174.34 \pm 0.37~\mbox{(stat.)} \pm 0.52~\mbox{(syst.)}\,$GeV 
\cite{Tevatron:2014cka}. 
The current values measured by the ATLAS and CMS collaborations 
at the Large Hadron Collider (LHC) are 
$m_t=172.69 \pm 0.25~\mbox{(stat.)} \pm 0.41~\mbox{(syst.)}\,$GeV 
\cite{ATLAS:2018fwq} (ATLAS) from runs at $7\,$TeV and $8\,$TeV 
centre-of-mass energy, and $m_t=171.77 \pm 0.37\,$GeV (with negligible statistical error) from CMS \cite{CMS:2023ebf} at 
$\sqrt{s}=13\,$TeV. A less precise value can be obtained from the 
total production cross section as illustrated for example in  
\cite{Beneke:2012wb}, and $t\bar{t}+$jet events. 
A further substantial reduction of the uncertainty 
of these measurements is very difficult} due to the complicated 
theoretical systematics of top jet mass reconstruction at hadron 
colliders. The above numbers should be 
compared to the precision of $30\,$MeV that can be achieved 
experimentally~\cite{Martinez:2002st,Seidel:2013sqa,Horiguchi:2013wra,Simon:2016pwp} 
from the $t\bar t$ threshold scan at a 
\blue{high-energy $e^+ e^-$ collider.} 
In particular, while it is not   
evident which renormalized top mass parameter is determined with the 
quoted accuracy at the Tevatron and LHC,\footnote{\blue{A review of the pertinent issues and recent theoretical progress can be found in \cite{Hoang:2020iah}.}} the threshold cross section 
in $e^+ e^-$ collisions provides an observable 
that can be unambiguously related to a particular top mass definition. 

Aside from determining a fundamental 
parameter of the Standard Model (SM), accurate top mass measurements are 
of interest for extrapolating the SM or its TeV scale extension to 
higher energies, either in the context of ``electroweak precision 
tests'' or a theory of the Yukawa couplings. Furthermore, the  
discovery of a Higgs boson \cite{Aad:2012tfa,Chatrchyan:2012ufa} 
with a mass of about 125$\,$GeV if interpreted 
as the SM Higgs boson determines the Higgs self-coupling and leads 
to the conclusion that the SM vacuum becomes metastable at scales 
above $10^{10}\,$GeV~\cite{Degrassi:2012ry}. The precise value of 
this scale turns out to depend very sensitively on the value of the 
top quark mass. Finally, the absence of any hint on physics beyond 
the SM in high-energy collisions at LHC has 
renewed the interest in performing precision measurements of 
properties of SM particles, including the top quark mass, 
width and Yukawa coupling. Measurements 
at the top pair production threshold are uniquely suited for 
this purpose.

The challenge is thus to calculate the heavy-quark\footnote{In the 
following we will often refer to the heavy quark as the 
``top quark'', since this covers the most general case. For 
charm and bottom quarks, one simply sets the decay width $\Gamma$ to 
zero in the top-quark expressions.} spectral functions 
precisely in the threshold region. This kinematic region is 
characterized by two features, which make the theoretical calculation 
of QCD corrections rather different from standard loop calculations: 
the small three-velocity $v$ of the heavy quarks, which allows to 
expand Feynman diagrams in $v$ rather than calculate them exactly, 
and the strong colour-Coulomb force, which on the other hand requires 
certain diagrams to be summed to all orders in the strong coupling 
$\alpha_s$ similar to bound-state calculations in quantum 
electrodynamics. The expansion of the cross section relative to the 
ultra-relativistic point-particle cross section $\sigma_0$ 
is then organized as
\begin{eqnarray}
\label{systematics}
R=\sigma_{t\bar t}/\sigma_0 &\sim& 
v \sum_k \left(\frac{\alpha_s}{v}\right)^k \{1 (\textrm{LO});
\,\alpha_s,v \, (\textrm{NLO}); 
\,\alpha_s^2,\alpha_s v,v^2(\textrm{NNLO});
\nonumber \\ 
&&\hspace*{1cm} 
\,\alpha_s^3,\alpha_s^2 v,\alpha_s v^2,v^3
(\textrm{NNNLO});
\,\ldots\},
\end{eqnarray}
where the overall factor of $v$ arises from the phase-space of the 
two produced massive particles, and the order of the various terms 
is indicated explicitly. To perform the expansion and the required 
summation of perturbation theory to all orders beyond the 
next-to-leading order (NLO), non-relativistic effective 
field theory~\cite{Thacker:1990bm,Lepage:1992tx,Bodwin:1994jh} 
and the threshold expansion of Feynman diagrams~\cite{Beneke:1997zp} are 
the methods of choice. The cross section is then obtained 
from the expression\footnote{Neglecting here the contributions 
from $Z$-boson exchange. See section~\ref{sec:EFT} for the full 
expression.} 
\begin{equation}
\label{R1}
R = 12\pi e_t^2\,\mbox{Im}
\left[\frac{N_{c}}{2m^{2}}\left(c_v 
\left[c_v-\frac{E}{m}\,\left(c_v+\frac{d_v}{3}\right)\right]
G(E) + \ldots\right) \right],
\end{equation}
where $c_v$, $d_v$ denote certain relativistic matching corrections, 
$E=\sqrt{s}-2 m$, $N_c=3$ the number of colours, 
and $G(E)$ represents a two-point function of heavy-quark currents 
in the non-relativistic effective theory.
The purpose of this paper is to present the results 
for the part of the third-order (NNNLO) QCD corrections to the heavy quark 
anti-quark pair production cross section near threshold related 
to the correlation function $G(E)$, which contains the all-order 
summation. Since this concludes the non-relativistic 
third-order calculation, we also present details of the methods 
and calculations that have been used but not given in earlier 
publications, \blue{together with the required expressions for the 
matching coefficients $c_v$, $d_v$.}

For the top pair production threshold the leading order (LO) and 
next-to-leading order (NLO) approximations to the cross section 
(\ref{systematics}) have 
been examined long
ago~\cite{Fadin:1987wz,Fadin:1988fn,Strassler:1990nw}. Several 
other aspects of the threshold such as top 
momentum distributions and polarization have been 
computed at this order~\cite{Jezabek:1992np,Sumino:1992ai,Fujii:1993mk,Harlander:1994ac}.
Beyond NLO, matching the non-relativistic approximation to QCD is 
non-trivial, because the separation between relativistic and 
non-relativistic physics is no longer unambiguous. A consistent field 
theoretical approach based on non-relativistic effective
QCD is now
required. The second-order (NNLO) QCD corrections to the total pair 
production cross section have been computed in this framework  
\blue{more than twenty} years
ago~\cite{Hoang:1998xf,Melnikov:1998pr,Beneke:1999qg,Hoang:1999zc,Yakovlev:1998ke,Nagano:1999nw,Penin:1998mx}
and turned out to be 
surprisingly large even for top quarks. While some of the large corrections 
can be understood as being due to mass 
renormalization~\cite{Beneke:1998rk}, and can be avoided by a change of 
renormalization convention, there remains an apparently slow convergence 
of successive approximations to the normalization of the cross 
section, which necessitates the calculation of the NNNLO term. 
An alternative approach that sums logarithms of $v$ has also been 
pursued, and an improvement of convergence has been found in 
a (still partially incomplete) next-to-next-to-leading logarithmic 
(NNLL) approximation~\cite{Hoang:2000ib,Hoang:2001mm,Pineda:2006ri,Hoang:2013uda}. 
Nonetheless, the NNNLO non-logarithmic terms not included in the 
NNLL approximation are required to be 
certain that the theoretical calculation is sufficiently accurate for 
the proposed mass measurement 
\blue{from the production threshold.} 
This is the main motivation 
for the present work. Over the past years a significant 
number of results relevant to the NNNLO calculation or partial results 
for third-order quarkonium energy levels and wave-functions at the 
origin have already
appeared~\cite{Brambilla:1999qa,Kniehl:1999ud,Brambilla:1999xj,Kniehl:1999mx,Manohar:2000kr,Kiyo:2000fr,Kniehl:2001ju,Kniehl:2002br,Penin:2002zv,Kniehl:2002yv,Hoang:2003ns,Beneke:2005hg,Penin:2005eu,Marquard:2006qi,Beneke:2007gj,Beneke:2007pj,Beneke:2007uf,Beneke:2008ec,Beneke:2008cr,Smirnov:2008pn,Marquard:2009bj,Anzai:2009tm,Smirnov:2009fh,Kiyo:2013aea,Beneke:2013kia,Marquard:2014pea}.
In the following we summarize the status of the 
NNNLO top cross section calculation. 

\subsubsection*{\it Matching calculations}

While the resummed cross section is calculated in an effective field 
theory (EFT), a number of matching calculations has to be performed to 
guarantee that the EFT reproduces QCD to the required accuracy. This is 
done in two steps. Hard matching (scale $k\sim m$) yields the 
coefficients of the non-relativistic QCD 
(NRQCD) interactions and heavy-quark currents; soft matching 
(scale $k\sim m v$) the quark-anti-quark potentials. At NNNLO the 
coefficients of several subleading NRQCD interactions must be determined 
with one-loop precision. This calculation has been performed 
in~\cite{Manohar:1997qy}. However, as shall be explained, 
the calculation of the cross section 
requires the $O(\epsilon)$ terms of the coefficient functions. We 
therefore repeated the NRQCD matching calculation, confirm the results 
of~\cite{Manohar:1997qy} and provide the expressions for the  
$O(\epsilon)$ terms in this paper (see also \cite{Wuester:2003}). 
Hard matching of the non-relativistic currents at NNNLO is needed 
at the one-loop level for the subleading currents and at the three-loop 
level for the leading current. The former are known~\cite{Luke:1997ys}, 
and will be rederived in the present paper, 
\blue{and the three-loop matching 
coefficient $c_3$ has been calculated 
in~\cite{Marquard:2006qi,Marquard:2014pea,Egner:2022jot}. }

As concerns soft matching, the potentials of order $1/(m^2 r^3)$ must be 
determined with one-loop precision, the $1/(m r^2)$ potential with 
two-loop precision, and the $1/r$ Coulomb potential at three loops,
since $r$ counts as the Bohr radius $1/(m \alpha_s)$. 
Except for the Coulomb potential, the coefficients of the other potentials 
have been calculated in~\cite{Kniehl:2001ju,Kniehl:2002br}, 
but again these results are not sufficient for the cross section 
calculations, since the $O(\epsilon)$ terms of all these potentials are 
needed.  We repeated the calculation of the one-loop $1/(m^2 r^3)$ 
potentials, confirm the results of~\cite{Kniehl:2002br} and provide the 
expressions for the  $O(\epsilon)$ terms in this paper (see 
also~\cite{Wuester:2003}). The $O(\epsilon)$ term of the two-loop 
$1/(m r^2)$ potential coefficient $b_2$ 
\blue{can be found in \cite{Beneke:2014qea}.}
As concerns the three-loop Coulomb potential, the fermionic
contributions to the three-loop coefficient $a_3$ have been 
calculated first in~\cite{Smirnov:2008pn} and 
the full result is now also known~\cite{Anzai:2009tm,Smirnov:2009fh}, 
\blue{including its fully analytic expression~\cite{Lee:2016cgz}.}
 
To summarize: the matching coefficients required for the NNNLO calculation 
of the heavy-quark production cross section near threshold are 
complete.\footnote{\blue{Except for the flavour-singlet contribution to the 
three-loop matching coefficient $c_3$, discussed in 
section~\ref{sec:momentumregions}, which is expected to be small.}}

\subsubsection*{\it Matrix element calculations}

After matching the heavy-quark currents and Lagrangians the cross section
calculation is mapped to the calculation of the imaginary part of the 
two-point function $G(E)$ of non-relativistic currents in (\ref{R1}). The 
leading-order colour Coulomb potential is now part of the leading-order 
effective Lagrangian, since the Coulomb interaction is strong near 
threshold. The propagators to be used in the perturbative calculation 
of the two-point function are the Coulomb Green functions, making 
this part of the computation similar to QED bound-state problems. 
The calculation can be divided into three parts: contributions up to 
the third order involving only the Coulomb potential, already completed 
in~\cite{Beneke:2005hg}; the ultrasoft contribution, appearing first 
at NNNLO, which has been computed in~\cite{Beneke:2008cr}; 
and finally, contributions involving at least one potential other than the 
Coulomb potential (``non-Coulomb potential contribution''), 
which are not yet known.

The main result of this paper is the missing non-Coulomb potential 
contribution. Compared to the Coulomb contributions the major complication 
is the singular nature of the potential insertions. The ultraviolet 
divergences must be regulated in dimensional regularization 
in a scheme consistent 
with the calculation of the matching coefficients order by 
order in the strong coupling, while retaining the resummation of 
infinitely many Coulomb gluon exchanges by the use of Coulomb Green 
functions, whose $d$-dimensional expression is unknown. The techniques 
we apply are an extension of those used in the NNLO 
calculation~\cite{Beneke:1999qg}. Since the method of that calculation 
was never written up (though some results are scattered in 
\cite{Pineda:2006ri,Beneke:2005hg}), we devote some effort to presenting 
the third-order calculation in some technical detail.

In addition to the dominant production of the top-quark pair 
through a virtual photon there is also a $Z$-boson contribution. 
The contribution from the vector-coupling of the $Z$ 
is trivially inferred from the photon-mediated cross section, while 
the axial-vector coupling contribution is suppressed near threshold and 
begins only at NNLO. Thus, only the first-order correction to the 
axial-vector in non-relativistic perturbation theory is needed. 
Some results at this order are 
available~\cite{Penin:1998mx,Penin:1998ik,Kuhn:1999hw}, but none
of these results are given in dimensional regularization. 
\blue{The NNNLO P-wave contribution in dimensional regularization was 
obtained in \cite{Beneke:2013kia}, hence in the present} 
paper we focus on the missing third-order terms 
in the S-wave vector current contribution. 

To summarize: together with the results of this paper, the matrix 
element calculation is complete to NNNLO.

\subsubsection*{\it Electroweak and electromagnetic corrections}

Less work has been done on electroweak and electromagnetic 
corrections. Counting the electromagnetic and electroweak coupling 
as two powers of the strong coupling, electromagnetic corrections
contribute from NLO through the electromagnetic Coulomb potential. 
This effect is easily included and has been discussed 
in~\cite{Pineda:2006ri,Hoang:2010gu,Beneke:2010mp}. Electroweak 
contributions to the matching coefficients of NRQCD currents and 
production operators, 
\blue{which count as NNLO,}  
have been calculated in~\cite{Guth:1991ab,Hoang:2004tg,Hoang:2006pd,Eiras:2006xm,Kiyo:2008mh} 
\blue{and were included into the 
top threshold calculation in \cite{Beneke:2017rdn} together with 
Higgs-Yukawa coupling effects up the NNNLO 
computed earlier \cite{Beneke:2015lwa}.}
The formalism for calculating 
initial-state radiation, and soft and collinear photon corrections 
in general, simultaneous with summing Coulomb exchange has originally been 
worked out in~\cite{Actis:2008rb} for $W$-boson pair production 
\blue{and was included for the top threshold in \cite{Beneke:2017rdn}. 
Thus, electroweak and electromagnetic effects are known to NNLO, 
and partially to NNNLO.}

For top quarks the sizeable decay width $\Gamma_t$ introduces further 
complications. In leading order the width is correctly 
included by evaluating the current two-point functions in PNRQCD at complex 
energy argument $E+i\Gamma_t$~\cite{Fadin:1987wz,Fadin:1988fn}, 
where $E=\sqrt{s}-2m_t$. We adopt this 
prescription as the {\em definition} of the pure QCD 
contributions to the cross 
section. Beginning at NLO there exist further contributions related 
to the finite width. Since the physical final state is 
$W^+ W^- b\bar b\,$,\footnote{To the extent that we focus on top 
width effects, the $W$ boson may be regarded as stable.} there 
exist irreducible ``backgrounds'' related to off-shell and single or 
non-resonant top-quark pair production. In fact, the QCD contribution
as defined above cannot be unambiguously separated from electroweak 
effects at this order -- perhaps not surprisingly, since the top quark
width itself is such an effect -- and the 
fact that a physical scattering cross section should refer only to 
stable (or sufficiently long-lived) particles in the final state, must 
be taken into account. The incompleteness of the QCD cross section is 
signaled explicitly by the presence of uncancelled singularities in 
dimensional regularization with coefficients proportional to 
$\Gamma_t$ starting at NNLO.\footnote{At NLO these finite-width
  divergences are linear and therefore do not show up as poles in
  dimensional regularization. Nevertheless, this implies an implicit
  dependence of the result of the regularization scheme, which is
  cancelled by computing the non-resonant contributions consistently
  in the same scheme, as was done in~\cite{Beneke:2010mp}.}  
The origin and consistent cancellation of 
these singularities is discussed in~\cite{Beneke:2008cr,Hoang:2004tg} 
and the corresponding calculations of electroweak effects and 
non-resonant contributions to a physical final state such as 
$W^+ W^- b\bar b$ can in principle be performed in the framework of 
unstable-particle effective field theory~\cite{Beneke:2003xh,Beneke:2004km} 
as already done for $W$ pair
production~\cite{Actis:2008rb,Beneke:2007zg}. 
The corresponding calculation of the non-resonant NLO correction 
for top production has been performed in~\cite{Beneke:2010mp} 
and confirmed by a different method in \cite{Penin:2011gg}. 
Rather than embarking on the rather difficult computation of 
non-resonant contributions up to NNNLO, however, 
\blue{one may} also suppress them by appropriate invariant 
mass cuts~\cite{Actis:2008rb,Hoang:2008ud}. Calculations of 
top quark pair production near threshold with cuts on the 
final-state $Wb$ invariant masses have 
appeared~\cite{Hoang:2010gu,Beneke:2010mp,Beneke:2017rdn} in the 
non-relativistic QCD and unstable particle effective theory 
frameworks. As should be expected, the non-resonant contributions to 
the $W^+ W^- b\bar b$ are sizeable below the nominal top 
pair production threshold, and
hence can change the shape of the threshold cross section in the
region of interest for the top-quark mass determination. 
\blue{A fully differential treatment of the threshold region can be 
found in \cite{Bach:2017ggt}, but is limited to NLO accuracy.}
In the present paper, we focus only on the QCD part of the problem. 
We shall, however, make the finite-width $1/\epsilon$ poles explicit, 
so that they can be cancelled analytically 
with future computations of electroweak effects. Indeed, the 
pole parts of the NNLO non-resonant contribution have
been computed~\cite{Jantzen:2013gpa,Ruiz-Femenia:2014ava} 
and the cancellation has been verified. 
\blue{Meanwhile, the full NNLO $W^+ W^- b\bar b$ cross section 
near $s\approx 4 m_t^2$ has been computed \cite{Beneke:2017rdn}, 
which includes the one-gluon correction to the non-resonant 
process.}

To summarize: while the formalism for computing electroweak and 
finite-width effects consistently is in place, 
\blue{the calculation 
of electroweak and non-resonant effects has been done up to 
now only to NNLO \cite{Beneke:2017rdn}. However, the results of 
this paper and its companion paper~II suggest that the NNNLO non-resonant correction, even though formally 
required to cancel residual singularities of the NNNLO QCD calculation, 
is numerically a small effect.}

\vskip0.3cm
Since our intention to present the concepts, techniques and calculations in 
some detail resulted in a rather lengthy text, we have split it into 
two parts. Part I presents the effective field theory set-up, 
the NRQCD and PNRQCD matching coefficients and ends with a 
master formula for the third-order heavy-quark pair production cross 
sections. This part could also be read as a review of non-relativistic 
effective theory in the weak-coupling regime complementary 
to~\cite{Pineda:2011dg}. Part II contains the actual 
PNRQCD matrix element calculation together with a 
\blue{detailed analysis} of the new contributions to the 
top cross section. A preliminary version was presented in~\cite{Beneke:2008ec} 
\blue{and the final NNNLO QCD result (excluding 
the small P-wave contribution, which was discussed before in 
\cite{Beneke:2013kia}), including the results of the present 
work, appeared in short form already in \cite{Beneke:2015kwa}.}

The outline of paper I is as follows: In section~\ref{sec:EFT}
we review the effective field theory framework and discuss the 
power-counting arguments that lead to the identification of the matching 
coefficients needed for the NNNLO calculation. The subsequent two 
sections deal with matching QCD to a sequence of two non-relativistic 
effective theories, NRQCD and PNRQCD.  Section~\ref{sec:nrqcd}
discusses the NRQCD aspects of the calculation. 
In particular, we calculate the relevant
one-loop matching coefficients including the new $O(\epsilon)$ terms 
and collect all other results that feed into the cross section 
calculation. \blue{We comment on the precise definition of the 
matching scheme in the context of the threshold expansion.} 
The second matching step from NRQCD to PNRQCD is 
discussed at length in section~\ref{sec:pnrqcd}, since a coherent 
summary is not yet available in the literature. Among the new results 
of this section are the path-integral derivation of the PNRQCD 
Lagrangian (neglecting ultrasoft gluons), the $O(\epsilon)$ terms 
of the one-loop potentials, and a discussion on the non-renormalization 
of currents in the NRQCD to PNRQCD matching.

PNRQCD perturbation theory in the pole-mass scheme provides a poor 
approximation to the top-quark pair production cross section near 
threshold. The top quark decay width is obviously an important effect, 
as is the conversion from the pole mass renormalization 
scheme that is employed in the primary calculations to renormalization
schemes that absorb large corrections into the mass counterterm, 
which is a prerequisite for reliable perturbative 
calculations~\cite{Beneke:1998rk,Beneke:1998jj}. Furthermore, a resummation 
of PNRQCD perturbation theory for the Green function is necessary in the 
vicinity of the bound state poles despite the sizeable top quark width. 
These refinements will be explained in paper II, \blue{together with 
their detailed numerical analysis and the size of the individual 
third-order QCD corrections to the cross section.}
In section~\ref{sec:master} we conclude paper I by providing the 
master formula for the computation of the third-order cross section 
in PNRQCD. \blue{The results of this paper and paper~II have also been made 
available 
in the code \texttt{QQbar\_threshold}~\cite{Beneke:2016kkb} based on 
the {\sc mathematica/C++} software.}

\section{Top pair production near threshold in 
effective field theory}
\label{sec:EFT}

In this section we present the relation of the 
pair production cross section to correlation 
functions of heavy quark currents together with the arguments, 
why this relation holds true at NNNLO. We review the scales and
momentum regions relevant to the problem both of which are central 
to the systematics of the effective theory approach.


\subsection{Heavy-quark correlation function} 
\label{sec:HQCF} 

\begin{figure}[t]
\begin{center}
\makebox[0cm]{ 
\scalebox{0.58}{\rotatebox{0}{\includegraphics{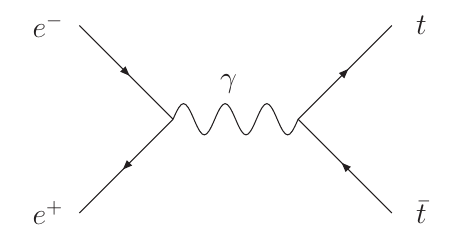}}}
\hspace{0.1cm} 
\scalebox{0.58}{\rotatebox{0}{\includegraphics{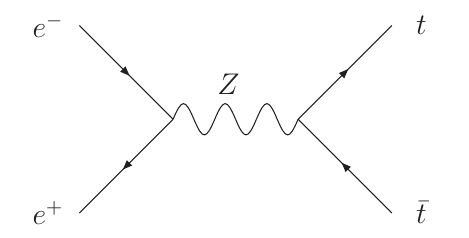}}}
}
\end{center}

\begin{center}
\makebox[0cm]{ 
\scalebox{0.58}{\rotatebox{0}{\includegraphics{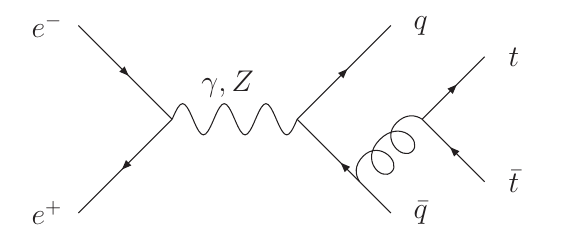}}}
\hspace{0.1cm} 
\scalebox{0.58}{\rotatebox{0}{\includegraphics{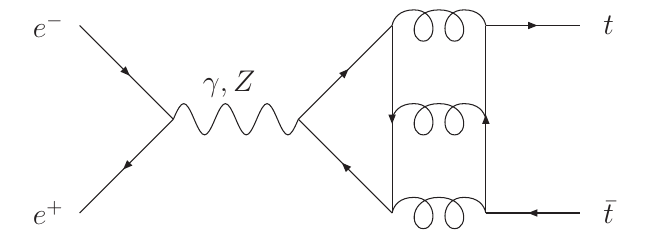}}}
}
\end{center}
\vspace*{-0.1cm}
\caption{\label{fig:eett} 
Basic electroweak  $t\bar{t}$ production processes (upper part of 
the figure) and production mechanisms (``$t\bar t$-radiation'' and 
``singlet production'') not contained in the heavy-quark correlation 
functions (lower part).}
\end{figure}

The basic top pair production mechanisms in $e^+ e^-$ annihilation 
are shown in the upper part of figure~\ref{fig:eett}. Since we work 
to lowest order in the electromagnetic and electroweak couplings, 
the optical theorem allows us to relate the total cross
section $\sigma_{t\overline{t}X}$ of the process $e^+ e^- \rightarrow t
\bar{t}X$ to the two point functions of the vector and 
axial-vector heavy quark current. We define 
\begin{eqnarray}
\label{eq:pi} 
\Pi^{(X)}_{\mu\nu}(q^2) &=& i\int d^4 x\,e^{i q\cdot x}\,\langle
0|T(j^{(X)}_\mu(x) j^{(X)}_\nu(0))|0\rangle 
\nonumber\\
&=& (q_\mu q_\nu-q^2 g_{\mu\nu})\,\Pi^{(X)}(q^2) + 
q_\mu q_\nu \Pi_L^{(X)}(q^2),
\end{eqnarray}
for the vector current $j_{\mu}^{(v)}=\bar{t}\gamma_{\mu}t$ and the 
axial vector current $j_{\mu}^{(a)}=\bar{t}\gamma_{\mu}\gamma_{5}t$. 
The cross section is then given by 
\begin{eqnarray}
\label{eq:sigtt} 
\sigma_{t\overline{t}X} &=& \sigma_0 \times 12\pi\,
\textrm{Im}\bigg[e_t^2 \Pi^{(v)}(q^2)-\frac{2q^2}{q^2-M_Z^2}v_e v_t
e_t \Pi^{(v)}(q^2) 
\nonumber \\ 
&&\hspace*{2.3cm}+\,\Bigg(\frac{q^2}{q^2-M_Z^2}\Bigg)^{\!2}\,
(v_e^2+a_e^2)(v_t^2\Pi^{(v)}(q^2)+a_t^2
\Pi^{(a)}(q^2))\bigg],
\end{eqnarray}
where $ \sigma_0=4\pi\alpha_{\rm em}^2/(3 s)$ is the high-energy limit
of the $\mu^+ \mu^-$ production cross section, $s=q^2$ the
center-of-mass energy squared, and $M_Z$ the $Z$-boson mass. 
$e_t=2/3$ denotes the top quark electric
charge in units of positron charge and $\alpha_{\rm em}$ is the 
electromagnetic coupling.  The vector- and
axial-vector couplings of fermion $f$ to the $Z$-boson are given by
\begin{equation}
v_f=\frac{T_3^f-2e_f \sin^2\theta_w}{2\sin\theta_w\cos\theta_w},
\qquad\quad
a_f=\frac{T_3^f}{2\sin\theta_w\cos\theta_w},
\end{equation}
with $\theta_w$ the weak mixing angle,
$e_f$ the electric charge of fermion $f$ and $T_3^f$ its third component of
the weak isospin. 

The dominant production mechanism is through the coupling to the 
virtual photon. The vector coupling of the $Z$-boson increases 
the photon-mediated cross section by only about 8\% in the 
threshold region $q^2\approx 4 m^2$. The axial-vector contribution is
even smaller, since the axial coupling is suppressed near threshold 
by the small velocity of the top quarks. $\Pi^{(a)}(q^2)$ contributes 
to (\ref{eq:sigtt}) only at NNLO relative $\Pi^{(v)}(q^2)$.

Eq.~(\ref{eq:sigtt}) which relates the inclusive top cross section 
to the spectral functions of heavy-quark currents is not exact. 
There exist top production mechanisms, shown in the lower part of 
figure \ref{fig:eett}, which are not captured by the heavy-quark 
current correlation functions, since the photon or $Z$-boson 
couples to light quarks. Vice versa, there exist cuts contributing 
to $\mbox{Im}\,\Pi^{(X)}(q^2)$ related to annihilation subdiagrams, 
see figure~\ref{fig:annihilation} below,  
which do not contain top quarks and 
hence should be excluded. We shall discuss in the next subsection 
that these contributions are either highly suppressed and not 
relevant at third order, or can easily be included.

\subsubsection*{Energy variables}

The characteristic non-relativistic energy in threshold production is 
much smaller than $\sqrt{s}\approx 2 m$. We define $E=\sqrt{s}-2 m$ and the 
top quark ``velocity'' $v=(E/m)^{1/2}$. This is related to another 
often used velocity parameter $\beta=(1-4 m^2/s)^{1/2}$ by 
\begin{equation}
v = \sqrt{\frac{E}{m}} = \left(\frac{2}{(1-\beta^2)^{1/2}}-2
\right)^{\!1/2} = \beta+\frac{3\beta^3}{8}+\ldots.
\end{equation}
Real values of $s$ and $E$ should be interpreted with a $+i\epsilon$ 
prescription. We will extensively use another variable $\lambda$ defined by 
\begin{equation}
\lambda = \frac{\alpha_s C_F}{2\sqrt{-\frac{E}{m}}}.
\label{eq:deflambda}
\end{equation}
Here and below $\alpha_s$ without any argument denotes the strong coupling 
in the $\overline{\rm MS}$ scheme \cite{Bardeen:1978yd} 
at the renormalization scale $\mu$, and $C_F=(N_c^2-1)/(2 N_c)=4/3$. 

The value of $\lambda$ determines when resummation to all orders 
in $\alpha_s$ is necessary. Above threshold, the variable $\lambda$ is 
purely positive-imaginary, below threshold it is real and positive. 
The threshold region is characterized by an absolute 
value of $\lambda$ of order 1 or larger. In particular, the Coulomb 
bound state poles are found at positive integer values of $\lambda$. 
Conventional fixed-order perturbation theory can be used only when 
$\lambda \ll 1$. As will be discussed later the width of the top 
quark is accounted for by substituting $E\to E+i\Gamma$. Thus, 
as $E$ varies from $-\infty$ to $\infty$ the variable $\lambda$ sweeps
through a curve in the complex plane that begins at the origin, 
moves out into the first quadrant into the direction of the positive
real axis and returns to the origin from above near the imaginary axis. 
The absolute value of $\lambda$ along this curve is always smaller than 
$\alpha_s C_F/2\times (m/\Gamma)^{1/2}$, which is about 1.2 for top 
quarks, with $\alpha_s(15\,\mbox{GeV})\approx 0.16$, but since 
it reaches order one in the threshold region, the perturbation 
expansion in $\alpha_s$ breaks down and 
resummation is necessary.

\subsection{Momentum regions and effective field theory}
\label{sec:momentumregions} 

Near the heavy-quark pair production threshold only a small 
kinetic energy $\sqrt{s}-2 m = E = m v^2$ is available to the 
final state. In the natural frame where $q^\mu = (2 m+E,\bff{0})$ 
this implies that the typical three momentum of 
a heavy quark is of order $m v$ (about 20~GeV for top quarks), 
while the energy and momentum of any other nearly massless 
particle can at most be $m v^2$ (about 2~GeV for tops). 
The presence of several small scales propagates into the 
loop diagrams that contribute to the spectral functions and causes 
a breakdown of the standard perturbation expansion in the 
strong coupling $\alpha_s$. However, since $v$ is small one does 
not have to compute the loop integrals exactly -- an expansion 
in $v$ suffices. This leads to a reorganized expansion as 
shown in (\ref{systematics}), in which $\alpha_s$ and $v$ 
are expansion parameters but $\alpha_s/v$ or, equivalently, 
$\lambda$ is of order one.

For a given Feynman diagram the expansion in $v$ can be 
constructed without first computing the full expression using 
the threshold expansion \cite{Beneke:1997zp}. The method uses 
that every diagram is the sum of terms, for which each loop 
momentum is in one of the following four regions:
\begin{eqnarray}\label{eq:scales}
  \mbox{hard (h)}: & \ell^0 \sim m, & \bm{\ell} \sim m \\ \nonumber
   \mbox{soft (s)}: & \ell^0 \sim m v, & \bm{\ell} \sim m v \\ \nonumber
   \mbox{potential (p)}: & \,\ell^0 \sim m v^2, & \bm{\ell} \sim m v \\ \nonumber
   \mbox{ultrasoft (us)}: & \,\ell^0 \sim m v^2, & \bm{\ell} \sim m v^2\,\,
\end{eqnarray}
When on-shell, only massless particles (gluons, light quarks and 
ghosts) can be ultrasoft, and only the heavy quarks can be 
potential.\footnote{
Here and in the following we set the masses of all quarks other 
than the heavy quark to zero. This is a good approximation 
for top quarks, but less so for bottom quarks, in which case 
the charm mass is of order of the soft scale.} In each region, 
the loop integrand is expanded in the terms which are small in
the corresponding region and the loop integration of the 
expanded integrand is carried out over the complete $d$-dimensional 
space-time volume. The expansion generates ultraviolet and 
infrared divergences which are regulated dimensionally 
($d=4-2\epsilon$) and subtracted according to the $\overline{\rm MS}$ 
prescription. However, the divergences generated by the 
separation of the diagram into regions cancel in the sum 
over all terms.

This procedure is largely equivalent to constructing 
appropriate effective Lagrangians within dimensional regularization, 
but it clarifies 
the correct matching procedure which is subtle in dimensional 
regularization if the effective theory contains more 
than one scale as in non-relativistic 
QCD \cite{Manohar:1997qy,Beneke:1997av}. The threshold expansion 
also synthesizes the non-relativistic velocity power counting 
rules developed for the different modes (momentum regions)
in \cite{Luke:1997ys,Labelle:1996en,Luke:1996hj,Grinstein:1997gv}.
In the construction of the non-relativistic, resummed expansion 
of the pair production cross section we derive effective 
Lagrangians in two steps by integrating out the large momentum modes 
according to the following scheme:

\begin{equation}
\begin{array}{cc}
{\cal L}_{\rm QCD}\,[Q(h,s,p),\,g(h,s,p,us)] \quad& \mu>m \\[0.3cm]
\Big\downarrow & \\[0.3cm]
{\cal L}_{\rm NRQCD}\,[Q(s,p),\,g(s,p,us)] \quad&  m v < \mu < m \\[0.3cm]
\Big\downarrow & \\[0.3cm]
{\cal L}_{\rm PNRQCD}\,[Q(p),\,g(us)] \quad
& \mu < m v\\[0.3cm]
\end{array}
\end{equation}
In square brackets we display the modes of the 
heavy quarks ($Q$) and massless particles ($g$) 
which are still contained in the effective 
Lagrangian; the others are integrated out when the energy 
cut-off $\mu$ is lowered as indicated on the right.
The first step leads to 
NRQCD~\cite{Thacker:1990bm,Lepage:1992tx,Bodwin:1994jh}, 
in which all interactions are local, since only the short-distance 
hard modes have been eliminated. The expansion rules of the threshold 
expansion {\em define} the dimensionally regularized NRQCD Lagrangian. 
The second step whereby soft modes and potential massless modes 
are integrated out was suggested in \cite{Beneke:1997zp,Pineda:1997bj} 
in the context of the effective Lagrangian and threshold 
expansion method. The result is the potential NRQCD (PNRQCD) 
Lagrangian~\cite{Beneke:1999qg,Beneke:1998jj,Pineda:1997bj,Pineda:1997ie,Brambilla:1999xf}. 
The PNRQCD Lagrangian is not local. It contains spatially 
non-local but temporally local, i.e. instantaneous interactions of the 
heavy quarks, since the three-momentum of the potential heavy quark 
field still present in PNRQCD is of the same order as the one of the modes 
integrated out. These interactions provide a precise definition 
of the concept of ``heavy-quark potentials.'' Perturbation theory 
in PNRQCD resembles quantum-mechanical perturbation theory closely, 
since the leading colour-Coulomb interaction is part of the 
unperturbed theory. Thus, the propagator of PNRQCD includes 
the leading Coulomb interaction exactly, which effects the required 
resummation of conventional perturbation theory to all orders.

To illustrate the velocity scaling of Feynman diagrams, we consider 
the power counting for the loop integrand. Eq.~(\ref{eq:scales}) 
implies that the integration measure
$d^4\ell$ scales as $v^0$, $v^4$, $v^5$ and $v^8$, when $\ell$ is hard,
soft, potential and ultrasoft, respectively. 
The denominator of a gluon (massless) propagator 
with momentum $\ell = (\ell^0,\bm{\ell})$ is approximated 
at leading order in a given region by:
\begin{equation}
\label{eq:gluonprop} \frac{1}{\ell^2} = \left\{
\begin{array}{ll}
\displaystyle \frac{1}{\ell^2} & \mbox{  hard }(v^0), \mbox{ soft
}(v^{-2}),
\mbox{ ultrasoft }(v^{-4}) \\[0.5cm]
\displaystyle -\frac{1}{\bm{\ell}^2}+\ldots & \mbox{  potential
}(v^{-2})
\end{array}
\right.
\end{equation}
The velocity scaling is given in brackets. For the heavy-quark propagator
with momentum $q/2+\ell = (m+E/2+\ell^0,\bm{\ell})$ the denominators 
are:
\begin{equation}
\label{eq:quarkprop} \frac{1}{\left(q/2+\ell \right)^2-m^2} = \left\{
\begin{array}{ll}
\displaystyle
\frac{1}{\ell^2+q\cdot \ell} + \ldots & \mbox{  hard }(v^0) \\[0.5cm]
\displaystyle
\frac{1}{2 m}\,\frac{1}{\ell^0} + \ldots & \mbox{  soft }(v^{-1}) \\[0.5cm]
\displaystyle \frac{1}{2 m}\,\frac{1}{E/2+\ell^0-\bm{\ell}^2/(2 m)} +
\ldots & \mbox{  potential }(v^{-2})
\end{array}
\right.
\end{equation}

\begin{figure}
\begin{center}
\includegraphics[width=0.9\textwidth,height=1.8cm]{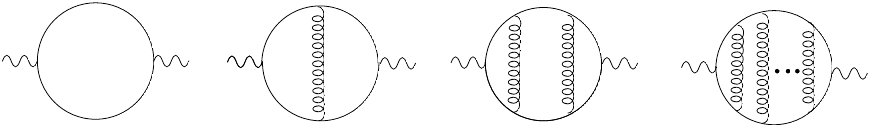}
\end{center}
\vspace*{-0.1cm}
\caption{\label{fig:ladder} Ladder diagrams.}
\end{figure}

\vskip0.2cm
Using these scaling rules, it is easy to see why an all-order 
resummation of Feynman diagrams is required in the threshold 
region. It will become clear from the later systematic derivation 
that the relevant diagrams are the ladder diagrams shown in 
figure~\ref{fig:ladder}, and that the dominant term in the velocity 
expansion arises from the loop momentum region when all loop 
momenta are in the potential region. Adding an additional rung to 
the ladder adds one potential gluon ($1/v^2$) and two potential 
heavy-quark propagators ($1/v^2\times 1/v^2$)  
to the diagram. The numerator of the diagram contains no 
velocity suppression factors, hence accounting for the potential 
loop measure ($v^5$) and strong coupling from the two additional 
vertices ($g_s^2$), we find 
that each rung provides a factor of order $\alpha_s/v$, 
which is unsuppressed in the threshold region. It will be seen 
below that only potential gluon exchange generates this $1/v$ enhancement,
which is equivalent to the statement that only the leading 
Coulomb interaction must be included in the unperturbed effective 
Lagrangian. 

We now return to the discussion of heavy-quark production mechanisms 
not captured by (\ref{eq:sigtt}), which expresses the cross section 
in terms of the heavy-quark current spectral functions. In the case 
of heavy-quark radiation (lower left in figure~\ref{fig:eett}) 
the final state consists of $Q\bar Q q\bar q$, and since the 
available energy at threshold is limited to $m v^2$, the light 
quarks must be ultrasoft. In the three-loop diagram that represents 
the square of the heavy-quark radiation amplitude, the $Q\bar Q$ 
loop must be potential and the two other loops ultrasoft, which 
leads to a factor of $v^{21}$ from the loop integration measure. 
The intermediate gluon and light-quark propagators in the amplitude 
must be hard to produce the $Q\bar Q$ pair and hence do not 
contribute inverse powers of $v$. The two potential heavy quark and 
the two ultrasoft light quark propagators 
($1/\mbox{$\hspace*{-0.03cm}\not\hspace*{-0.04cm} \ell\,$}$) supply 
a factor of $1/v^2$ each, so the heavy-quark radiation 
contribution to the cross section scales at least as 
$\alpha_s^2 v^{13}$ which should be compared to $v$ for the 
leading term. Inspection of the analytic
expression \cite{Hoang:1994it} confirms this result, hence this 
contribution can be safely neglected. In the case of 
singlet production (lower right in figure~\ref{fig:eett}) through 
the coupling of the virtual photon or $Z$-boson to light quarks 
the dominant term comes from three hard loops, leading to 
the counting $\alpha_s^3 v$, which represents a third-order 
correction to the cross section. 
While not part of $\mbox{Im}\,\Pi^{(v)}(q^2)$ this 
mechanism can be included in the three-loop short-distance coefficient 
$c_v^{(3)}$ of the non-relativistic heavy-quark current, 
which is discussed below, although it is not known at present. 
Note that the singlet contribution to $c_v^{(3)}$ is complex, but the 
imaginary part should be discarded, since it corresponds to 
the three-gluon and light-quark cut, which is not part of the heavy-quark 
production cross section. A similar singlet-production diagram exists for 
the axial-vector coupling with only two gluons coupling to the 
light-quark triangle, but due to velocity suppression this 
contribution begins only at the fourth order (N4LO). 

\begin{figure}[t]
\begin{center}
\makebox[0cm]{ \scalebox{0.45}{\rotatebox{0}{
     \includegraphics{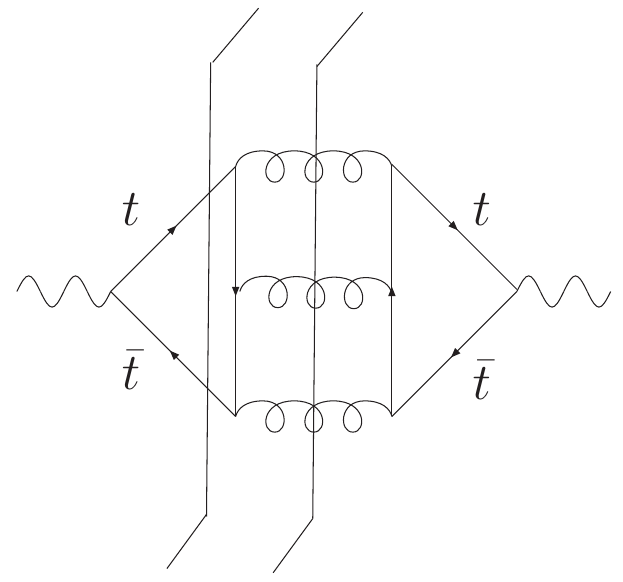}}}}
\end{center}
\vspace*{-0.1cm}
\caption{\label{fig:annihilation} Diagram containing cuts not 
related to top quark production.}
\end{figure}

We have thus argued that the production mechanisms not 
included in $\mbox{Im}\,\Pi^{(v,a)}(q^2)$ are either suppressed 
or easily included at third order. Consider now 
figure~\ref{fig:annihilation}, which shows a diagram contained 
in $\mbox{Im}\,\Pi^{(v)}(q^2)$, but whose three-gluon cut should 
not be part of the heavy-quark cross section. The possible 
loop momentum regions for this diagram are h-h-h-h, p-h-h-p, 
p-h-h-h and h-h-h-p, where the first and last letter refers 
to the left and right heavy quark loop, respectively. In 
the all-hard configuration only the three-gluon cut contributes 
to $\mbox{Im}\,\Pi^{(v)}(q^2)$, so the correct prescription is 
to simply not include this configuration. The p-h-h-p configuration 
may be interpreted as heavy-quark production followed 
by rescattering through annihilation. Annihilation is 
suppressed by $\alpha_s^2 v^2$ relative to Coulomb exchange as 
can be seen by counting loop integration and propagator 
factors, so this configuration is relevant only from fourth order as 
a contribution to the $\alpha_s^3/m^2$ potential. 
The remaining two configurations with only one potential 
heavy quark loop are analogous to the singlet production 
mechanisms. That is, one drops the imaginary part of the h-h-h 
subdiagram, which 
comes from the three-gluon cut and associates its real 
part to $c_3^{(v)}$. 
There exists a three-loop diagram similar to figure~\ref{fig:annihilation} 
but with two gluon lines only for the axial-vector coupling, 
but as in the singlet diagram with a light-quark triangle 
discussed above, the axial-vector 
coupling implies another factor of $v^2$, so this diagram 
is never relevant at third order.

\section{Non-relativistic QCD}
\label{sec:nrqcd} 

In this section we discuss the matching of the  
vector current correlation function for $q^2\approx 4 m^2$ to 
its equivalent representation in non-relativistic QCD. This amounts to 
integrating out the hard modes, which correspond to ``relativistic 
effects'' involving the scale of the heavy-quark mass. 
Non-relativistic QCD is expressed in terms of a two-component 
quark field $\psi$ and the corresponding anti-quark field\footnote{Our 
convention is to use the anti-particle field from the four-component 
Dirac field. Alternatively, we could treat particles and
anti-particles on the same footing and 
introduce a particle field in the anti-triplet colour representation 
for the anti-quark, which corresponds to the charge-conjugate 
of the convention adopted in this paper.} $\chi$ to represent the 
remaining soft and potential fluctuations of the original 
quark field. The effective gluon field
$A_{\mu}=A_{\mu}^A T^A$ can be soft, potential and ultrasoft.

Before going into the details of the Lagrangian and power counting 
we briefly sketch the result. As will be shown below 
the expansion of the vector 
current $j^{(v)\,\mu}$ in terms of the non-relativistic fields is 
given by 
\begin{eqnarray}
\label{eq:QCDVectorCurrent}
j^{(v) \,i}=c_v\, \psi^\dag\sigma^i\chi
+ \frac{d_v}{6m^2}\,\psi^\dag\sigma^i\,{\bf D^2}\chi
+\ldots,
\end{eqnarray}
where the hard matching coefficients $c_v$, $d_v$ have perturbative 
expansions in $\alpha_s$. In the ``rest frame'' 
$q^\mu=(2 m+E,\bff{0})$, eq.~(\ref{eq:pi}) implies $\Pi^{(v)}_{ij} = 
q^2\delta_{ij}\,\Pi^{(v)}(q^2)$, so 
\begin{equation}
\label{eq:pitoNRQCD}
\Pi^{(v)}(q^2) = \frac{1}{(d-1) q^2} \,\Pi^{(v)}_{ii} 
= \frac{N_{c}}{2m^{2}}\,c_v 
\left[c_v-\frac{E}{m}\,\left(c_v+
\frac{\cal{E}}{E}\frac{d_v}{3}\right)\right]
G({\cal E}) + \ldots,
\end{equation}
where ${\cal E} = E+i\Gamma$, and   
the neglected terms on the right-hand side include a subtraction
term that does not contribute to the imaginary part of
$\Pi^{(v)}(q^2)$ as well as terms beyond the third order 
(NNNLO).
The important quantity is the 
two-point function of the non-relativistic current
\begin{equation}
\label{eq:G}
G(E) =\frac{i}{2 N_c (d-1)} \int d^{d} x\, e^{iEx^0}\,
\langle 0| \,T(\,
[\chi^{\dag}\sigma^i\psi](x)\,
[\psi^{\dag}\sigma^i\chi](0))
|0\rangle_{| \rm NRQCD}\,,
\end{equation}
where now the matrix element must be evaluated in 
non-relativistic QCD (NRQCD). The terms proportional to $E$ in 
(\ref{eq:pitoNRQCD}) arise from expanding the prefactor $1/q^2$ and 
from the $1/m^2$ suppressed current in (\ref{eq:QCDVectorCurrent}), 
whose matrix element can be reduced to the one of the leading 
current by an equation-of-motion relation derived 
later.\footnote{\blue{By including the factor ${\cal E}/E$ in 
(\ref{eq:pitoNRQCD}) we extend the validity of 
the above equation to the case when the top decay width is accounted 
for by the substitution $E\to E+i\Gamma$. In this case, the factor 
$E$ in front of $c_v$ arises from the expansion of 
$q^2=(2 m+E)^2$, where $E$ is real, while the factor ${\cal E}$ 
in front of $d_v$ arises from the equation of motion identity, 
for which the complex energy 
value ${\cal E}$ must be used.}
\label{footnoteonE}} Thus the main
ingredients to the non-relativistic representation 
are the calculation of $G(E)$ and the current 
matching coefficients.

Similar relations hold for the axial-vector contribution to the 
cross section (\ref{eq:sigtt}), which arises from $Z$-boson exchange. 
The axial-vector current $j^{(a)\,\mu} = \bar t\gamma^\mu\gamma_5 t$
is represented in NRQCD by the expansion  
\begin{eqnarray}
\label{eq:QCDAxialVectorCurrent}
j^{(a) \,i}=\frac{c_a}{2 m}\,\psi^\dag\Big[\sigma^i,
(-i)\bm{\sigma}\cdot\bff{D}\Big]\chi
+\ldots,
\end{eqnarray}
with hard matching coefficient $c_a$. As is the case for the vector
current, only the spatial components of the current contribute to 
the cross section, since the lepton tensor from the $e^+ e^-$ initial 
state is transverse to both initial state momenta when the 
electron mass is neglected. Only the leading term in the $1/m$
expansion is needed for NNNLO accuracy, since the derivative in 
the leading current implies the well-known P-wave velocity 
suppression. The QCD correlation function is then given by the expression
\begin{eqnarray}
\label{eq:piAtoNRQCD}
\Pi^{(a)}(q^2) &=& \frac{1}{(d-1) q^2} \,\Pi^{(a)}_{ii} 
\\
&& \hspace*{-1.6cm} = \,\frac{N_{c}}{8m^{4}}\,c_a^2\times
\frac{i}{2 N_c (d-1)} \int d^{d} x\, e^{iEx^0}\,
\langle 0| \,T(\,
[\psi^{\dag}\Gamma^i\chi]^\dagger(x)\,
[\psi^{\dag}\Gamma^i\chi](0))
|0\rangle_{| \rm NRQCD}
+ \ldots,\qquad
\nonumber
\end{eqnarray}
where $\Gamma^i = (-i) [\sigma^i,
\bm{\sigma}\cdot\bff{D}]$. 
\blue{The NNNLO result for the P-wave contribution is 
available from \cite{Beneke:2013kia} and will not be discussed further 
in this paper and paper~II.}

\subsection{Lagrangian and Feynman rules}
\label{sec:nrqcdlagrangian}

For the present purpose the non-relativistic effective 
Lagrangian can be divided into five parts,
\begin{eqnarray}
 \label{nrqcd}
{\cal L}_{\rm NRQCD} &=& {\cal L}_{\psi}+{\cal L}_{\chi}+{\cal
L}_{\psi\chi}+{\cal L}_{g}+{\cal L}_{\rm light}.
\end{eqnarray}

The gluon field is contained in the gauge-covariant 
derivative $D^\mu=\partial^\mu-i g_s A^\mu$, 
field strength tensor $G_{\mu\nu}=(i/g_s)\,\left[D_\mu,D_\nu\right]$, 
and the chromoelectric and chromomagnetic fields defined as
\begin{eqnarray}
E^i &\equiv& G^{i0}=-\nabla^i A^0-\frac{\partial}{\partial t}\,A^i-
ig_s\left[A^i,A^0\right], \nonumber
\\
\bm{\sigma}\cdot\bff{B} &\equiv& -\frac{1}{2}\,\sigma^{ij} G^{ij},
\label{eq:magneticfield}
\end{eqnarray}
where $\sigma^{ij} = (-i/2) \,[\sigma^i,\sigma^j]$ and $D^i =
-\nabla^i$. With these
definitions the bilinear heavy-quark Lagrangian is given by\footnote{
Note that in four dimensions 
$\sigma^{ij}\{D^i,E^j\} = 
\bm{\sigma} \cdot (\bm{D}\times\bm{E}-\bm{E}\times \bm{D})$.
The $d_3$ term is misprinted in eq.~(8) of \cite{Beneke:1999qg}, 
where $\psi^\dagger \sigma^{ij} [D^i, E^j]\psi$ should read 
$\psi^\dagger \sigma^{ij} \{D^i, E^j\}\psi$.
}
\begin{eqnarray}
\label{eq:nrqcdqq}
{\cal L}_{\psi}&=&
\psi^{\dag} \bigg( i D^0 + \frac{{\bf D}^2}{2m} + 
\frac{{\bf D}^4}{8 m^3} \bigg)\psi
-\frac{d_1 g_s}{2m}\psi^{\dag} {{\bm{ \sigma}} \cdot {\bf B}}\,\psi
\nonumber \\[0.1cm] 
&& +\,
\psi^{\dag}\bigg(\frac{d_2 g_s}{8 m^2}[D^i,E^i] +
 i \, \frac{d_3 g_s}{8 m^2}\sigma^{ij}\{D^i,E^j\}\bigg) \psi
\nonumber \\[0.1cm] 
\nonumber &&+ \,\psi^{\dag}\bigg(-d_W g_s\frac{\{{\bf D}^2,{\bm{ \sigma}} 
\cdot {\bf
B}\}}{8m^3} +d_A g_s^2 \frac{{\bf B}^2-{\bf E}^2}{8m^3}+ d_B
g_s^2\frac{\sigma^{ij}(B^i B^j- E^i E^j)}{8m^3}\,\bigg) \psi
\nonumber\\[0.1cm]
&& +\,O\bigg(\frac{1}{m^4}\bigg),
\\[0.2cm] 
{\cal L}_{\chi}&=& - {\cal L}_{\psi}\quad\mbox{with 
$\psi\to\chi, i D^0\to -i D^0, 
E^i\to - E^i$}\,.
\end{eqnarray}
Note that in $d$ dimensions we cannot define the $B^i$ individually,
since they do not represent the components of a $d-1$ dimensional
vector. However, all we need are scalars such as
(\ref{eq:magneticfield}) and ${\bf B}^2$, $\sigma^{ij}B^i B^j$, which
can be consistently defined through the $d$ dimensional field strength
tensor: 
\begin{equation}
{\bf B}^2\equiv\frac{1}{2} \,G^{ij} G^{ij},
\quad\qquad 
\sigma^{ij}B^i B^j \equiv -\frac{1}{2}\sigma^{ij} [G^{ik},G^{kj}]\,. 
\end{equation}

\begin{figure}[p]
\vspace*{-4.5cm}\hskip-1cm\includegraphics[height=25.5cm]{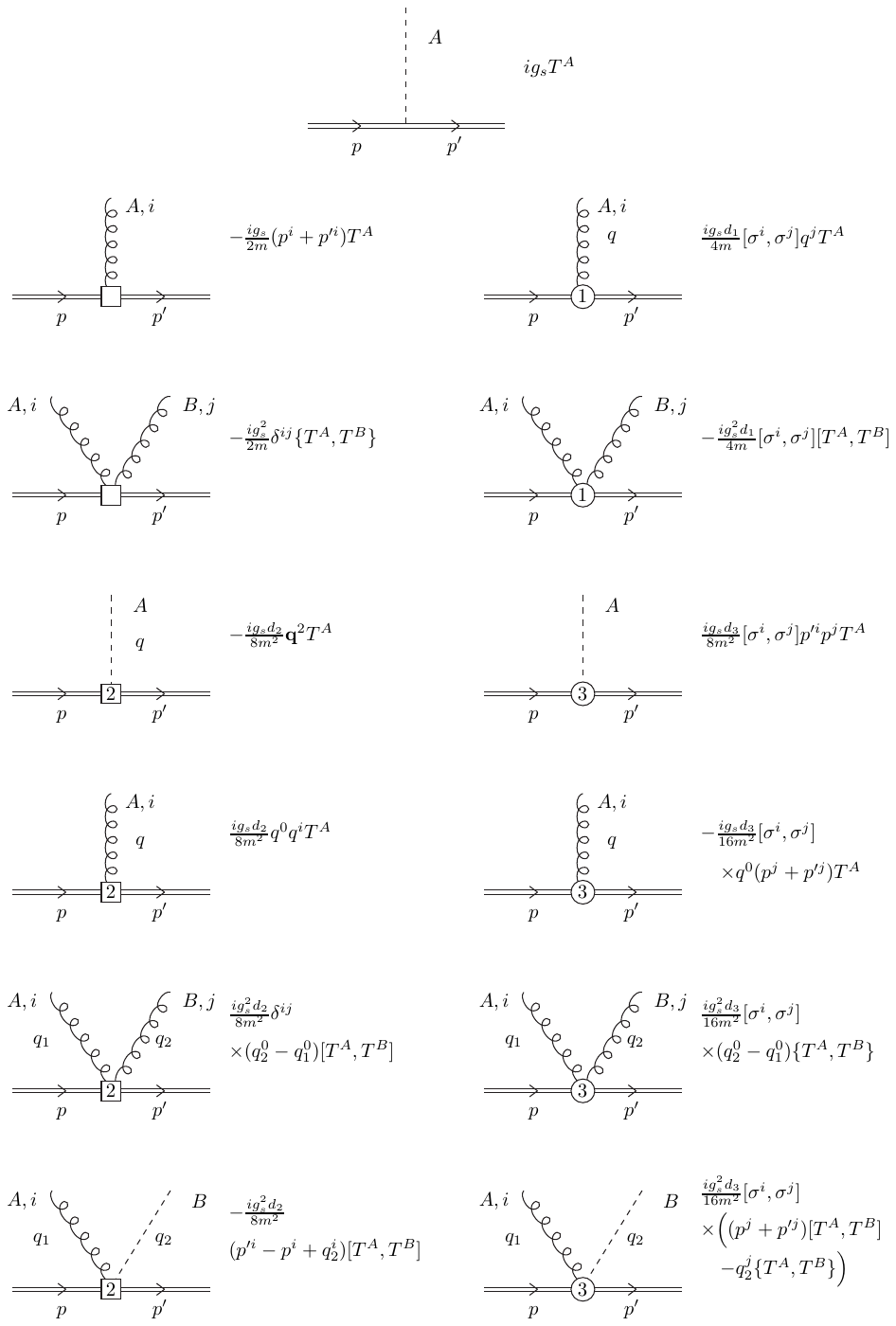}
\vskip-0.7cm
\caption{\label{fig:NRQCD_FR} NRQCD Feynman rules for 
two-quark vertices up to order $1/m^2$. Dashed (curly) lines denote 
the $A^0$ ($A^i$) gluon field. $q = p^\prime -p$.}
\end{figure}

\vskip-0.1cm\noindent
The four-fermion quark-antiquark terms in the effective Lagrangian read
\begin{eqnarray}
{\cal L}_{\psi\chi}&=& \frac{d_{ss}}{m^2}\psi^{\dag} \psi \,
\chi^{\dag} \chi - \frac{d_{sv}}{8 m^2}\psi^{\dag} [\sigma^i,\sigma^j]
\psi \, \chi^{\dag}[\sigma^i,\sigma^j]  \chi +
\nonumber
\\ \label{eq:nrqcdqqqq} && + \,\frac{d_{vs}}{m^2}
\psi^{\dag} T^a \psi \, \chi^{\dag} T^a \chi -
\frac{d_{vv}}{8 m^2} \psi^{\dag} T^a [\sigma^i,\sigma^j] \psi \,
\chi^{\dag} T^a [\sigma^i,\sigma^j]
\chi+O\bigg(\frac{1}{m^3}\bigg),
\end{eqnarray}
where the factor $-1/8$ in the definition of the spin-triplet 
operators has been inserted, since in four dimensions 
$[\sigma^i,\sigma^j]\otimes [\sigma^i,\sigma^j] = 
-8 \,\sigma^i\otimes\sigma^i$. 
The pure gluon Lagrangian takes the form
\begin{eqnarray}
 \label{eq:nrqcdg}
{\cal L}_{g}&=&-\frac{d_4}{4}G^{A}_{\mu \nu}G^{A\mu
\nu}+\frac{d_5}{m^2}G^{A}_{\mu \nu}D^2G^{A\mu
\nu}+\frac{d_6}{m^2}g_sf^{ABC}G^{A}_{\mu \nu}G^{B\mu}_{\phantom{B\mu}
\alpha}G^{C\nu \alpha}+O\bigg(\frac{1}{m^4}\bigg).
\qquad
\end{eqnarray}
Finally, ${\cal L}_{\rm light}$ is the same as the light-quark Lagrangian 
in full QCD. The Feynman rules for the $1/m$ and $1/m^2$ terms 
in ${\cal L}_{\psi}$ are given in figure~\ref{fig:NRQCD_FR}.\footnote{
The vertices involving $q^0$ in the fifth row can be eliminated 
using the heavy-quark equation of motion. This generates $1/m^3$ 
terms not shown in the figure and modifies the $1/m^2$ four-point 
vertices in the last row of the figure as follows: drop the 
$\bff{p}$ and $\bff{p^\prime}$ terms and multiply the remaining 
terms by two.}
The hard matching coefficients $d_{1-4}, d_W, d_A, d_B$ equal one 
at tree level, while the others vanish. All coefficients 
obtain one-loop corrections, which must be determined by matching the
QCD diagrams to the NRQCD diagrams order by order in the
non-relativistic expansion. For reasons which will be explained
later, the matching coefficients are needed to order $\epsilon$ 
in the dimensional regularization parameter. In writing 
(\ref{eq:nrqcdqq}) we did not include $1/m^3$ interactions 
with vanishing tree-level coefficients as well as mixed heavy-light 
quark operators of the 
form $\psi^\dagger\psi \bar q q$ in ${\cal L}_\psi$, since 
they are irrelevant for the NNNLO calculation, 
as we discuss now. 
\blue{The complete Lagrangian bilinear in the fermion fields at 
$O(1/m^3)$ can be found in \cite{Manohar:1997qy}.}

The velocity scaling of the fields depends on the momentum region. 
Heavy quark fields can be potential or soft. From the scaling
rules of the propagator (compare (\ref{eq:quarkprop})) and
the integration measure, it follows that in both cases the field
scales as $v^{3/2}$. The same power counting can be done with the
gluon field, which is either soft, potential or ultrasoft and one 
finds that $g_sA$ scales as $v^{3/2}$, $v^{2}$ and $v^{5/2}$
respectively, which includes a factor of $v^{1/2}$ from the coupling
constant $g_s$.

Turning to the effective Lagrangian, 
we first consider the bilinear terms in the heavy Lagrangian 
${\cal L}_\psi$. For potential quarks, the bilinear terms in the 
kinetic term $\psi^{\dag} ( i
D^0 + {\bf D}^2/(2m))\psi$ are both of order $v^5$. 
The relativistic correction $\psi^{\dag}( \bff{
\partial}^4/(8m^3))\psi$ scales as $v^7$. Being suppressed by 
$v^2$ relative to the leading terms it contributes from NNLO. The
next term in the expansion of the relativistic energy-momentum relation 
would be of order $v^9$ and is beyond
NNNLO. For soft quarks, the quadratic kinetic energy term is an
order $v$ correction to the leading-order static Lagrangian
$\psi^\dagger i \partial^0 \psi$, which scales as $v^4$. 
This explains why static heavy-quark propagators can be used in 
the calculation of the heavy-quark potentials. For soft heavy quarks the
quartic kinetic energy correction is a NNNLO effect. 

Consider now the interactions of the heavy quark with the gauge field, 
i.e. terms of the form $\psi^\dagger\psi (g_s A)^n$, potentially with 
derivatives on the
quark and gluon fields. The $g_s\psi^\dagger\psi A^0$ interaction
that arises from $\psi^\dagger iD^0 \psi$ scales as $v^5$ when all
fields are potential. Therefore it is not suppressed relative to the
bilinear terms that persist as $g_s\to 0$. In the potential region, this
interaction has to be treated non-perturbatively; this is why the
cross section near threshold requires a summation of some loop 
momentum contributions to all orders in $\alpha_s$. When the gluon 
field is soft, the velocity scaling of
the $g_s\psi^\dagger\psi A^0$ interaction is $v^{9/2}$, but since in
this case the leading term $\psi^\dagger i\partial^0\psi$ scales as
$v^4$, this interaction is now a perturbation. It follows that all soft
interactions can be treated in conventional perturbation theory.
Further three-point interactions of the form above carry
derivatives. Each derivative gives at least a suppression of $v$ 
so interactions with up to three derivatives in $g_s\psi^\dagger\psi A$ 
may contribute at NNNLO. The requirement of gauge invariance restricts 
the possible interaction terms to the so-called chromomagnetic interaction 
$g_s \psi^{\dag} {{\bm{ \sigma}} \cdot {\bf B}}\psi$ at order $v^6$ (NLO), 
and the Darwin and spin-orbit interactions at order $v^7$ (NNLO), 
multiplied by the short-distance
coefficients $d_1$, $d_2$ and $d_3$, respectively. However, a single  
chromomagnetic interaction contributes only in connection with a 
$v$ suppressed quark-gluon vertex from the $\psi^\dagger \,{\bf D}^2/(2
m)\psi$ interaction, and not at all to the current 
correlation function, since the trace over an odd number of Pauli 
matrices vanishes. Thus the chromomagnetic, Darwin and 
spin-orbit interaction all start to contribute at NNLO (with 
two insertions of the chromomagnetic term). Hence, at NNNLO one 
needs the coefficient functions $d_1$, $d_2$ and $d_3$ in the 
one-loop approximation. Beyond order $v^7$ only the interactions with 
non-vanishing tree-level coefficient functions given in (\ref{eq:nrqcdqq})
can potentially contribute to NNNLO. None of them does, however, 
since single insertions of interactions with Pauli matrices 
vanish, as discussed above, while the terms with two electric or 
magnetic field strengths are of the form $\psi^\dagger 
\psi\,(g_s A)^2 $ with two derivatives, which is smaller than NNNLO 
in both, the potential and soft region.

Bilinear heavy quark operators in conjunction with light quarks, 
$\psi^\dagger\psi \bar q q$, are of order $v^6$. For these operators 
to contribute to the heavy-quark current correlation function 
at least one interaction of the form $g_s\bar q q A$ is required, 
which costs a factor of $v$ or, equivalently $\alpha_s$. Thus, 
a $\psi^\dagger\psi \bar q q$ operator is relevant at NNNLO, if its 
short-distance coefficient function is of order $\alpha_s$. 
The only operator that may have a tree-level coefficient is 
$\psi^\dagger T^A \psi \,\bar q \gamma^0 T^A q$, since in this 
case the intermediate potential gluon propagator can be 
cancelled, making the operator local. This operator is generated 
at tree level from the Darwin term in the Lagrangian (\ref{eq:nrqcdqq}) 
by the use of the field equation for the chromoelectric field. 
Our convention is that we 
do not eliminate the Darwin term by the field equation, hence 
$\psi^\dagger\psi \bar q q$ operators must be added to the Lagrangian 
only with coefficients of order $\alpha_s^2$ producing corrections 
to the heavy-quark current correlation function beyond NNNLO.

We therefore conclude that only the terms in the first two lines 
of (\ref{eq:nrqcdqq}) are needed for the NNNLO calculation, the 
same as in NNLO. The only 
difference to NNLO is that the short-distance coefficients  $d_1$, $d_2$ 
and $d_3$ are required at the one-loop order. We also see that 
only the  $g_s\psi^\dagger\psi A_0$ interaction in the potential 
region is non-perturbative, 
and this explains why only ladder diagrams of Coulomb gluons must 
be summed to all orders.

The four-quark operator Lagrangian ${\cal L}_{\psi\chi}$ 
(\ref{eq:nrqcdqqqq}) is generated 
by hard scattering of quarks and anti-quarks, or by quark anti-quark 
annihilation. Hard scattering with momentum exchange of order $m$ 
requires the exchange of at least two 
gluons, corresponding to one-loop diagrams so the coefficient 
functions are of order $\alpha_s^2$.\footnote{Once again, this holds 
only because we do not eliminate the Darwin term by the chromoelectric 
field equation, which would otherwise generate a local four-quark 
operator with a tree-level coefficient function.} 
The four-quark operator counts 
as $v^6$. Including the one-loop coefficient function gives the 
counting $\alpha_s^2 v^6$, which is a NNNLO effect relative to 
the leading-order Lagrangian of order $v^5$ for the scattering of 
potential quarks. The annihilation 
contribution is present already at order $\alpha_s$ (tree-level), 
but the operator has the colour structure $\psi^\dagger T^a
\chi\,\chi^\dagger T^a\psi$, which does not contribute 
to the current correlation function due to $\mbox{tr}\,T^a=0$. 
Similarly, annihilation 
into two gluons does not contribute to the vector correlation 
function, since it leads to fermion loops with three vector couplings 
that vanish by charge conjugation. Thus, at NNNLO, we can restrict 
ourselves to the four-fermion operators generated by hard quark 
anti-quark scattering. For this reason we write the operators 
in the ``scattering ordering'' $(\psi^\dagger\psi) (\chi^\dagger\chi)$ 
rather than the ``annihilation ordering''   
$(\psi^\dagger\chi) (\chi^\dagger\psi)$. Although the two orderings 
are related by a Fierz transformation in four space-time dimensions, 
the two are inequivalent in dimensional regularization. In general, 
we would have to introduce the difference of the two as evanescent 
operators. This complication is avoided here, since there are 
no annihilation contributions at NNNLO. Adopting the ``scattering 
ordering'' in the Lagrangian, we do not need to perform any 
Fierz transformations.

The necessity to avoid relations that hold only in four dimensions is 
also the reason for introducing the definition (\ref{eq:magneticfield}) 
that does not make use of the three-dimensional $\epsilon^{ijk}$ 
symbol, which is not defined in $d-1$ dimensions. In particular, 
the commutator
\begin{equation}
\sigma^{ij} = \frac{1}{2 i}\left[\sigma^i,\sigma^j\right]\, ,
\end{equation}
must be considered as an independent element of the $d$-dimensional 
algebra of Pauli matrices. This poses no difficulty in the 
calculation of the vector current correlation function, since 
in the end all expressions can be evaluated using the $d$-dimensional 
identities
\begin{eqnarray}
\sigma^i\sigma^i &=& d-1, \\
\sigma^i\sigma^j\sigma^i &=& (3-d)\,\sigma^j,
\end{eqnarray}
where $d=4-2\epsilon$ and $\mbox{tr}\,(1)=2$.

The pure gauge field Lagrangian (\ref{eq:nrqcdg}) follows from 
integrating out heavy-quark loops with small momentum gluon 
lines attached. The renormalization of the standard kinetic term 
by the coefficient $d_4$ is well-known to be related to the 
matching of the strong coupling from the $n_f+1$ flavour 
theory including the heavy quark to the theory with $n_f$ light flavours. 
In the following we express all results in terms of the 
strong coupling in the $\overline{\rm MS}$ scheme in the 
$n_f$ flavour theory, which is the appropriate coupling for 
calculations in NRQCD, where the heavy quark short-distance 
fluctuations have been integrated out. 
After redefining the strong coupling, $d_4$ should be set to 
one in (\ref{eq:nrqcdg}), so that the kinetic term is canonically 
normalized.
The next term, $G^{A}_{\mu \nu}D^2G^{A\mu\nu}$, in the gauge field 
Lagrangian involves two derivatives and a coefficient function 
$d_5\propto\alpha_s\sim v$. Therefore it must be included at NNNLO. 
On the other hand, the term involving three gluon field strengths 
can be neglected at this order. 

Having collected the relevant terms in the effective Lagrangian, 
we are now in the position to discuss the matching calculations. 
Most of the results required at NNNLO are available in the 
literature. However, many of the matching coefficients multiply 
NRQCD correlation functions, which exhibit $1/\epsilon$ poles. 
Thus, as will be explained in section~\ref{sec:matchcurrent} below,
we also need the $O(\epsilon)$ terms of the matching 
coefficients, which have not been calculated or presented up 
to now. We therefore had to repeat these matching calculations 
and extend them to the $O(\epsilon)$ terms. 
The matching calculation is performed in the center-of-mass frame, so that
the three-momenta of the heavy quark and anti-quark are of opposite
sign. The external heavy quark spinors in QCD are given by
\begin{equation}
\label{eq:spinors} u(p) =
\frac{1}{(E_p+m)^{1/2}}\left(\begin{array}{c} (E_p+m)\,
\xi \\[0.2cm] \bm{\sigma}\cdot\bff{p}\,\xi \end{array}\right),
\quad v(p) = \frac{1}{(E_p+m)^{1/2}}\left(\begin{array}{c}
\bm{\sigma}\cdot\bff{p}\,\eta \\[0.2cm]  (E_p+m)\,
\eta \end{array}\right),
\end{equation}
for external momentum $p=(E_p,\bff{p})$ with 
$E_p\equiv (m^2+\bff{p}^2)^{1/2}$.
The variables $\xi$ and $\eta$ denote the quark and anti-quark
two-spinors, respectively. They are normalized according to
$\xi^\dagger\xi= \eta^\dagger\eta =1$.

\subsection{Bilinear heavy-quark operators}

\begin{figure}[t]
\begin{center}
\includegraphics[height=3cm]{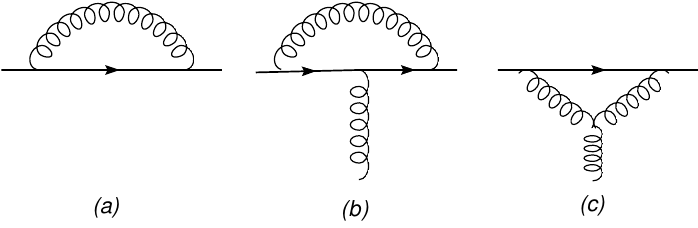}
\vspace*{-0.5cm}
\end{center}
\caption{\label{fig:formfactors} One-loop form factor diagrams: 
wave-function renormalization and vertex corrections.}
\end{figure}

The coefficient functions of the interactions of heavy quarks 
with a single gauge field in ${\cal L}_\psi$ can be deduced from the 
heavy quark form factors in background field gauge. The finite 
part of the one-loop form factors was 
calculated in \cite{Manohar:1997qy}. However, the
order $\epsilon$ coefficients were not computed there. The on-shell form 
factors can be brought into the general form
\begin{eqnarray}\label{eq:formfactdef}
ig_s T^a
\bar{u}(p^{\prime})\bigg[\gamma^{\mu}F_1(q^2)+\frac{i\sigma^{\mu
\nu}q_{\nu}}{2m}F_2(q^2)\bigg]u(p),
\end{eqnarray}
where $q=p^\prime -p$. For the one-loop diagrams shown in 
figure \ref{fig:formfactors} we obtain the following result for 
the expansion of the ultraviolet-renormalized 
form factors in $\alpha_s$ and $q^2$ up to 
order $\alpha_s q^2/m^2$:
\begin{eqnarray}
\nonumber
F_1 &=&1+\frac{\alpha_s}{\pi}\bigg(\frac{\mu}{m}\bigg)^{2\epsilon}
\frac{\Gamma(\epsilon)e^{\gamma_E\epsilon}}{48(4\epsilon^2-1)}\frac{
q^2}{m^2}\Bigg[C_A(-12\epsilon^3+4\epsilon^2+3\epsilon+5)
\nonumber\\
&& +\,2C_F(12\epsilon^3-4\epsilon^2+3\epsilon+4)\Bigg],
\\ \nonumber
F_2 &=&\frac{\alpha_s}{\pi}\bigg(\frac{\mu}{m}\bigg)^{2\epsilon}
\frac{\Gamma(\epsilon)e^{\gamma_E\epsilon}}{24(4\epsilon^2-1)}
\Bigg[C_A\bigg(6(2\epsilon+1)(2\epsilon^2-1)+\frac{
q^2}{m^2}(4\epsilon^4+8\epsilon^3+5\epsilon^2-2\epsilon-6)\bigg)
\\ 
&&  +\,C_F\bigg(\!-12\epsilon(2\epsilon+1)^2-\frac{
q^2}{m^2}2\epsilon(\epsilon+1)(2\epsilon+1)^2\bigg)\Bigg].
\end{eqnarray}
Here and below we use the standard colour factors $T_F=1/2$, 
$C_F=4/3$, $C_A=3$.
The remaining divergences are infrared divergences of the on-shell 
form factors. By calculation of the corresponding form factors in the 
effective theory we obtain the relations between the coefficients $d_i$ 
and the form factors:
\begin{eqnarray}
d_1&=&F_1(0)+F_2(0),
\\
d_2&=&F_1(0)+2F_2(0)+8F^{\prime}_1(0),
\\
d_3&=&F_1(0)+2F_2(0),
\end{eqnarray}
where $F_i(0)=F_i|_{q^2=0}$ and
$F^{\prime}_1(0)=dF_1/d(q^2/m^2)|_{q^2=0}$. Since $F_1(0)=1$ exactly, 
this implies the well-known relation $d_3=2 d_1-1$. The $\overline{\rm MS}$ 
renormalized coefficient functions follow by subtracting the 
$1/\epsilon$ poles from the above expressions. The 
$O(\epsilon)$ terms of the above expressions 
are in agreement with \cite{Wuester:2003}. \blue{We note that 
the two-loop expressions for $d_1$ \cite{Grozin:2007fh} 
(hence also $d_3$) and $d_2$ \cite{Gerlach:2019kfo} are also known, 
but contribute only from N4LO.}

\subsection{Gauge field operators}

To obtain the bilinear pure gauge field Lagrangian ${\cal L}_g$, the 
gluon self energy has to be matched. In the one-loop order the 
relevant diagram is the heavy-quark loop, which gives:
\begin{eqnarray}
d_4&=&1+\frac{\alpha_s}{\pi}\bigg(\frac{\mu}{m}\bigg)^{2\epsilon}
\frac{T_F\Gamma(\epsilon)e^{\gamma_E\epsilon}}{3},
\\
d_5&=&\frac{\alpha_s}{\pi}\bigg(\frac{\mu}{m}\bigg)^{2\epsilon}
\frac{T_F\Gamma(1+\epsilon)e^{\gamma_E\epsilon}}{60}.
\end{eqnarray}
The results agree for $d=4$ with the ones in \cite{Manohar:1997qy}.
As mentioned above, the operator with coefficient $d_6$ is not needed, 
because it can contribute to the heavy quark correlation function 
only with an additional loop, which is beyond NNNLO. Recall that 
$d_4=1$ should be used after having normalized the fields canonically.

\subsection{Four-fermion operators}

\begin{figure}[t]
\begin{center}
     \includegraphics[height=1.8cm]{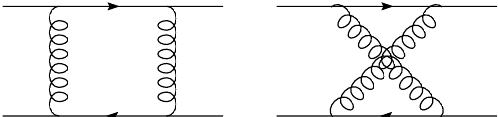}
\end{center}
\caption{\label{fig:4quark} QCD diagrams for the four-fermion operators
at one loop.}
\end{figure}

The remaining part of the Lagrangian to be matched to QCD 
is the one containing the four-fermion operators. As discussed in 
section~\ref{sec:HQCF} we do not need the annihilation contributions 
and therefore restrict ourselves to the scattering diagrams 
shown in  figure~\ref{fig:4quark}. The results in $d=4$ can be 
obtained from the equal mass limit of the unequal mass case 
given in \cite{Pineda:1998kj}. Here we present the $d$-dimensional 
matching coefficients:
\begin{eqnarray}
d_{ss}&=&\alpha_s^2C_F(C_A-2C_F)\bigg(\frac{\mu}{m}\bigg)^{2\epsilon}
\frac{e^{\gamma_E\epsilon}(2\epsilon-3)(2\epsilon^2+\epsilon+1)
\Gamma(2+\epsilon)}
{2\epsilon(8\epsilon^3+12\epsilon^2-2\epsilon-3)},
\\
d_{sv}&=&\alpha_s^2C_F(C_A-2C_F)\bigg(\frac{\mu}{m}\bigg)^{2\epsilon}
\frac{e^{\gamma_E\epsilon}\Gamma(1+\epsilon)}
{2(1+2\epsilon)},
\\
d_{vs}&=&\alpha_s^2\bigg(\frac{\mu}{m}\bigg)^{2\epsilon}
\frac{e^{\gamma_E\epsilon}(3-2\epsilon)
(C_A(\epsilon(2\epsilon+1)(4\epsilon+3)+5)-
8C_F(1+\epsilon)(2\epsilon^2+\epsilon+1))\Gamma(\epsilon)}
{4(2\epsilon-1)(2\epsilon+1)(2\epsilon+3)}, 
\nonumber\\
\\
d_{vv}&=&\alpha_s^2\bigg(\frac{\mu}{m}\bigg)^{2\epsilon}
\frac{e^{\gamma_E\epsilon}(-C_A (1+4\epsilon)+8C_F\epsilon)
\Gamma(\epsilon)}{4(1+2\epsilon)}.
\end{eqnarray}
This agrees with the finite part of the unequal mass case in
\cite{Pineda:1998kj}. The 
$O(\epsilon)$ terms of the above expressions 
are in agreement with \cite{Wuester:2003}.\footnote{The 
short-distance coefficients $d_{sv}$ and $d_{vv}$ defined in 
\cite{Wuester:2003} are $(1-\epsilon)$ times those above.}
\blue{We note that the two-loop expressions for the above 
heavy four-quark operator coefficients
are also known \cite{Gerlach:2019kfo}, 
but contribute only from N4LO.}

\subsection{Matching of the vector current} 
\label{sec:matchcurrent}

We finally need an expression for the heavy quark vector 
current $j_{\mu}^{(v)}$ to NNNLO accuracy in the effective 
theory. The perturbative matching coefficients of the NRQCD 
currents come from diagrams where the hard loop connects to one of the
external current vertices. Since the zero component of the 
vector current is irrelevant, we focus on matching the 
operator $\bar t\gamma^i t$. 

At leading order in the velocity expansion the unique NRQCD 
vector current is $\psi^\dagger\sigma^i\chi$ with coefficient 
function $c_v$ as given by the first term on the right-hand side 
of (\ref{eq:QCDVectorCurrent}). The precise definition of the 
matching coefficient is \cite{Beneke:1997jm}
\begin{equation}
\label{eq:newmatch}
Z_{2,{\rm QCD}}\,\Gamma_{\rm QCD} =
c_v\,Z_{2,{\rm NRQCD}}\,Z_J^{-1}\,\Gamma_{\rm NRQCD} \,,
\end{equation}
where $Z_2$ are the on-shell wave function renormalization constants
in QCD and NRQCD, respectively. $\Gamma$ represents the amputated, bare 
electromagnetic current vertex function evaluated for on-shell 
heavy quarks directly at threshold, i.e.~with zero relative momentum, 
expressed in terms of the renormalized QCD coupling and pole mass. 
In dimensional regularization, $Z_{2,{\rm NRQCD}}=1$, and 
$\Gamma_{\rm NRQCD}$, the corresponding NRQCD vertex function,  
equals its tree-level expression 
$\xi^\dagger\sigma^i\eta$, since the NRQCD integrals for 
zero external relative momentum are scaleless. Here it is important 
that the threshold expansion is employed to define NRQCD in dimensional 
regularization. Thus, $c_v$ 
equals the UV renormalized on-shell QCD vertex directly at 
threshold with infrared divergences subtracted recursively 
by the NRQCD renormalization factor $Z_J$. This definition 
is equivalent at order $\alpha_s^n$ to extracting the purely 
hard (h-h-\ldots -h) momentum regions in the threshold 
expansion of the $n$-loop vertex function with external heavy quark 
momenta in the potential region. 
\blue{The definition also requires the use of the $d$-dimensional 
coefficients of the NRQCD Lagrangian as discussed at the end 
of this subsection.}

The coefficient $c_v$ is needed to three-loop accuracy to achieve 
NNNLO precision. 
\blue{While the two-loop expression has been known for some 
time \cite{Beneke:1997jm,Czarnecki:1997vz}, as have been the three-loop 
terms with at least one fermion loop \cite{Marquard:2006qi,Marquard:2009bj}
and the logarithmic terms related to the 
anomalous dimension of the current and strong coupling 
renormalization~\cite{Kniehl:2002yv,Hoang:2003ns,Beneke:2007pj}, 
the full three-loop 
correction has been computed relatively  
recently~\cite{Marquard:2014pea,Egner:2022jot}, except for
the ``singlet diagrams'' where the fermion loop attaches to the 
external vertex (see lower right 
figure~\ref{fig:eett}).} Defining
\begin{equation}
L_m = \ln(\mu/m),
\end{equation}
the coefficients of perturbative expansions of any quantity $S$ 
in $\alpha_s = \alpha_s(\mu)$ through
\begin{equation}
S=1+ \sum_{n} S^{\,(n)} \left(\frac{\alpha_s}{4\pi}\right)^n,
\end{equation}
and the coefficients of the QCD $\beta$ function in the 
$\overline{\rm MS}$ scheme with $n_f$ light flavours (not including 
the heavy quark) \cite{Tarasov:1980au}
\begin{eqnarray}
\beta_{0} &= &\frac{11}{3}C_{A}-\frac{4}{3}T_{F}n_f,
 \nonumber\\ 
\beta_{1} &=&
\frac{34}{3}C_{A}^{2}-\frac{20}{3}C_{A}T_{F}n_f-4C_{F}T_{F}n_f,
\nonumber\\ 
\nonumber \beta_{2} &=&
\frac{2857}{54}C_{A}^{3}-\frac{1415}{27}C_{A}^{2}T_{F}n_f
-\frac{205}{9}C_{A}C_{F}T_{F}n_f +2C_{F}^{2}T_{F}n_f
\\ && +\frac{158}{27}C_{A}T_{F}^{2}n_f^{2}
+\frac{44}{9}C_{F}T_{F}^{2}n_f^{2},
\end{eqnarray}
the known results for the vector current matching coefficients 
are given by: 
\begin{eqnarray}
c_v^{(1)}(\mu) 
&=&
-8C_F - \blue{\epsilon\,\big[16 C_F L_m\big] + O(\epsilon^2)} ,
\label{eq:cv1} \\[0.2cm]
c_v^{(2)}(\mu)
&=&
2 \beta_0 L_m  c_v^{(1)}(m) 
+ L_m 
C_F \pi^2
\bigg[- \frac{16}{3} C_F-8 C_A \bigg]
+c_v^{(2)}(m),
\\[0.2cm]
c_v^{(3)}(\mu)
&=&
\left(4\beta_0^2 L_m^2 +2\beta_1 L_m\right)c_v^{(1)}(m)
\nonumber \\
&&
+4\beta_0 L_m 
\bigg\{ 
\blue{\,\frac{1}{2}}L_m C_F \pi^2
\bigg[- \frac{16}{3} C_F-8 C_A \bigg]
+c_v^{(2)}(m)
\bigg\}
\nonumber
\\
&&
+L_m^2 C_F \pi^2
\bigg[
-40C_F^2
\blue{-60C_F C_A
-16C_A^2}\bigg]
\nonumber
\\
&&
+L_m C_F \pi^2
\bigg[
 \left(-72+192\ln{2}\right) C_F^2
+\left(-\frac{1888}{27}-96\ln{2}\right) C_F C_A
\nonumber 
\\
&&\hspace{2cm}
+\left( -\frac{256}{9}-96\ln{2}\right)  C_A^2
+\frac{800}{27} C_F T_F n_f
\nonumber 
\\
&&\hspace{2cm}
+\frac{296}{9} C_A T_F n_f
-\frac{32}{5} C_F T_F
\bigg]
+c_v^{(3)}(m),
\label{eq:cv3} 
\end{eqnarray}
where $c_v^{(i)}(m)$ is the matching coefficient
evaluated at $\mu=m$. 
\blue{For the one-loop coefficient, we have also given the 
$O(\epsilon)$ term in agreement with \cite{Kiyo:2008mh}.}
The two-loop 
\blue{and three-loop non-logarithmic terms 
read: 
\begin{eqnarray}
c_{v}^{(2)}(m)
&=& 
16 \,\Bigg[C_F^2\left(
\frac{23}{8}-\frac{1}{2}\zeta_3-\frac{79}{36}\pi^2+\pi^2\ln{2}
\right)
\nonumber \\
&&
+\,C_F C_A \left(
-\frac{151}{72}-\frac{13}{4}\zeta_3+\frac{89}{144}\pi^2-\frac{5}{6}\pi^2\ln{2}
\right)
\nonumber \\
&&
+\,\frac{11}{18} C_F T_F n_f
+C_F T_F \left(\frac{22}{9}-\frac{2}{9}\pi^2\right)\Bigg],
\\[0.2cm]
c_{v}^{(3)}(m)
&=& 64\,\Bigg[
36.49486246 \,C_F^3 - 188.0778417\,C_F^2 C_A - 97.73497327 \,C_A^3
\nonumber\\ 
&& 
+ \, C_F T_F n_f 
\bigg[
46.69169291\,  C_F 
+39.62371855\,C_A
\nonumber \\
&&\hspace{1.5cm}
+ T_F n_f \left(-\frac{163}{162}-\frac{4}{27}\pi^2\right)
+ T_F \left(-\frac{557}{162}+\frac{26}{81}\pi^2\right) 
\bigg]
\nonumber \\
&& 
+\,
C_F T_F 
\bigg[
-0.8435622912\,  C_F
-0.1024741615\,C_A
\nonumber \\
&&\hspace{1.5cm}
+\, T_F
\left(-\frac{427}{162}+\frac{158}{2835}\pi^2+\frac{16}{9}\zeta_3\right) 
\bigg]\,\Bigg]
+
c_{v,\rm singlet}^{(3)}(m)
\nonumber\\
&=& 64\,\Big[-2090.332863+ 120.661081 n_f -0.822779 n_f^2\Big]+
c_{v,\rm singlet}^{(3)}(m)\,,\quad
\end{eqnarray}
where the last term $c_{v,\rm singlet}^{(3)}(m)$ is the unknown singlet} 
contribution and $\zeta_3$ is a short-hand for the Riemann 
zeta function value $\zeta(3)$. We recall our convention that 
$\alpha_s$ denotes the strong coupling 
in the $\overline{\rm MS}$ scheme with $n_f$ light flavours. 

Turning to the next orders in the velocity expansion, we find 
the operators 
\begin{eqnarray}
O_a&=&\frac{1}{2m^2}\psi^\dag{\bm\sigma}\cdot{\bf D}\,D^i\chi\, ,
\nonumber\\
O_b&=&\frac{1}{m^2}\psi^\dag\sigma^i\,{\bf D^2}\chi\,,
\label{eq:currentsv2}
\end{eqnarray}
which are suppressed by $O(v^2)$ relative to the leading NRQCD 
current. Further operators of dimension five contain the ultrasoft 
gauge field strength $g_s F_{\mu\nu}$ of order $v^{9/2}$. 
Thus up to NNNLO all production vertices contain only the 
quark-antiquark pair. The on-shell heavy 
quark-antiquark production vertex 
in full QCD can be decomposed into the expression
\begin{equation}
V^\mu = \bar u(p_1) \left[\gamma^\mu \hat F_1(q^2) + 
\frac{i\sigma^{\mu\nu} q_\nu}{2 m} \hat F_2(q^2)\right] v(p_2),
\end{equation}
where now $p_1=(E_p,\bff{p})$, $p_2=(E_p,-\bff{p})$, and 
$q=p_1+p_2=(2 E_p,\bff{0}) = (2 m+E,\bff{0})$. Inserting
(\ref{eq:spinors}) for the external spinors, we obtain the 
exact expression 
\begin{equation}
V^i = 2 E_p \left[\hat F_1(q^2)+\hat F_2(q^2)\right]
\xi^\dagger\sigma^i\eta
-2 E_p \left[\frac{m}{E_p} \hat F_1(q^2)-\hat F_2(q^2)\right] 
\frac{p^i}{m (E_p+m)} 
\,\xi^\dagger\bff{\sigma}\cdot\bff{p}\,\eta.\quad
\end{equation}
This shows explicitly that only quark-antiquark operators with a 
sigma matrix can appear as assumed in (\ref{eq:currentsv2}). 
Expanding this expression in $q^2-4 m^2 = 4\bff{p}^2$ we find
\begin{eqnarray}
c_v &=& [\hat F_1+\hat F_2]_{\,|q^2=4 m^2, \,\mbox{\scriptsize hard}},
\nonumber\\[0.2cm] 
d_{va} &=&  [\hat F_1-\hat F_2]_{\,|q^2=4 m^2, \,\mbox{\scriptsize hard}},
\nonumber\\ 
d_{vb} &=& (-4) \,\frac{d}{d(q^2/m^2)} 
[\hat F_1+\hat F_2]_{\,|q^2=4 m^2, \,\mbox{\scriptsize hard}}.
\end{eqnarray}
The subscript ``hard'' means that only the hard regions should 
be included in the computation. The one-loop hard form factors 
can be extracted from \cite{Bernreuther:2004ih}, by dropping 
the non-analytic terms in the expansion in $\bff{p}^2$, which 
originate from the potential region. We obtain $c_v^{(1)}$ 
given in (\ref{eq:cv1}) above and 
\blue{
\begin{eqnarray}
d_{va}&=& 1 + \frac{\alpha_s C_F}{4 \pi} 
\left[- 4 +\epsilon \left(4\ln\frac{m^2}{\mu^2} +8\right)\right]+ O(\alpha_s^2)\,, 
\nonumber \\
d_{vb}&=&\frac{\alpha_s C_F}{4 \pi} 
\left[\frac{8}{3}\ln\frac{m^2}{\mu^2}-\frac{2}{9} 
+\epsilon\left(-\frac{4}{3}\ln^2\frac{m^2}{\mu^2}+
\frac{2}{9}\ln\frac{m^2}{\mu^2}-\frac{428}{27}-\frac{2\pi^2}{9}
\right)\right]\nonumber \\
&&+\,O(\alpha_s^2)\,. 
\label{eq:dvab}
\end{eqnarray}
Again we have provided the $O(\epsilon)$ terms. The unrenormalized 
coefficient $d_{vb}$ contains the pole part $-8/(3\epsilon)$ in 
the square bracket, which has been minimally subtracted.}
The logarithm in $d_{vb}$ arises as the consequence of mixing with 
the leading order current through subleading NRQCD interactions, see 
(\ref{eq:usdiv}) below. 
These results agree with the computation of the one-loop corrected matching 
coefficients of the subleading current operators through explicit 
NRQCD matching \cite{Luke:1997ys}.

At NNNLO the correlation functions of velocity-suppressed currents 
will be evaluated only with the leading and 
next-to-leading order Coulomb potential, which is spin-independent. 
Hence, only the traces 
$\mbox{tr}\,(O_{a,b}\sigma^i)$ appear. This allows us to 
combine 
\begin{equation}
d_{va} O_a + d_{vb} O_b \;\to\;  \frac{d_v}{6m^2}\,\psi^\dag\sigma^i\,{\bf
D^2}\chi
\label{eq:currentspinprojection}
\end{equation}
such that the QCD vector current is now represented by 
\begin{eqnarray}
\label{eq:QCDVectorCurrent2} j_{\,i}^{(v)}=c_v\,
\psi^\dag\sigma_i\chi + \frac{d_v}{6m^2}\psi^\dag\sigma_i\,{\bf
D^2}\chi +O(1/m^4),
\end{eqnarray}
as anticipated in (\ref{eq:QCDVectorCurrent}). From (\ref{eq:dvab}) 
and (\ref{eq:currentspinprojection}) we obtain 
\begin{eqnarray}
d_v(\mu)&=& \frac{3}{3-2\epsilon} \,d_{va} + 6 d_{vb} 
\nonumber\\
&=& 
\blue{\frac{3}{3-2\epsilon}
+\frac{\alpha_sC_F}{4\pi}\left[-32 L_m-\frac{16}{3}
+\epsilon\left(-32 L_m^2-\frac{32}{3} L_m
-\frac{808}{9}-\frac{4\pi^2}{3}\right)
+O(\epsilon^2)\right]}
\nonumber\\
&& +\,
O(\alpha_s^2)\,.\;
\label{eq:dv1loop}
\end{eqnarray}

The explicit scale dependence of the matching coefficients is due 
to evolution of the strong coupling and the factorization of the 
hard scale. It must cancel when 
all contributions to the cross section are combined. We have 
checked explicitly that this is indeed the case.

Note that we do not need the $O(\epsilon)$ terms of the 
coefficient functions $c_v$ and $d_v$ to compute the heavy-quark
current correlation function, since they multiply the 
finite, renormalized NRQCD correlation function in
(\ref{eq:pitoNRQCD}). One may wonder then what is the 
difference between the NRQCD current and the NRQCD Lagrangian 
matching coefficients $d_1$, $d_2$ etc., since for the latter we need the 
$O(\epsilon)$ terms as stated and given above. The reason is 
the particular definition (\ref{eq:newmatch}) of the current matching 
coefficient. Imagine that we calculate the QCD and NRQCD vertex 
functions $\Gamma$ with non-vanishing external relative 
momentum. Then the NRQCD diagrams are no longer scaleless and 
$\Gamma_{\rm NRQCD}$ is the sum of potential, soft and ultrasoft 
loop momentum contributions. Because of the $1/v$ factors from 
potential gluon exchange, the higher-dimensional NRQCD 
interactions contribute to the leading current matching 
equation at some order in perturbation theory. As a result 
$\Gamma_{\rm NRQCD}$ will be different whether one 
uses the NRQCD Lagrangian with four-dimensional or with 
$d$-dimensional short-distance coefficients. The difference is, 
however, a local term that can be absorbed into the matching
coefficients of the external currents. Which definition does 
(\ref{eq:newmatch}) correspond to? Suppose we follow the more 
conventional path to define the renormalized effective Lagrangian 
with $d=4$ short-distance coefficients. In this case 
$\Gamma_{\rm NRQCD}$ does {\em not} represent the sum of 
all potential, soft and ultrasoft loop momentum regions plus 
the hard ones not connected to the external vertex (encapsulated in 
the Lagrangian matching coefficients $d_1$ etc.), since one 
misses some $O(\epsilon)$ terms from hard subgraphs that multiply 
divergent soft, potential or ultrasoft loops. These missing 
local contributions can be and must be added back by adapting the 
external current matching coefficient. Hence the 
matching coefficient $c_v$ defined by this prescription does 
not correspond to (\ref{eq:newmatch}). On the other hand, 
the NRQCD Lagrangian with $d$-dimensional short-distance coefficients 
reproduces these missing terms directly, so the matching 
coefficient corresponding to this case is simply the 
contribution from the purely hard (h-h-...-h) regions as it was 
defined in  (\ref{eq:newmatch}). Moreover, the purely hard 
regions can now be computed directly at zero external relative  
momentum, which simplifies the calculation.

The same discussion applies to the matching of the potentials in 
PNRQCD, to which we turn next. However, while the $O(\epsilon)$ 
terms of the NRQCD Lagrangian are relevant only at NNNLO, the 
difference between four- and $d$-dimensional potentials in the 
\mbox{PNRQCD} Lagrangian matters already at NNLO. The 
$d$-dimensional ones must be used in conjunction with 
(\ref{eq:newmatch}) as was done in \cite{Beneke:1999qg}.

\blue{We further note that the statement that the $O(\epsilon)$ 
terms of the coefficient functions $c_v$ and $d_v$ are not needed 
holds as long as the quantity that is computed, here the 
production cross section of a pair of stable heavy quarks, is finite. 
When the decay of the heavy quark is consistently included, 
as is required for top pair production near threshold, the 
observable is the cross section for the final state $W^+ W^- 
b\bar{b}$. This cannot be fully computed in NRQCD, since the 
observable contains non-resonant contributions, which lie outside 
the scope of NRQCD. As a result, $R$ as defined in (\ref{R1}) 
exhibits finite-width $1/\epsilon$ poles, which cancel with 
the non-resonant contributions, as mentioned in the introduction 
and discussed in detail in paper~II. In this case the 
$O(\epsilon)$ terms of the hard matching coefficients multiply 
the uncancelled finite-width poles of the non-relativistic 
two-point function $G(\mathcal{E})$ resulting in finite terms, 
which must be included for a consistent addition of the 
resonant and non-resonant contributions. It is for this reason 
that we have given the $O(\epsilon)$ terms of the matching 
coefficients $c_v^{(1)}$ and $d_v$ above.}

\subsection{Matching of the axial-vector current}

Due to the $v^2$ suppression of the P-wave correlation function 
(\ref{eq:piAtoNRQCD}) relative to the S-wave case (\ref{eq:G}), 
the hard matching coefficient $c_a$ of the axial-vector current 
is needed only with one-loop accuracy for the NNNLO calculation of 
the top-quark pair production cross section near threshold. 
The relevant expression is
\blue{
\begin{equation}
c_a = 1-4C_F\cdot\frac{\alpha_s}{4\pi}
\left[1-\epsilon\ln\frac{m^2}{\mu^2}+O(\epsilon^2)\right]+
O(\alpha_s^2)\,.
\end{equation}
The $O(\epsilon)$ term is taken from \cite{Beneke:2013kia}.}
The two-loop correction is also known \cite{Kniehl:2006qw}.

\section{Potential NRQCD} 
\label{sec:pnrqcd} 

As discussed in Section~\ref{sec:momentumregions} to perform the all-order 
resummation, a second matching procedure is required, by which the
soft region and potential light fields (gluons and light quarks) are
integrated out. This results in the potential NRQCD (PNRQCD) effective 
field theory~\cite{Beneke:1999qg,Beneke:1998jj,Pineda:1997bj,Pineda:1997ie,Brambilla:1999xf}. 
In PNRQCD the light fields are purely ultrasoft and the heavy quarks 
are potential, hence the terms in the effective Lagrangian can be 
assigned a unique scaling in the velocity expansion. The effective 
Lagrangian relevant to 
the third-order calculation takes the simple form
\begin{eqnarray}
{\cal L}_{\rm PNRQCD} &=& \psi^\dag
\Big(i\partial_0+g_s A_0(t,\bff{0})+\frac{\bffmath{\partial}^2}{2m}+ 
\frac{{\bffmath\partial^4}}{8m^3}\,\Big)\psi +\chi^\dag
\Big(i\partial_0+g_s A_0(t,\bff{0})-\frac{{\bffmath
\partial^2}}{2m}-\frac{ {\bffmath\partial^4}}{8m^3}\Big)\chi
\nonumber \\ \nonumber &&+ \int d^{d-1} {\bf r} \, \Big[ \psi^\dag_a
\psi^{}_b \Big](x+{\bf r}) \, V_{ab;cd} (r,{\bffmath \partial})\,
\Big[\chi^\dag_c \chi^{}_d\Big](x)
\\
&&-g_s\psi^\dagger(x)\bff{x}\cdot\bff{E}(t,\bff{0})\psi(x)-
g_s\chi^\dagger(x)\bff{x}\cdot\bff{E}(t,\bff{0})\chi(x),
\label{eq:pnrqcd}
\end{eqnarray}
where 
\begin{equation}
V_{ab;cd} (r,{\bffmath \partial})=T^A_{ab} T^A_{cd}V_0 (r)+
\delta V_{ab;cd} (r,{\bffmath \partial})
\end{equation}
with $V_0=-\alpha_s/r$ the tree-level colour Coulomb potential.
The PNRQCD Lagrangian consists of kinetic terms 
(first line; including the relativistic
corrections proportional to ${\bffmath\partial^4}/m^3$), heavy-quark 
potential interactions (second line) and an ultrasoft interaction 
that contributes first at third order 
\blue{to the top production cross section near threshold.} 
The heavy-quark potentials 
generated in the matching to PNRQCD should be considered as 
short-distance coefficients of the \mbox{PNRQCD} interactions. They are split 
into the tree-level Coulomb potential, which must be treated 
non-perturbatively and a remainder 
$\delta V_{ab;cd} (r,{\bffmath \partial})$, 
which represents a perturbation. To achieve 
a homogeneous velocity scaling the position argument of 
ultrasoft fields should be multipole-expanded in interactions with heavy 
quarks \cite{Labelle:1996en,Grinstein:1997gv,Beneke:1999zr}, 
which explains the space-time argument of $A_0$ and the chromoelectric 
field in the ultrasoft interaction terms. 

As will be discussed below no further matching of the non-relativistic 
vector current is needed, that is $\psi^{\dag}\sigma^i\chi_{|\rm NRQCD} = 
\psi^{\dag}\sigma^i\chi_{|\rm PNRQCD}$ to the required accuracy. 
Thus, instead of (\ref{eq:G}), we have to calculate 
\begin{equation}
\label{eq:GPNRQCD}
G(E) =\frac{i}{2 N_c (d-1)} \int d^{d} x\, e^{iEx^0}\,
\langle 0| \,T(\,
[\chi^{\dag}\sigma^i\psi](x)\,
[\psi^{\dag}\sigma^i\chi](0))
|0\rangle_{| \rm PNRQCD}\,,
\end{equation}
where now the matrix element must be evaluated to third-order in PNRQCD 
perturbation theory. 

The dimensionally regulated PNRQCD Lagrangian required for 
second-order calculations of heavy-quark pair production near 
threshold was provided in \cite{Beneke:1999qg}, and 
the explicit derivation of the ultrasoft interaction in the third line 
of (\ref{eq:pnrqcd}) from NRQCD was given in \cite{Beneke:1999zr}. 
The only new piece that is needed is the third-order heavy-quark 
potential in $\delta V_{ab;cd} (r,{\bffmath \partial})$. 
In the remainder of this subsection we first give the PNRQCD 
Feynman rules (when the ultrasoft interactions are neglected) and 
then sketch several ways of deriving these rules and the form of 
the propagator. Subsequently, we 
summarize the heavy-quark 
potentials. We also derive equation-of-motion relations that allow 
us to reduce the number of potential insertions to be calculated 
and briefly discuss the ultrasoft contribution already calculated 
in~\cite{Beneke:2008cr}.

\subsection{Feynman rules}

\begin{figure}[t]
\begin{center}
\makebox[0cm]{\scalebox{0.7}{\rotatebox{0}{
\includegraphics{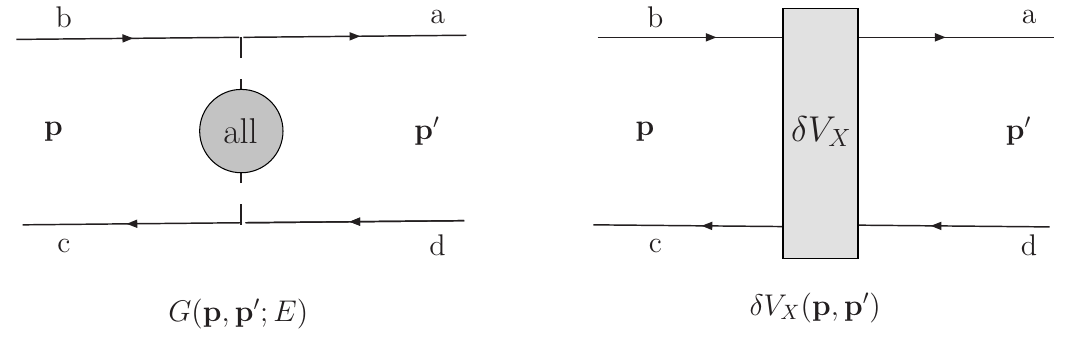}}}}
\end{center}
\vspace*{-0.3cm}
\caption{\label{fig:pnrqcd} PNRQCD Feynman rules.}
\end{figure}

We begin by summarizing the rules for calculating PNRQCD diagrams with 
insertions of potential interactions, but no ultrasoft interactions. 
Since the leading-order 
Lagrangian includes the Coulomb potential $V_0(r)$, the propagator 
is the one for a heavy quark anti-quark {\em pair}. We draw the 
propagator as in the left diagram of figure~\ref{fig:pnrqcd}, where 
the blob stands for the sum of all potential (Coulomb) ladder diagrams,
which is included in the propagator. For the pair in a colour-singlet 
state each propagator gives a factor
\begin{equation}
\frac{1}{N_c}\delta_{bc}\delta_{da} \,i G_0^{(1)}(\bff{p},\bff{p}^\prime;E),
\end{equation}
where $E=\sqrt{s}-2 m$ is the non-relativistic energy 
of the pair and $\bff{p}$ 
($\bff{p}^\prime$) the three-momentum of the in-coming (out-going) quark. 
For the colour-octet state the propagator is 
$2\,T^A_{bc}T^A_{da} \,i G_0^{(8)}({\bf p},{\bf p}^\prime;E)$. The 
function $G_0^{(R)}({\bf p},{\bf p}^\prime;E)$ is the solution to 
the $d$-dimensional Lippmann-Schwinger equation for the pair 
in the irreducible SU(3) colour representation~$R$,
\begin{eqnarray}
&& \left(\frac{\bff{p}^2}{m} - E\right) G_0^{(R)}(\bff{p},\bff{p}^\prime;E)
+ \tilde{\mu}^{2\epsilon}\int\frac{d^{d-1}\bff{k}}{(2\pi)^{d-1}}\,
\frac{4\pi D_R\alpha_s}{\bff{k}^2}\, G_0^{(R)}(\bff{p}-\bff{k},\bff{p}^\prime;E) 
\nonumber\\
&& \qquad = (2\pi)^{d-1}\,\delta^{(d-1)}(\bff{p} - \bff{p}^\prime)\,,
\label{eq:lippmann-schwinger}
\end{eqnarray}
where $D_R=-C_F$ and $D_R = -(C_F-C_A/2)$ for the colour-singlet 
and colour-octet representation, respectively.  Explicit expressions 
\blue{for the solution} will be given below. The scale 
$\tilde\mu = \mu \,[e^{\gamma_E}/(4\pi)]^{1/2}$ 
is defined such that minimal subtraction 
of $1/\epsilon$ poles corresponds to the $\overline{\rm MS}$ rather than 
MS scheme \cite{Bardeen:1978yd}. The Fourier transform 
\begin{equation}
G_0^{(R)}(\bff{r},\bff{r}^\prime;E) = 
\int \frac{d^{d-1}\bff{p}}{(2\pi)^{d-1}}
\frac{d^{d-1}\bff{p}^\prime}{(2\pi)^{d-1}} 
\,e^{i \bff{p}\cdot\bff{r}}
\,e^{-i \bff{p}^\prime\cdot\bff{r}^\prime}
\, G_0^{(R)}(\bff{p},\bff{p}^\prime;E)
\end{equation}
satisfies in $d=4$ dimensions the Schr\"odinger equation 
\begin{equation}
\left(-\frac{\bff{\nabla}^2_{(r)}}{m}+\frac{D_R\alpha_s}{r}-E\right) 
G_0^{(R)}(\bff{r},\bff{r}^\prime;E) = \delta^{(3)}(\bff{r} - \bff{r}^\prime)\,.
\label{eq:schroedinger}
\end{equation}
In the following, when the 
superscript is left out, the propagator refers to the colour-singlet 
representation.

The vertex associated with the insertion of a perturbation 
potential $\delta V_{ab;cd} (\bff{p},\bff{p}^\prime)$ in momentum space is 
given by 
\begin{equation}
i \delta V_{ab;cd} (\bff{p},{\bff p}^\prime)\,,
\end{equation}
and internal relative momenta $\bff{p}_i$ are integrated over with measure 
$\tilde{\mu}^{2\epsilon}\int d^{d-1}\bff{p}_i/(2\pi)^{d-1}$. Note that the 
insertion of a potential does not change the colour state of the 
quark anti-quark pair, when it is in an irreducible representation, 
that is the colour-singlet or the colour-octet state. The reason for 
this is that $\frac{1}{N_c}\delta_{bc}\delta_{da}$ and $2\,T^A_{bc}T^A_{da}$ 
constitute a complete set of orthogonal projectors.\footnote{See 
\cite{Beneke:2009rj,Beneke:2010da} for a general discussion of the 
colour decomposition in arbitrary representations.} If the in-coming 
pair is, for example, in the colour-singlet state, all propagators 
will be colour-singlet propagators and the potential insertions 
are effectively projected to the colour-singlet potential
\begin{equation}
\delta V(\bff{p},{\bff p}^\prime) = 
\frac{1}{N_c}\,\delta_{bc}\delta_{da}
\,\delta V_{ab;cd} (\bff{p},{\bff p}^\prime).
\label{eq:singlet-potential}
\end{equation}
This is different when an ultrasoft gluon is emitted, in which 
case the pair changes its colour state, as explicitly seen 
in (\ref{eq:gfcorr}) below.

A general PNRQCD diagram with multiple insertions of perturbation 
potentials is therefore an expression of the form
\begin{equation}
\int \Bigg[\prod_{i}\frac{d^{d-1}{\bf
p}_i}{(2\pi)^{d-1}}\Bigg] 
i{G_0}({\bf p}_1,{\bf p}_2;E)
i\delta{V}_1({\bf p}_2,{\bf p}_3) i{G_0}({\bf p}_3,{\bf p}_4;E)\, 
i\delta{V}_2({\bf p}_4,{\bf p}_5)  i{G_0}({\bf p}_5,{\bf p}_6;E)\, 
\ldots
\label{multipleinsertions}
\end{equation}
(for the case of a colour-singlet state). For convenience of notation 
here and below we often leave out factors 
of $\tilde{\mu}^{2\epsilon}$ required in 
dimensional regularization to restore the proper dimension of the 
given expression.

\subsection{Three derivations of the PNRQCD rules}

In this section we sketch three derivations of the rules for 
PNRQCD perturbation theory: diagrammatic, quantum-mechanical, and 
by path-integral methods. For simplicity we assume the colour-singlet 
representation, but the derivation is easily generalized to an 
arbitrary irreducible representation.

\subsubsection{Diagrammatic} 
Consider the amputated amplitude of the heavy-quark scattering 
process $Q(p_1)\bar Q(p_2) \to Q(p_1^\prime)\bar Q(p_2^\prime)$
with non-relativistic 
external momenta $p_1=(E/2,\bff{p})$, $p_2=(E/2,-\bff{p})$ 
and $p_1^\prime=(E/2,\bff{p}^\prime)$, $p_2^\prime=(E/2,-\bff{p}^\prime)$ 
in the rest frame of the $Q\bar Q$ pair. The sum of all (ladder) diagrams with 
any number (greater than zero) of leading-order potential insertions 
$V_0(r)$ is given in momentum space by 
\begin{eqnarray}
H(\bff{p},\bff{p}^\prime;E) &=& \sum_{n=0}^\infty\,C_F^{n+1}\,
\int \left[\prod_{i=1}^n \frac{d^d k_i}{(2\pi)^d}\right]
\frac{(i g_s)^2 i}{(\bff{k}_1-\bff{k}_0)^2}\,
\frac{(i g_s)^2 i}{(\bff{k}_2-\bff{k}_1)^2}\, \ldots
\frac{(i g_s)^2 i}{(\bff{k}_{n+1}-\bff{k}_n)^2}\,
\nonumber\\
&&\cdot \,\prod_{i=1}^n\,
\frac{i}{\frac{E}{2}+k_i^0-\frac{(\bff{p}+\bff{k}_i)^2}{2m}+
i\epsilon}\,
\frac{-i}{\frac{E}{2}-k_i^0-\frac{(\bff{p}+\bff{k}_i)^2}{2m}+
i\epsilon} \,,
\end{eqnarray}
where we define $\bff{k}_{n+1} = \bff{p}^\prime-\bff{p}$ and 
$\bff{k}_0\equiv 0$. We perform 
the integrations over the loop momentum zero components $k_i^0$ by 
closing the contour in the upper half plane, and pick up the 
residues from the poles at $k_i^0 = E/2 - (\bff{p}+\bff{k}_i)^2/(2m) +
i\epsilon$, which results in 
\begin{eqnarray}
H(\bff{p},\bff{p}^\prime;E) &=& i \,
\sum_{n=0}^\infty\,(-g_s^2 C_F)^{n+1}\,
\!\int \,\left[\prod_{i=1}^n \frac{d^{d-1} \bff{k}_i}{(2\pi)^{d-1}}\right]
\,\frac{1}{\bff{k}_1^2}\;
\nonumber\\
&&\hspace*{0.0cm}\times 
\prod_{i=1}^n\,
\frac{1}{(\bff{k}_{i+1}-\bff{k}_i)^2 
(E-\frac{(\bff{p}+\bff{k}_i)^2}{\blue{m}}+i\epsilon)}\,.
\qquad
\label{eq:Hc}
\end{eqnarray}
The $n=0$ term in this and the previous sum is understood as $(-i g_s^2 C_F)/
(\bff{p}^\prime-\bff{p})^2$, which is the expression for the exchange 
of a single potential (Coulomb) gluon. $H(\bff{p},\bff{p}^\prime;E)$ 
sums the leading p-p-...-p region to all orders.

Next we multiply the propagator factors $(-i)/(E+i\epsilon -\bff{p}^2/m)$ 
for the external pairs of lines and add 
the zero-Coulomb exchange graph. Multiplying by $(-i)$ this 
defines 
\begin{equation}
G_0(\bff{p},\bff{p}^\prime;E) =
-\frac{(2\pi)^{d-1}\delta^{(d-1)}(\bff{p}^\prime-\bff{p})}
{E+i\epsilon-\frac{\bff{p}^2}{m}} + 
\frac{1}{E+i\epsilon-\frac{\bff{p}^2}{m}}\,
iH(\bff{p},\bff{p}^\prime;E) \,
\frac{1}{E+i\epsilon-\frac{\bff{p}^{\prime\,2}}{m}}\,.
\label{eq:coulombfn}
\end{equation}
It is straightforward to show that this expression satisfies the 
$d$-dimensional Lippmann-Schwinger equation (\ref{eq:lippmann-schwinger}), 
and hence represents the colour-singlet Coulomb Green function. The summation of ladder diagrams is therefore 
accomplished by associating the quantity $i G_0(\bff{p},\bff{p}^\prime;E)$ 
with the propagator of the quark anti-quark pair and the vertex 
$i\delta V(\bff{p},\bff{p}^\prime)$ with the interaction potentials. 
It is understood that the integrations over the zero-components of 
loop momenta are already done. 

Note that while no closed expression for the Green function is 
known in $d$ dimensions, it is important that the above expression 
is defined in dimensional regularization, and that it can be 
expanded perturbatively in $g_s^2$ in $d$ dimensions. This guarantees the 
consistency of the dimensional regularization procedure, which 
requires subtracting terms with a finite number of 
Coulomb exchanges from $G_0(\bff{p},\bff{p}^\prime;E)$ in $d$ 
dimensions as will be seen in part II of the paper.

The PNRQCD correlation function (\ref{eq:GPNRQCD}) describes a 
quark anti-quark pair created in a colour-singlet, spin-triplet 
state at point 0, which propagates and is destroyed locally at 
point $x$. In terms of the momentum-space propagator it is given 
by
\begin{eqnarray}
G(E)&=& 
\int \frac{d^{d-1}\bff{p}}{(2\pi)^{d-1}}
\frac{d^{d-1}\bff{p}^\prime}{(2\pi)^{d-1}} 
\,\bigg[\,G_0(\bff{p},\bff{p}^\prime;E) 
\nonumber\\
&&+\, 
\int \frac{d^{d-1}\bff{p}_1}{(2\pi)^{d-1}}
\frac{d^{d-1}\bff{p}_1^\prime}{(2\pi)^{d-1}} \,
G_0(\bff{p},\bff{p}_1;E)
\,i\delta V(\bff{p}_1,\bff{p}_1^\prime)\,
iG_0(\bff{p}_1^\prime,\bff{p}^\prime;E)
+\ldots\bigg]\,,\qquad
\label{eq:Gexp}
\end{eqnarray}
where the potential refers to the colour-singlet potential 
(\ref{eq:singlet-potential}), and the ellipses to terms with 
multiple potential insertions. Both the propagator and the potential 
carry spin-indices in general. However, the unperturbed PNRQCD 
Lagrangian is spin-independent, so the propagator is diagonal 
in the spin indices and we drop the $\delta_{\alpha\beta^\prime}
\delta_{\alpha^\prime\beta}$ 
spin factor. The spin-dependence of the perturbation potential enters 
only at NNLO, hence up to N$^3$LO only one of the 
$\delta V$ insertions can carry a non-trivial spin-dependence. 
When this insertion appears in (\ref{eq:Gexp}), $\delta V$ is 
understood as
\begin{equation}
\delta V = \frac{1}{2 (d-1)} \,\sigma^i_{\alpha\alpha'}\, 
\delta V_{\alpha\beta';\alpha'\beta} \,\sigma^i_{\beta\beta'}\,,
\label{eq:triplet-projection}
\end{equation}
and the trace must be carried out in $d$ dimensions. Greek indices of 
the potential refer to spin (rather than colour). The 
normalization factor and Pauli matrices result from the definition 
of (\ref{eq:GPNRQCD}) and correspond to the spin-triplet projection 
of the potential. For the insertions of spin-independent 
potentials the spin-factor 
\begin{equation}
\frac{1}{2 (d-1)} \,\sigma^i_{\alpha\alpha'}\, 
\delta_{\alpha\beta^\prime}
\delta_{\alpha^\prime\beta}\,\sigma^i_{\beta\beta'} = 1 
\end{equation}
is included in (\ref{eq:Gexp})

\subsubsection{Quantum-mechanical} 
\label{sec:qm}
With only potential interactions  
the PNRQCD Lagrangian can be projected onto the Fock states 
with a single quark and a single anti-quark without loss of content, 
since potential interactions do not change particle number. We 
define the centre-of-mass wave function of a quark anti-quark 
state $|\psi\rangle$ in the position-space representation via
\blue{
\begin{equation}
\psi(t,\bff{r}) = 
\langle 0|[\psi(t,\bff{r}/2)\chi^\dagger(t,-\bff{r}/2)]
|\psi\rangle\,,
\end{equation}
}
which is a matrix in colour and spin indices. For simplicity, we 
assume again a projection on the colour-singlet representation. 
By reversing the steps that lead from the Schr\"odinger 
equation to a second-quantized Schr\"odinger field theory, 
making use of the field equation and the canonical commutation 
relations, we find that 
\begin{equation}
i\partial_t \psi(t,\bff{r}) = H \psi(t,\bff{r}) 
= \left[\,-\frac{\bff{\nabla}^2}{m} + C_F V_0(r) + \delta V(r)\,\right] 
 \psi(t,\bff{r})\,.
\end{equation}
The Green function of the Schr\"odinger operator is given 
by $\langle \bff{r}|[H-E-i\epsilon]^{-1}|\bff{r}'\rangle$ 
with $|\bff{r}\rangle$ a quark anti-quark separation eigenstate 
with eigenvalue $\bff{r}$. In operator notation the Green function is 
$\hat G_H(E) = [H-E-i\epsilon]^{-1}$ such that
\begin{equation}
\hat G_H(E) = \hat G_{0}(E) + 
\hat G_{0}(E) i\delta V i \hat G_{0}(E) + \ldots\,,
\label{eq:Goperatorexp}
\end{equation}
where $\hat{G}_{0}(E) = [H_0-E-i\epsilon]^{-1}$ and $H=H_0+\delta V$. Since 
$G_0^{(1)}(\bff{r},\bff{r}^\prime;E)=
\langle \bff{r}| \hat{G}_{0}(E)|\bff{r}'\rangle$, the previous equation 
verifies the PNRQCD Feynman rules. In this notation the correlation 
function (\ref{eq:GPNRQCD}) is given by 
\begin{equation}
G(E) = \langle \bff{0}| \hat{G}_{H}(E)|\bff{0}\rangle\,,
\end{equation}
which is equivalent to (\ref{eq:Gexp}) upon using (\ref{eq:Goperatorexp}) 
and inserting complete sets of momentum eigenstates,
\begin{equation}
1 = \int \frac{d^{d-1}\bff{p}}{(2\pi)^{d-1}}
\,|\bff{p}\rangle\langle\bff{p}|\,,
\end{equation}
leaving the spin-average implicit.

\subsubsection{Path integral derivation}
In \cite{Pineda:1997bj,Pineda:1997ie,Brambilla:1999xf} the PNRQCD 
Lagrangian is expressed in terms of composite colour-singlet and 
colour-octet fields. Here we provide a path-integral derivation of 
this formulation. We focus on the colour-singlet field and drop the 
colour and spin indices of the composite field 
\blue{
\begin{equation}
\left[ S(x,y)\right]_{x^0=y^0} = 
\left[ \psi(x)\chi^\dag(y) \right]_{x^0=y^0}\,,
\qquad 
\left[ S^\dag(y,x)\right]_{x^0=y^0} = 
\left[ \chi(y)\psi^\dag(x) \right]_{x^0=y^0}\,.
\end{equation}
}

The partition function $Z_{\rm PNRQCD}$ of PNRQCD is defined as
\begin{eqnarray}
Z_{\rm PNRQCD} =\int {\cal D} \psi  {\cal D} \psi^\dag {\cal D} \chi  
{\cal D} \chi^\dag 
\exp\bigg\{\,i\!
\int d^4x\, {\cal L}_{\rm PNRQCD}(x)
\bigg\}\,,
\end{eqnarray}
where we use the leading-order PNRQCD Lagrangian
\blue{
\begin{eqnarray}
{\cal L}_{\rm PNRQCD}(x)
&=&
\psi^\dag(x)\bigg[i\partial_0 + 
\frac{\bffmath{\partial}^2}{2m}\bigg]\psi(x)
+\chi^\dag(x)\bigg[i\partial_0 - \frac{\bffmath{\partial}^2}{2m}\bigg]\chi(x)
\nonumber\\
&&
-\int d^4y \left[\chi(y)\psi^\dag(x)\right] V(x,y) 
\left[ \psi(x)\chi^\dag(y)\right],
\end{eqnarray}
}
with the quark anti-quark potential  
$V(x,y) = V_0(x-y) \delta (x^0-y^0)$. The derivation remains however valid 
with an arbitrary potential, not just the leading-order Coulomb potential. 
The different sign of the 
potential term relative to (\ref{eq:pnrqcd}) is due to the different 
order of fermion fields. 
The composite field is introduced by means of 
\blue{
\begin{eqnarray}
1 &=& \int {\cal D}S \,
{\delta}\left(\left[ S(x,y)\right]_{x^0=y^0} - 
\left[\psi(x)\chi^\dag(y)\right]_{x^0=y^0}\right) 
\nonumber\\
&=&
\int {\cal D}S \,{\cal D}\sigma\,
\exp\bigg\{ \,i \!\int d^4x \int d^4y \, \sigma^\dag(y,x) \delta(x^0-y^0) 
\left( S(x,y)- \left[\psi(x)\chi^\dag(y)\right]\right)
\bigg\},\;\;
\qquad
\end{eqnarray}
}
where the second line is the Fourier representation of the 
functional delta-function which introduces the auxiliary field 
$\sigma$. With a similar formula for the complex conjugate field 
we can rewrite  the partition function as
\begin{eqnarray}
Z_{\rm PNRQCD} = \int {\cal D} \psi  {\cal D} \psi^\dag 
{\cal D} \chi  {\cal D} \chi^\dag
{\cal D} S  {\cal D} S^\dag {\cal D} \sigma  {\cal D} \sigma^\dag
\exp\bigg\{\,i \!\int d^4x \int d^4y \,{\cal L}_{\psi,\chi,S,\sigma}(x,y) 
\bigg\}\,,
\qquad
\end{eqnarray}
where 
\blue{
\begin{eqnarray}
{\cal L} _{\psi,\chi,S,\sigma}(x,y) 
&=&
\psi^\dag(x) K_Q (x,y) \psi(y) +\chi^\dag(x)K_{\bar{Q}} \chi(y)
-S^\dag(y,x) V(x,y) S(x,y)
\nonumber\\
&&+\,\Sigma^\dag(y,x)\left( S(x,y)- \left[\psi(x)\chi^\dag(y)\right]\right)
\nonumber\\
&&+\,\Sigma(x,y)\left( S^\dag(y,x)- \left[\chi(y)\psi^\dag(x)\right]
\right)\,.
\end{eqnarray}
}
To make equations concise we introduced 
$\Sigma(x,y)=\sigma(x,y)\delta(x^0-y^0)$ and 
\begin{eqnarray}
K_{Q}(x,y) &=& \delta^4 (x-y) \left( i\partial_{0} +
\frac{\bffmath\partial^2}{2m}\right)_y \,,
\nonumber\\
K_{\bar{Q}}(x,y)&=& \delta^4 (x-y) \left( i\partial_{0} - 
\frac{\bffmath\partial^2}{2m}\right)_y \,.
\end{eqnarray}

Our task is to integrate over $\psi,\chi, \sigma$ and their 
conjugates to obtain the Lagrangian for the $S$ field. In the 
following we do this step by step obtaining the sequence of 
Lagrangians 
\begin{eqnarray}
{\cal L}_{\psi,\chi,\sigma,S} ~
\stackrel{\psi}{\to} ~
{\cal L}_{\chi,\sigma,S}~
\stackrel{\chi}{\to}~
{\cal L}_{\sigma,S}~
\stackrel{\sigma}{\to}~
{\cal L}_{S}\,.
\end{eqnarray}
We drop field-independent factors which can be absorbed into the 
path-integral measure. First we integrate out $\psi$ and its 
conjugate by completing squares in the exponent:
\blue{
\begin{eqnarray}
&& 
\int d^4x \int d^4y \,{\cal L}_{\psi,\chi, S,\sigma} (x,y) =
\int d^4x \int d^4y \left[ {\cal L}_{\psi,\chi, S,\sigma} (x,y)
\right]_{\psi=0}
\nonumber\\
&&\qquad +\,
\left( \psi^\dag+\chi^\dag\cdot\Sigma^\dag\cdot K_Q^{-1} \right) 
\cdot K_Q\cdot
\left(\psi+K_Q^{-1} \cdot\Sigma\cdot \chi \right)
-\chi^\dag\cdot\Sigma^\dag\cdot K_Q^{-1} \cdot \Sigma \cdot \chi\,.
\qquad
\end{eqnarray}
}
In the second line the abbreviation $(\cdot )$ stands for the 
integration of the adjoining variables, such that, for example, 
the last term reads explicitly
\begin{equation}
\chi^\dag\cdot\Sigma^\dag\cdot K_Q^{-1} \cdot \Sigma \cdot \chi 
= \int d^4x \int d^4z_1 \int d^4z_2 \int d^4y \,
\chi^\dag(x)\Sigma^\dag(x,z_1) K_Q^{-1}(z_1,z_2)\Sigma(z_2,y)\chi(y)\,. 
\end{equation}
The inverse operators are given by 
\begin{eqnarray}
K_Q^{-1}(x,y)&=&\int \frac{d^4p}{(2\pi)^4} 
\frac{e^{-ip(x-y)}}{p^0-\frac{\bff{p}^2}{2m}+i\epsilon}\,,
\nonumber\\
K_{\bar{Q}}^{-1}(x,y)&=&
\int \frac{d^4p}{(2\pi)^4} 
\frac{e^{-ip(x-y)}}{p^0+\frac{\bff{p}^2}{2m}-i\epsilon}\,.
\label{eq:inverseKQ}
\end{eqnarray}
Changing variables to 
\blue{ $\psi'=\psi+K_Q^{-1}\cdot \Sigma\cdot \chi$}
(similarly for $\psi^\dag$) and integrating over $\psi$, $\psi^\dag$ 
we obtain 
\begin{eqnarray}
\int d^4x \int d^4y  \,
{\cal L}_{\chi, S,\sigma} (x,y) =
\int d^4x \int d^4y \, 
\left[ {\cal L}_{\psi,\chi, S,\sigma} (x,y)\right]_{\psi=0}
-\chi^\dag\cdot\Sigma^\dag\cdot K_Q^{-1} \cdot \Sigma \cdot \chi.
\qquad
\end{eqnarray}
The integration over $\chi, \chi^\dag$ is done analogously resulting 
in the partition function
\blue{
\begin{eqnarray}
&&
Z_{\rm PNRQCD} = 
\int {\cal D} S {\cal D} S^\dag  {\cal D} \sigma  {\cal D}\sigma^\dag
\det(1-K_{\bar{Q}}^{-1}\cdot \Sigma^\dag\cdot K_{Q}^{-1}\cdot \Sigma)
\\
&&\hspace*{0.5cm} \times
\exp
\left( i \!\int \!d^4x d^4 y \,
\bigg\{
\Sigma^\dag(y,x) S(x,y) +  S^\dag(y,x)\Sigma(x,y) 
- S^\dag(y,x)V(x,y)S(x,y) 
\bigg\}
\right).
\nonumber
\end{eqnarray}  
}
The determinant contains the $\sigma$ field. To write it as a term in 
the Lagrangian we use 
\begin{eqnarray}
&& \det(1 - K_{\bar{Q}}^{-1}\cdot \Sigma^\dag\cdot K_{Q}^{-1}\cdot \Sigma)
=
\exp \left\{ {\rm Tr} \ln\left(1 - K_{\bar{Q}}^{-1}\cdot 
\Sigma^\dag\cdot K_{Q}^{-1}\cdot \Sigma \right)\right\}
\nonumber\\
&& \hspace*{0.5cm} = 
\exp \left\{ -{\rm Tr}  
\,K_{\bar{Q}}^{-1}\cdot \Sigma^\dag\cdot K_{Q}^{-1}\cdot 
\Sigma +\ldots \right\},
\label{eq:expdet}
\end{eqnarray}
and keep only the bilinear term in the $\Sigma$ field in the expansion 
of the logarithm. We comment on the other terms below. The 
\blue{effective action} after this procedure is 
\blue{
\begin{eqnarray}
&&\int d^4x \, \int d^4y \, {\cal L}_{S, \sigma}(x,y) =
\nonumber\\
&&\hspace*{0.5cm}
\int d^4x \int d^4y \,
\bigg(\Sigma^\dag(y,x) S(x,y) +  S^\dag(y,x)\Sigma(x,y) 
- S^\dag(y,x)V(x,y)S(x,y)\bigg)
 \nonumber\\
 && \hspace*{0.5cm}+ \,  
\int d^4x \int d^4y \int d^4x' \int d^4y' \,\Sigma^\dag(y',x') 
\,i K_{Q}^{-1}(x',x) K_{\bar{Q}}^{-1}(y,y') \Sigma(x,y)\,.
\end{eqnarray}
}
Now we integrate over the delta-functions in the time coordinates implicit 
in the definition of the $\Sigma$ field and obtain 
\blue{
\begin{eqnarray}
&& \int d^4x \, \int d^4y \, {\cal L}_{S, \sigma}(x,y) =
\int d^7z \,
\bigg(
 \sigma^\dag(z)S(z) +S^\dag(z)\sigma(z) - S^\dag(z)V(z)S(z) \bigg)
 \nonumber\\ 
&&\hspace*{0.5cm} 
+ \int d^7z d^7z' \,\sigma^\dag(z') \,i K_\sigma(z'; z) \sigma(z)\,,
\qquad
\end{eqnarray}
}
where $z=(t, \bff{x}, \bff{y})$ represents the coordinates of the quark and 
anti-quark at coincident time $t=x^0=y^0$. The fields with argument $z$ 
are defined as 
\begin{eqnarray}
S(z)&=& \left[ S(x,y)\right]_{x^0=y^0}
\nonumber\\
\sigma(z)&=&\left[ \sigma(x,y)\right]_{x^0=y^0},
\nonumber\\
K_\sigma(z',z)&=&\left[K_{Q}^{-1}(x',x) K_{\bar{Q}}^{-1}(y,y')
\right]_{t=x^0=y^0, \,t'=x^{0'}=y^{0'}}
\label{eq:Ksig}
\end{eqnarray}
The last step consists of performing the Gaussian integral over $\sigma$. 
The inverse of $K_\sigma$ is defined by 
\begin{equation}
\delta^{(7)}(z_1-z_2)
= \int d^7z \,K_\sigma^{-1}(z_1,z) K_\sigma(z,z_2),
\end{equation}
resulting in 
\blue{
\begin{equation}
\int d^7z \,{\cal L}_{S}(z) = 
\int d^7z d^{\,7}z' \,S^\dag(z')\,iK^{-1}_\sigma(z',z)S(z) 
- \int d^7z \,S^\dag(z)V(z)S(z) \,.
\end{equation}
}
To compute $iK^{-1}_\sigma(z',z)$ we use (\ref{eq:Ksig}) and the 
definitions (\ref{eq:inverseKQ}). The integrals over the zero-components 
of the two momenta \blue{from (\ref{eq:inverseKQ})} 
can be written as integrals over relative momentum $q^0$ 
and total momentum $P^0$. Since the exponentials are independent of 
$P^0$, the $P^0$ integral can be performed by contour integration 
which gives 
\begin{equation}
K_\sigma(z',z) = \int\frac{d^7K}{(2\pi)^7}\,
e^{-i K(z^\prime-z)}\,\frac{i}{q^0-\frac{{\bf p}^2}{2m} -
\frac{{\bf p}^{\prime 2}}{2m}}\,,
\end{equation} 
where $K=(q^0,{\bf p},{\bf p}^\prime)$. It follows that  
\begin{equation}
K_\sigma^{-1}(z, z') = \delta^{(7)}(z-z')\, 
(-i)\,\bigg[
i\partial_0 +\frac{\bffmath{\partial}_x^2}{2m}+
\frac{\bffmath{\partial}_y^2}{2m}
\bigg]\,,
\end{equation} 
and therefore 
\begin{equation}
Z_{\rm PNRQCD}=\int {\cal D}S{\cal D} S^\dag \,
\exp \left\{\,i \!\int d^7z \,{\cal L}_{S}(z)\right\}
\end{equation}
with 
\begin{eqnarray}
{\cal L}_{S}(z) &=&
\int d^7z \,S^\dag(z) 
\bigg\{\,
i\partial_0 +\frac{\partial_z^2}{2m} - V(z)
\bigg\} S(z).
\end{eqnarray}
After separating the free centre-of-mass motion this represents the 
PNRQCD Lagrangian expressed in terms of the composite quark anti-quark 
field.

When one keeps the higher-order terms in the expansion of the logarithm 
in (\ref{eq:expdet}) the path-integral over $\sigma$ can no longer be 
done exactly. Expanding the quartic and higher-order terms in the 
exponential, we obtain vertices involving four and more 
$S$ fields, which describe scattering of composite fields. These terms 
are clearly not relevant to the threshold dynamics of a single 
quark anti-quark pair.

\subsection{Explicit forms of the propagator 
(Coulomb Green function)}
\label{sec:CoulGreen} 

In four dimensions explicit solutions for the 
the Schr\"odinger equation (\ref{eq:schroedinger}) can be found, 
equivalent to the sum of diagrams (\ref{eq:coulombfn}). We 
quote the results for the colour-singlet Green function. 
The general case is obtained by substituting $C_F\to -D_R$ 
everywhere.

The momentum space PNRQCD propagator (Coulomb Green function)
can be expressed in the form 
\cite{Schwinger:1964zzb} 
\begin{eqnarray}
G_0(\bff{p},\bff{p}^\prime;E) &=&
-\frac{(2\pi)^3\delta^{(3)}(\bff{p}^\prime-\bff{p})}
{E -\frac{\bff{p}^2}{m}} + 
\frac{1}{E-\frac{\bff{p}^2}{m}}\,
\frac{g_s^2 C_F}{(\bff{p}-\bff{p}^\prime)^2} \,
\frac{1}{E-\frac{\bff{p}^{\prime\,2}}{m}}
\nonumber\\
&& \hspace*{-1.5cm}+\,
\frac{1}{E-\frac{\bff{p}^2}{m}}\,
\int_0^1 dt\,\frac{g_s^2 C_F\,\lambda \,t^{-\lambda}}
{(\bff{p}-\bff{p}^\prime)^2 \,t - \frac{m}{4 E} 
(E-\frac{\bff{p}^2}{m})(E-\frac{\bff{p}^{\prime\,2}}{m}) (1-t)^2} \,
\frac{1}{E -\frac{\bff{p}^{\prime\,2}}{m}}\,,
\qquad
\label{eq:schwingerrep}
\end{eqnarray}
which closely resembles (\ref{eq:coulombfn}) and shows that 
the sum from $n=1$ to infinity in (\ref{eq:Hc}) can be transformed into a 
remarkably simple integral.\footnote{Note the sign change 
compared to \cite{Schwinger:1964zzb}, since 
Schwinger defines the Green function 
with an opposite sign.} At this point we omit the $+i\epsilon$ 
prescription on $E$ and regard $G_0(\bff{p},\bff{p}^\prime;E)$ 
as a function of a complex 
energy variable, which has a cut for $E>0$ and isolated poles 
on the negative real axis. The variable $\lambda$ equals  
$\alpha_s C_F/(2\sqrt{-E/m})$ as defined 
in (\ref{eq:deflambda}). The first line of (\ref{eq:schwingerrep}) 
separates the zero- and one-Coulomb gluon exchange terms. 
In practice, we find it simpler to perform the all-order 
summation in the position space representation, where the 
potential insertions take a simple multiplicative (rather than 
convolutive) form, and therefore we do 
not make use of the above representation in the calculation in paper~II.

An integral representation for the position space Coulomb Green 
function is 
\begin{eqnarray}
G_0(\bff{r},\bff{r}^\prime;E) &=& 
-\frac{m}{4\pi\Gamma(1+\lambda)\Gamma(1-\lambda)}
\int_0^1dt \int_1^\infty ds \,[s(1-t)]^\lambda [t(s-1)]^{-\lambda}\,
\nonumber\\
&&\times 
\frac{\partial^2}{\partial t\partial s} 
\left(\frac{ts}{|s\bff{r}-t\bff{r}^\prime|} \,
e^{-\sqrt{-mE}\,((1-t)r^\prime+(s-1)r+|s\bff{r}-t\bff{r}^\prime|)}
\right),
\end{eqnarray}
valid for $r>r^\prime$, where $r=|\bff{r}|$, 
$r^\prime=|\bff{r}^\prime|$~\cite{Wichmann1961}. 
For $r<r^\prime$ exchange $\bff{r}\leftrightarrow \bff{r}^\prime$ in the 
above expression. Putting one of the arguments to zero, this simplifies 
to  
\begin{equation}
G_0(0,r;E)=\frac{m
\sqrt{-mE}}{2\pi}\,e^{-\sqrt{-mE} \,r}\int_{0}^{\infty}ds\,
e^{-2rs\sqrt{-mE}}\left(\frac{1+s}{s}\right)^{\!\lambda}\,,
\label{eq:greenint} 
\end{equation}
which depends only on $r=|\bff{r}|$. We use this form of the Coulomb 
Green function mainly for propagators connecting to the external 
current vertex, in which case (\ref{eq:greenint}) applies.

For the general case of a propagator in between two potential insertions 
the representation of the position-space Green function in terms 
of Laguerre polynomials 
$L_n^{(2l+1)}(x)$ \cite{Voloshin:1985bd,Voloshin:1979uv} turns out 
to be most useful. In this representation one first performs 
a partial wave expansion 
\begin{equation}
G_0(\bff{r},\bff{r}^\prime;E) = 
\sum_{l=0}^{\infty}\,(2l+1)\,
P_l\!\left(\frac{\bff{r}\cdot \bff{r}^\prime}{r r^\prime}\right)
G_{[l]}(r,r^\prime;E)\,,
\label{eq:gplexpand}
\end{equation}
where $P_l(z)$ are the Legendre polynomials. The partial-wave 
Green functions read
\begin{equation}
 G_{[l]}(r,r^\prime;E) = 
\frac{m p}{2\pi} \, (2 p r)^l (2 p r^\prime)^l\,
e^{-p(r+r^\prime)} 
\sum_{s=0}^{\infty}\,
\frac{s!\,L_s^{(2l+1)}(2pr) L_s^{(2l+1)}(2 p r^\prime)}
{(s+2l+1)! (s+l+1-\lambda)}\,,
\label{eq:gpartial}
\end{equation}
where $p=\sqrt{-m E}$, and the Laguerre polynomials are defined 
by 
\begin{equation}
L_s^{(\alpha)}(z) = \frac{e^z z^{-\alpha}}{s!}\,
\left(\frac{d}{dz}\right)^{\!s} \left[e^{-z} z^{s+\alpha}\right].
\end{equation}
Since at NNNLO accuracy the potential insertions cannot 
change the angular momentum of 
the quark anti-quark pair and since the production current 
$\psi^\dagger\sigma^i\chi$ creates an S-wave state, we only need 
the $l=0$ Green function to compute the potential contributions to the 
PNRQCD correlation function (\ref{eq:GPNRQCD}). The P-wave Green 
function is required to compute the ultrasoft 
contribution~\cite{Beneke:2007pj,Beneke:2008cr} and the contribution 
\blue{(\ref{eq:piAtoNRQCD}) from 
the P-wave production current \cite{Beneke:2013kia}.}

A general property of the Coulomb interaction is that the ultraviolet 
behaviour of the ladder diagrams improves with the number of exchanges. 
Thus, when the external current or potential insertions cause 
UV divergences, it is necessary to subtract only the first few terms 
in the sum of ladder diagrams. The divergent diagrams must be 
done in $d$ dimensions using standard methods, while for the 
convergent remainder one of the above expressions, properly subtracted, 
can be used. We therefore use the notation\footnote{In an abuse of notation 
we now use a superscript on the Green function to denote a) the colour 
representation and b) the number of Coulomb 
exchanges. What is meant should be clear from the context.}
\begin{eqnarray}
\label{eq:greensplit}
G_0(...;E)&=&G_0^{(0ex)}(\ldots;E)+G_0^{(1ex)}(...;E)+ \ldots +
G_0^{(n\,ex)}(...;E)+G_0^{(>n\,ex)}(...;E)\,.\;\qquad
\end{eqnarray}
For example, the three terms in (\ref{eq:schwingerrep}) correspond to 
$G_0^{(0ex)}(\ldots;E)+G_0^{(1ex)}(...;E)+G_0^{(>1\,ex)}(...;E)$.

From (\ref{eq:Gexp}) it follows that the leading term in 
$G(E)$ equals $G_0(\bff{r}=0,\bff{r}^\prime = 0;E)$, which is, 
however, divergent as can be seen from (\ref{eq:greenint}). 
To compute $G_0(0,0;E)$ in dimensional regularization, we note 
that the zero-Coulomb exchange term is linearly divergent, the 
one-Coulomb exchange logarithmically, and the remainder is 
convergent. We therefore compute $G_0^{(0ex)}(0,0;E)+G_0^{(1ex)}(0,0;E)$ 
from the first line of (\ref{eq:schwingerrep}), 
which yields 
\begin{eqnarray}
G_0^{(0+1ex)}(0,0;E) &=&
\int\frac{d^{d-1}\bff{p}}{(2\pi)^{d-1}}
\frac{-1}{E -\frac{\bff{p}^2}{m}} + 
\int\frac{d^{d-1} \bff{p}}{(2\pi)^{d-1}}
\frac{d^{d-1} \bff{p}^\prime}{(2\pi)^{d-1}}
\frac{1}{E-\frac{\bff{p}^2}{m}}\,
\frac{g_s^2 C_F}{(\bff{p}-\bff{p}^\prime)^2} \,
\frac{1}{E-\frac{\bff{p}^{\prime\,2}}{m}}
\nonumber\\
&& \hspace*{-2cm}= 
\frac{m^2}{4\pi}\left[-\sqrt{-\frac{E}{m}} -\alpha_s C_F 
\left\{-\frac{1}{4\epsilon}+\frac{1}{2}\ln\left(\frac{-4 m E}{\mu^2}\right) -\frac{1}{2}
\right\}+ O(\epsilon)\right].
\label{eq:G01ex}
\end{eqnarray}
The remaining terms can 
be calculated from a modified version of (\ref{eq:greenint}) with $r=0$ 
and integrand $(1+s)^\lambda \,s^{-\lambda+\delta}$. After subtracting 
the first two terms of the $\alpha_s$ expansion the result is finite 
as the regulator $\delta\to 0$ and gives the remaining 
contribution $G^{(>1ex)}(0,0;E)$. The final result for the 
$\overline{\rm MS}$ subtracted zero-distance Green 
function \cite{Beneke:1999zr,Eiras:1999xx} is 
\begin{equation}
G_0^{\overline{\rm MS}}(0,0;E)=
\frac{m^2}{4\pi}\left[-\sqrt{-\frac{E}{m}} -\alpha_s C_F 
\left\{
\frac{1}{2}\ln\left(\frac{-4 m E}{\mu^2}\right) -\frac{1}{2} 
+\gamma_E+\Psi(1-\lambda)
\right\}\right]\,.
\label{eq:G00MSbar}
\end{equation}
The poles of the Euler Psi-function at positive integer $\lambda$ 
correspond to the S-wave quark anti-quark bound states.  
Near the bound-state poles the Green function takes the form 
\begin{equation}
G_0(0,0;E) \stackrel{E\rightarrow
E_n}{=}\frac{|\psi_n(0)|^2}{E_n-E-i \epsilon} + 
\mbox{regular}\,,
\label{eq:Gpole} \end{equation}
where $\psi_n(0)$ is the wave-function at the origin of the 
$n$th bound state with energy 
$E_n =- (m\alpha_s^2 C_F^2)/(4 n^2)$. 
The imaginary part of the Green function for $E>0$ 
is known as the Sommerfeld factor~\cite{Sommerfeld:1931}. 
Explicitly, the imaginary part below and above threshold is 
given by 
\begin{equation}
\mbox{Im}\,G_0(0,0;E)  = \sum_{n=1}^\infty \frac{1}{8}\,
\left(\frac{\alpha_s C_F m}{n}\right)^{\!3} \delta(E-E_n) 
+\theta(E)\,\frac{m^2}{4\pi}\,
\frac{\pi\alpha_s C_F}{1-e^{-\frac{\pi\alpha_s C_F}{v}}}\,,
\end{equation}
for real energies $E=m v^2$. This expression can be used in (\ref{R1}) to 
obtain the leading-order approximation to the resummed 
top pair production cross section in the threshold region for 
vanishing decay width of the top quark.

\subsection{Potentials}
\label{sec:potentials}

We now summarize the momentum-space potentials required for the 
NNNLO calculation with the PNRQCD Lagrangian  (\ref{eq:pnrqcd}). 
Only the colour-singlet projection
\begin{equation}
V(\bff{p},{\bff p}^\prime) = 
\frac{1}{N_c}\,\delta_{bc}\delta_{da}
\,V_{ab;cd} (\bff{p},{\bff p}^\prime) 
= V_1 (\bff{p},{\bff p}^\prime) +C_F V_ T(\bff{p},{\bff p}^\prime)
\label{eq:singlet-potential2}
\end{equation}
of the general quark anti-quark potential 
$V_1 1_{ab} 1_{cd} +V_T T^A_{ab} T^A_{cd}$ is relevant 
to top-quark pair production through the electromagnetic and electroweak  
current, and this will be given in the following.

The various potential terms can be ordered in a $1/m$ expansion, 
beginning with the Coulomb potential of order $1/m^0$.  
Allowing for the spin-dependence from order $1/m^2$, we 
write the singlet-potential in the general form,
\begin{eqnarray}
V({\bf p},{\bf p}^{\prime})&=& 
-{\cal V}_C(\alpha_s)\frac{4\pi C_{F}\alpha_s}{{\bf
q}^2} + 
{\cal V}_{1/m}(\alpha_s) \frac{\pi^2 (4\pi)C_F\alpha_s} {m|{\bf q}|}  
\nonumber  \\ 
&& +\, {\cal V}_{\delta}(\alpha_s)\frac{2\pi C_F \alpha_s}{m^2}
\nonumber
 -  {\cal V}_{s}(\alpha_s)\frac{\pi C_F
 \alpha_s}{4m^2}[\sigma_i,\sigma_j]\otimes[\sigma_i,\sigma_j] 
\nonumber  \\ 
&&   -\,{\cal V}_{p}(\alpha_s)\frac{2\pi C_F
 \alpha_s({\bf p}^2+{\bf p}^{\prime 2})}{m^2{\bf{q}}^2}
+{\cal V}_{hf}(\alpha_s)\frac{\pi C_F \alpha_s}{4m^2{\bf q}^2}
 [\sigma_i,\sigma_j]q_j\otimes [\sigma_i,\sigma_k]q_k
\nonumber  \\ 
&&  - {\cal V}_{so}(\alpha_s)\frac{3\pi C_F \alpha_s}{2m^2 {\bf q}^2}
 \Bigg([\sigma_i,\sigma_j]q_i p_j
 \otimes 1
 -1\otimes [\sigma_i,\sigma_j]q_i p_j\Bigg)
+ \ldots,
\label{eq:potentialbeforespin} 
\end{eqnarray}
\blue{defining ${\bf q} = \bf{p}^\prime - {\bf p}$.}
The coefficients ${\cal V}_{X}$ of the potentials are $\alpha_s$
dependent:
\begin{eqnarray}
{\cal V}_{i}(\alpha_s)={\cal V}_{i}^{(0)}+\frac{\alpha_s}{4\pi}{\cal
V}_{i}^{(1)}+\bigg(\frac{\alpha_s}{4\pi}\bigg)^{\!2}{\cal
V}_{i}^{(2)}+ O(\alpha_s^3).
\label{eq:calVexp}
\end{eqnarray}
The above representation of the potential therefore corresponds to an 
expansion in  $\alpha_s$ and $v$, since the potential momenta 
$\bff{p}$, $\bff{p}^\prime$ and $\bff{q}$ are of order $mv$. 
With $v\sim \alpha_s$ the leading term is of order $\alpha_s/{\bf q}^2 
\sim 1/v$. The ellipses denote terms of order $\alpha_s^2 {\bf |q|}/ m^3, 
\alpha_s {\bf q}^2/m^4 \sim v^3$, which would contribute from 
N${}^4$LO. A notation has been used which is valid in $d$ dimensions
by avoiding the use of vector products or the totally antisymmetric 
$\epsilon$ tensor that would arise from using the three-dimensional 
identity for the commutator of Pauli matrices. 
The coefficients of the potentials are chosen 
such that the leading-order coefficients are either one or
zero. The tensor products $a\otimes b$ refer to the spin matrices 
on the quark ($a$) and anti-quark line ($b$). For the first three 
and the fifth terms of (\ref{eq:potentialbeforespin}), which are 
spin-independent, 
we omitted the trivial $1\otimes 1$ factor.

The on-shell matching calculation of the potential 
coefficients ${\cal V}_{C}^{(3)}$,
${\cal V}_{1/m}^{(2)}$, ${\cal V}_{\delta}^{(1)}$, ${\cal
V}_{p}^{(1)}$ results in infrared (IR) divergences
\cite{Brambilla:1999xj,Kniehl:2002br,Appelquist:1977es}, which are related to 
ultraviolet divergences in the calculation of the  ultrasoft correction. 
It is convenient to subtract these divergences in the results given below 
and add them back to the
ultrasoft calculation (see \cite{Beneke:2007pj,Beneke:2008cr} 
and section \ref{sec:ultrasoft}). The
subtraction term, which is {\em added}\footnote{
Note that in \cite{Beneke:2007gj} it was incorrectly stated that this 
term should be subtracted from the potential (rather than added to it).} 
to the potential, is
\begin{eqnarray}
\delta V_{c.t.} &=& \frac{\alpha_s C_F}{6\epsilon}
\Bigg[ C_A^{3}\frac{\alpha_s^{3}}{{\bf q}^2} + 4\left(C_A^{2}+2C_A
C_F\right)\, \frac{\pi\alpha_s^{2}}{m |{\bf q}|}
\nonumber \\
&& +16\left(C_F-\frac{C_A}{2}\right)\frac{\alpha_s}{m^2} +
16C_A\frac{\alpha_s}{m^2}\, \frac{{\bf p}^2+{\bf
p}^{\prime\,2}}{2 {\bf q}^2} \Bigg]. 
\label{eq:counter}
\end{eqnarray}

In addition to potential insertions, the relativistic correction 
to the kinetic energy term  $\pm \bffmath{\partial}^4/(8m^3)$ in 
(\ref{eq:pnrqcd}) needs to be included in PNRQCD perturbation 
theory. Formally, this can be done by adding 
\begin{equation}
V_{\rm kin}=-\frac{\bff{p}^4}{4m^3}(2\pi)^{d-1}\delta^{(d-1)}({\bf
p}-{\bf p}^{\prime})\, .
\end{equation}
to the potential. Since $V_{\rm kin}\sim v$, it counts as a NNLO 
potential. The delta function eliminates the momentum integration 
that is associated with a potential insertion, but which is not 
present for the kinetic energy correction to a (anti-)quark 
propagator.

We now present the results for the potential coefficients, starting
with the Coulomb potential.

\subsubsection{The Coulomb potential}
\label{sec:coulombpotential}

The coefficient ${\cal V}_{C}(\alpha_s)$ encodes the quantum corrections 
to the Coulomb potential, which are needed up to the three-loop order.
The insertions of Coulomb potentials are finite, so we do not need 
the $d$-dimensional expression of the potential, as long as only
Coulomb potential insertions are considered. This reflects the fact 
that the Schr\"odinger equation with the $1/r$ potential is non-singular 
and could be solved exactly, without referring to PNRQCD perturbation 
theory, as was done, for instance, in \cite{Beneke:2005hg}. 
However, in the third-order computation of the top anti-top production 
cross section also the double insertion of the NLO Coulomb potential together 
with the singular insertion of a NNLO non-Coulomb potential has to be 
taken into account; hence the order $\epsilon$ part of the one-loop 
Coulomb potential multiplies a divergent quantity and contributes to the
final result. The coefficient ${\cal V}_{C}^{\,(1)}$ is therefore 
given with the full $\epsilon$ dependence.

The first four terms in the expansion of the Coulomb potential 
can be represented in the form 
\begin{eqnarray}
\label{eq:vcoulomb}
{\cal V}_{C}^{\,(0)} &=&1,
\\
{\cal V}_{C}^{\,(1)} &=&
   \bigg[\bigg(\frac{\mu^2}{\bff{q}^2} \bigg)^{\!\epsilon} -1
   \bigg]\, \frac{\beta_0}{\epsilon}\,
+\bigg(\frac{\mu^2}{\bff{q}^2} \bigg)^{\!\epsilon}\, a_{1}(\epsilon),
\label{eq:vcoulombNLO}\\
{\cal V}_{C}^{\,(2)} &=& a_2+ 
\left(2a_1\beta_0+\beta_1\right)\ln\frac{\mu^2}{\bff{q}^2}
+\beta_0^2\ln^2\frac{\mu^2}{\bff{q}^2},
\\
{\cal V}_{C}^{\,(3)} &=& a_3 + 
\left(2a_1\beta_1+\beta_2+3a_2\beta_0+8\pi^2C_A^3\right)
\ln\frac{\mu^2}{\bff{q}^2}
\nonumber\\ 
&&+\left(\frac{5}{2}\beta_0\beta_1+3a_1\beta_0^2\right)
\ln^2\frac{\mu^2}{\bff{q}^2}+\beta_0^3\ln^3\frac{\mu^2}{\bff{q}^2},
\label{eq:vcoulombN3LO}
\end{eqnarray}
with
\blue{
\begin{eqnarray}
&& a_{1}(\epsilon) = \bigg(C_A\,[11-8\epsilon]-4\,T_Fn_f\bigg)\,
\frac{e^{\gamma_E\epsilon}\,\Gamma(1-\epsilon)\,\Gamma(2-\epsilon)\,
\Gamma(\epsilon)\, }{(3-2\epsilon)\,\Gamma(2-2\epsilon)}
-\frac{\beta_0}{\epsilon}
\\
&&  \hspace*{1cm} 
= \frac{31}{9} C_A - \frac{20}{9} T_F n_f + O(\epsilon)
\\
&& a_{2} = \left(\frac{4343}{162}+4\pi^2 -\frac{\pi^4}{4}
+\frac{22}{3}\zeta_{3}\right)C_{A}^{2}
-\left(\frac{1798}{81}+\frac{56}{3}\zeta_{3}\right) C_{A}T_{F}n_f
\nonumber\\
&& \hspace*{1cm} 
-\,\left(\frac{55}{3}-16\zeta_{3}\right) C_F T_{F}n_f + 
\frac{400}{81}\,(T_F n_f)^2.
\end{eqnarray}
}
The one-loop correction $a_1=a_1(\epsilon)$ has been known in $d=4$ 
for some time \cite{Fischler:1977yf,Billoire:1979ih}. We computed 
the $d$-dimensional expression and confirmed the result first shown
in~\cite{Schroder:1999sg}. The term $\beta_0/\epsilon$ without a 
momentum factor arises from the charge renormalization counterterm 
and the square bracket in (\ref{eq:vcoulombNLO}) produces the 
associated logarithm. However, the expansion in $\epsilon$ can only 
be done after the momentum integrals of the potential insertion 
are performed, if these integrals are divergent. The two-loop 
coefficient $a_2$ has first been calculated in
\cite{Peter:1996ig,Peter:1997me} and correctly in \cite{Schroder:1998vy}. 

\blue{The three-loop coefficient $a_3$ was first computed in a partially 
numerical form~\cite{Anzai:2009tm,Smirnov:2009fh}. We quote the analytic 
result from \cite{Lee:2016cgz}: 
\begin{eqnarray}
a_3 &=&
-\left(\frac{20}{9}n_f T_F\right)^3 \!
\nonumber \\
&&
+\,\bigg\{
       \left( \frac{12541}{243}
                +\frac{64\pi^4}{135}
                +\frac{368}{3}\zeta_3\right) C_A
   +\left( \frac{14002}{81}
                -\frac{416}{3}\zeta_3\right) C_F
 \bigg\} \,(n_f T_F)^2 
\nonumber \\
&&
+\,\bigg\{\,
\bigg(
-\frac{58747}{486}
+\pi^2 \left[\frac{17}{27} + \ln 2
\left(-\frac{4}{3}-14\zeta_3\right)-\frac{4}{3}\ln^4 2
-32\,\mbox{Li}_4(1/2)-\frac{19}{3}\zeta_3\right]
\nonumber\\
&&\hspace{1cm}
+\,\pi^4 \left[-\frac{157}{54}-\frac{5}{9}\ln 2+\ln^2 2\right]
+\frac{761 \pi^6}{2520}
-356\zeta_3+\frac{57}{2}\zeta_3^2+\frac{1091}{6}\zeta_5 
-48 s_6
\bigg)\,C_A^2
\nonumber\\
&&\hspace{1cm}
+\,\left(-\frac{71281}{162}+264\zeta_3+80\zeta_5\right) C_A C_F
\nonumber\\
&&\hspace{1cm}
 +\,
\left(\frac{286}{9}+\frac{296}{3}\zeta_3-160\zeta_5\right) C_F^{\,2}
\bigg\}\,n_f T_F
\nonumber \\
&&
+\,\bigg\{ 
\pi^2 \left(\frac{1264}{9}-\frac{976}{3}\zeta_3+\ln 2\left[
64+672\zeta_3\right]
\right)
+\pi^4\left(-\frac{184}{3}+\frac{32}{3}\ln 2-32\ln^2 2\right)
\nonumber\\
&&\hspace{1cm}
+\,\frac{10 \pi^6}{3}
\bigg\}\,
\biggl(\frac{d_F^{abcd}d_F^{abcd}}{N_A}\biggr)\,
n_f
\nonumber \\
&&
+\,\bigg\{\,
\frac{385645}{2916}
+\pi^2 \bigg[-\frac{953}{54} + \ln 2
\left(-\frac{922}{9}+\frac{217}{3}\zeta_3\right)+\frac{73}{9}\ln^4 2
+\frac{584}{3}\,\mbox{Li}_4(1/2)\nonumber\\
&&\hspace{1cm}
+\,\frac{175}{2}\zeta_3\bigg]
+\pi^4 \left[\frac{1349}{270}-\frac{20}{9}\ln 2-\frac{40}{9}\ln^2 2\right]
-\frac{4621 \pi^6}{3024}
+\frac{584}{3}\zeta_3-\frac{143}{2}\zeta_3^2
\nonumber \\
&&\hspace{1cm}
-\,\frac{1927}{6}\zeta_5 
+144 s_6
\bigg\}\, C_A^{\,3}
\nonumber \\
&&
+\,\bigg\{
\pi^2 \left(\frac{7432}{9}-\frac{6616}{3}\zeta_3+\ln 2\left[
\frac{14752}{3}-3472\zeta_3\right]-\frac{592}{3}\ln^4 2
-4736\,\mbox{Li}_4(1/2)
\right)
\nonumber\\
&&\hspace{1cm}
+\,\pi^4\left(-156+\frac{560}{3}\ln 2+\frac{496}{3}\ln^2 2\right)
+\frac{1511 \pi^6}{45}
\bigg\}\,
\biggl(\frac{d_A^{abcd}d_F^{abcd}}{N_A}\,\biggr)
\,.
\label{eq:a3}
\\
&=& 13432.5648565 -  3289.9052968 n_f + 185.9900266 n_f^2 - 
1.3717421 n_f^3.
\end{eqnarray}
The colour factors $d_X^{abcd}d_Y^{abcd}$ for $SU(N_c)$ 
are given by
\begin{equation}
\frac{d_F^{abcd}d_F^{abcd}}{N_A} =
\frac{N_c^4-6N_c^2+18}{96N_c^2},
\qquad\quad
\frac{d_A^{abcd}d_F^{abcd}}{N_A} =
\frac{N_c\left(N_c^2+6\right)}{48}.
\end{equation}
and the constant $s_6$ containing a multiple zeta value equals
$s_6 = \zeta(-5,-1) + \zeta_6 = 0.98744142640329971377$.}
Parts of the N4LO Coulomb potential are also known
\cite{Brambilla:2006wp}, but are not needed for the third-order cross 
section calculation.

The third-order Coulomb potential has an IR 
divergence~\cite{Brambilla:1999qa,Kniehl:2002br,Appelquist:1977es},
which cancels against a divergence in the calculation of
the NNNLO ultrasoft calculation. The corresponding $1/\epsilon$ pole is 
subtracted and therefore does not appear in (\ref{eq:vcoulombN3LO}), 
but the logarithmic part multiplied by $C_A^3$ in this equation 
comes from this divergence. Note that the coefficient of this 
logarithm agrees with \cite{Kniehl:2002br}, but is three times 
larger than the one in \cite{Brambilla:1999qa}. The reason for this 
is that here as in  \cite{Kniehl:2002br} all three loops are 
computed in $d$ dimensions not just the divergent one, as is required 
for consistency with the ultrasoft calculation. Hence the divergence 
related to the Coulomb potential in (\ref{eq:counter}) is multiplied 
by $(\mu^2/{\bf{q}}^2)^{3\epsilon}$ rather than  
$(\mu^2/{\bf{q}}^2)^{\epsilon}$.

\subsubsection{The $1/m$ potential}
The coefficient of the $O(1/m^1)$ potential is generated first at the 
one-loop order, where it is suppressed by $\alpha_s v$ relative to the 
leading Coulomb potential. Hence the two-loop coefficient is required 
for the NNNLO calculation of the cross section. The insertions of this 
potential cause ultraviolet divergences such that we need 
the one-loop coefficient to $O(\epsilon^2)$ and the two-loop one 
to $O(\epsilon)$. Up to the two-loop order we can represent the 
$O(1/m)$ potential in the form
\begin{eqnarray}
\label{eq:vNA} {\cal V}_{1/m}^{\,(0)}&=&0\,,
\\
{\cal V}_{1/m}^{\,(1)} &=& \bigg(\frac{{\mu }^2}{\bff{q}^2}
\bigg)^{\!\epsilon } \, b_{1}(\epsilon)\,,
\label{eq:vb1}
\\
{\cal V}_{1/m}^{\,(2)} &=&
  \Bigg[\bigg(\frac{{\mu }^2}{\bff{q}^2} \bigg)^{\!2 \epsilon }-1
  \Bigg]\left(-\frac{8}{3\epsilon}\right)
           \bigg(2 C_F C_A + C_A^2 \bigg)\,
\nonumber \\
&& +\, \Bigg[\bigg(\frac{\mu^2}{\bff{q}^2}\bigg)^{\!2\epsilon}
    - \bigg( \frac{{\mu }^2}{\bff{q}^2}\bigg)^{\!\epsilon}\,
  \Bigg]\, \frac{2\beta_0}{\epsilon} \, b_{1}(\epsilon)\,
+ \bigg(\frac{\mu^2}{\bff{q}^2} \bigg)^{\! 2\epsilon }\, 4 b_{2}(\epsilon)\,,
\label{eq:vb2}
\end{eqnarray}
with
\begin{eqnarray}
b_{1}(\epsilon) &=& \left(\frac{C_F}{2}\,[1-2\epsilon]
      -C_A\,[1-\epsilon]
\right)\, \frac{
e^{\gamma_E\epsilon}\,\Gamma(\frac{1}{2}-\epsilon)^{2}
\Gamma(\frac{1}{2}+\epsilon)}{\pi^{\frac{3}{2}}\,\Gamma(1-2\epsilon)}\,,
\label{b1eps}\\
b_{2}(\epsilon) &=&
 \left[\frac{65}{18}-\frac{8 }{3}\ln{2}\right] C_AC_F
-\left[\frac{101}{36}+\frac{4 }{3}\ln{2}\right] C_A^{2}
\nonumber \\
&& + \left[\frac{49}{36}\,C_A -\frac{2}{9} C_F\right] T_F n_f +
\epsilon \,b_{2}^{(\epsilon)}+O(\epsilon^2)\,.
\end{eqnarray}
Once again we subtracted the $1/\epsilon$ 
IR pole that remains after charge renormalization 
by adding the relevant part of 
$\delta V_{c.t.}$ from (\ref{eq:counter}). Hence (\ref{eq:vb2}) is 
finite; however, the expansion in $\epsilon$ must be performed only 
after the computation of the potential insertion.

The one-loop expression  $b_1(\epsilon)$ has been computed in 
$d$ dimensions~\cite{Beneke:1999qg}, and the four-dimensional 
value $b_2(\epsilon=0)$ of the two-loop coefficient is 
quoted from \cite{Kniehl:2001ju}. 
\blue{The $O(\epsilon)$ term 
of $b_{2}(\epsilon)$, which is also needed at NNNLO, has been 
parameterized above by $b_2^{(\epsilon)}$. Its analytic 
expression reads \cite{Beneke:2014qea}
\begin{eqnarray}
b_2^{(\epsilon)} &=&
  \left[-\frac{631}{108} - \frac{15\pi^2}{16} + \frac{65\ln2}{9}
    - \frac{8\ln^2 2}{3}\right] C_FC_A
  \nonumber\\
&&\mbox{}
  + \left[ -\frac{1451}{216} - \frac{161\pi^2}{72} 
    - \frac{101\ln2}{18} - \frac{4\ln^2 2}{3}\right]C_A^{\,2} 
  \nonumber\\
&&\mbox{}
  +\left\{
\left[\frac{115}{54} + \frac{5\pi^2}{18} + \frac{49\ln 2}{18}\right] 
C_A
+ \left[\frac{17}{27} - \frac{11\pi^2}{36} - \frac{4\ln 2}{9}\right]
C_F\right\}  T n_f 
  \,.
\label{eq:b2epsnew}
\end{eqnarray}
Note that for $n_f=5$, $b_2^{(\epsilon)} = -303.63$, which is 
significantly larger than the estimate $b_{2}^{(\epsilon)} 
=0 \pm 2 b_{2}(0)= 0 \pm 34$ employed in \cite{Beneke:2007gj} 
prior to the exact computation.}  

\subsubsection{The $1/m^2$ potential}
The coefficients of the $O(1/m^2)$ potentials are generated 
at tree level, where they are suppressed by $v^2$ relative to the 
leading Coulomb potential. Hence the one-loop coefficients are required 
for the NNNLO calculation of the cross section. The tree-level 
coefficients are:
\begin{eqnarray} {\cal V}_{\delta}^{\,(0)}=1, 
\qquad {\cal V}_{p}^{\,(0)}=1, 
\qquad{\cal V}_{so}^{\,(0)}=1,
\qquad {\cal V}_{hf}^{\,(0)}=1, 
\qquad {\cal V}_{s}^{\,(0)}=0.
\end{eqnarray}
As can be seen from (\ref{eq:potentialbeforespin}), spin-dependence 
arises first within the $O(1/m^2)$ potentials. The insertions of 
these potentials are again ultraviolet divergent. We therefore need the 
$O(\epsilon)$ term of the one-loop coefficients. These are available
only from  \cite{Wuester:2003}. We computed the $d$-dimensional 
expressions and confirmed the previous result. The spin-projected 
expression has already been given in our previous work 
\cite{Beneke:2007gj}.

\begin{figure}[t]
\begin{center}
     \includegraphics[height=3.5cm]{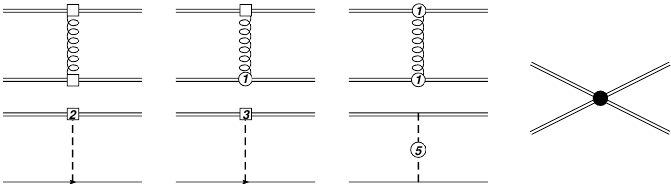}
\end{center}
\caption{\label{fig:NRQCDtree} NRQCD tree level diagrams of order
$1/m^2$.  Dashed (curly) lines denote 
the $A^0$ ($A^i$) gluon field. The number $i$ at the vertex refers 
to the NRQCD interaction with coefficient function $d_i$, see also 
figure~\ref{fig:NRQCD_FR} and (\ref{eq:nrqcdqq}), (\ref{eq:nrqcdqqqq}),
(\ref{eq:nrqcdg}). 
Symmetric diagrams are not shown.}
\end{figure}

\begin{figure}[t]
\begin{center}
     \includegraphics[height=10cm]{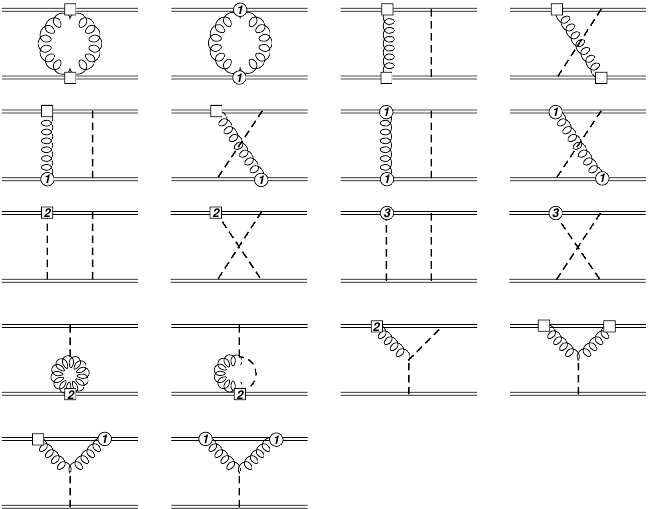}
\end{center}
\caption{\label{fig:NRQCDloop} NRQCD one-loop diagrams of order
$1/m^2$ (non-vanishing diagrams shown only, 
symmetric diagrams not displayed).}
\end{figure}

The complete one-loop coefficients consist of two different 
contributions. The first arises from the one-loop correction 
to the NRQCD couplings $d_i$, already discussed in the previous 
section. We call this the ``hard'' contribution. It can be 
extracted from the tree diagrams in figure \ref{fig:NRQCDtree} 
with one-loop corrected NRQCD vertices. The second contribution 
arises from explicitly integrating out soft loops.\footnote{And, 
in general, the part of potential loops not reproduced by the PNRQCD 
interactions of lower order. This happens for the $1/m$ potential 
as discussed in section~\ref{subsec:vectormatch} below.
However, there is no such contribution 
to the $1/m^2$ potentials at one loop. The reason for this is that 
in the case of the $1/m$ potential the relevant contribution 
arises from the box integral with all vertices of the leading-order 
$\psi^\dagger\psi A^0$ type, expanded to subleading order 
in the potential region. The next correction is always suppressed 
by {\em two} powers of $v$ (for instance, from replacing one of the 
vertices by a $O(v^2)$ vertex from the NRQCD Lagrangian), 
and hence can contribute only at order $1/m^3$.}
The corresponding 
one-loop NRQCD diagrams are shown in figure~\ref{fig:NRQCDloop}.
These contributions will be called ``soft''. The total
potential including tree and one-loop correction is 
then ${\cal V}_{X}(\alpha_s)={\cal
V}_{X}^{(hard)}(\alpha_s)+{\cal V}_{X}^{(soft)}(\alpha_s)$. 

The hard one-loop contributions can be easily calculated with the
Feynman rules presented in figure \ref{fig:NRQCD_FR}, since all
required $d$-dimensional 
NRQCD matching coefficients are already known at one-loop order from
the previous section. The result reads:
\begin{eqnarray}
\label{eq:vm2}
{\cal
V}_{\delta}^{(hard)}(\alpha_s)&=&\frac{1}{2}(1+d_2-16d_5)+\frac{1}{2\pi
C_F \alpha_s}(d_{ss}+C_F d_{vs})+O(\alpha_s^2)
\\ &=& \nonumber 1+\frac{\alpha_s}{\pi}
\bigg(\frac{\mu^2}{m^2}e^{\gamma_E}\bigg)^{\epsilon}\Gamma(\epsilon)
\bigg[C_A\frac{12\epsilon^3-44\epsilon^2+21\epsilon-13}{-96\epsilon^2+24}
\\
&+&C_F\frac{12\epsilon^4-32\epsilon^3-63\epsilon^2-4\epsilon+3}
{6(2\epsilon-1)(2\epsilon+1)(2\epsilon+3)}-\frac{2T_F}{15}\epsilon\bigg]
+O(\alpha_s^2),
\nonumber\\
{\cal V}_{p}^{(hard)}(\alpha_s) &=&1+O(\alpha_s^2),
\\
{\cal V}_{so}^{(hard)}(\alpha_s)
&=&\frac{1}{3}(2d_1+d_3)+O(\alpha_s^2)
\\ &=& \nonumber
1+\frac{\alpha_s}{\pi}\bigg(\frac{\mu^2}{m^2}e^{\gamma_E}\bigg)^{\epsilon}
\Gamma(\epsilon)
\bigg[\frac{C_A(2\epsilon^2-1)-2C_F\epsilon(2\epsilon+1)}{6\epsilon-3}\bigg]
+O(\alpha_s^2),
\\
{\cal V}_{hf}^{(hard)}(\alpha_s) &=&d_1^2+O(\alpha_s^2)
\\ &=& \nonumber
1+\frac{\alpha_s}{\pi}\bigg(\frac{\mu^2}{m^2}e^{\gamma_E}\bigg)^{\epsilon}
\Gamma(\epsilon)
\bigg[\frac{C_A(2\epsilon^2-1)-2C_F\epsilon(2\epsilon+1)}{4\epsilon-2}\bigg]
+O(\alpha_s^2),
\\
{\cal V}_{s}^{(hard)}(\alpha_s) &=&\frac{1}{2\pi C_F
\alpha_s}(d_{sv}+C_F d_{vv})+O(\alpha_s^2)
\label{eq:vm2hlast}
\\ &=& \nonumber
\frac{\alpha_s}{\pi}\bigg(\frac{\mu^2}{m^2}e^{\gamma_E}\bigg)^{\epsilon}
\Gamma(\epsilon)
\bigg[-\frac{C_A}{8}+\frac{\epsilon}{4\epsilon+2}C_F\bigg]+O(\alpha_s^2)\, .
\end{eqnarray}
In this notation the tree-level value of the potential coefficient 
is included and assigned to the hard contribution. 

The soft contributions come from diagrams of order $1/m^2$ 
shown in figure \ref{fig:NRQCDloop}, from diagrams of order
$1/m^0$, where the denominator of a propagator has been expanded 
to higher orders as appropriate to the soft region, 
from soft self-energy insertions containing gluons and 
light quarks into the tree diagrams, and from charge renormalization 
counterterms. The final result is:
\begin{eqnarray}
{\cal V}_{\delta}^{(soft)}(\alpha_s)
&=& \frac{\alpha_s}{4\pi}
\Bigg[\bigg(\frac{\mu^2}{{\bf q}^2}\bigg)^{\!\epsilon}
\frac{\epsilon\Gamma(-\epsilon)^2\Gamma(\epsilon)e^{\gamma_E\epsilon}}
{12(4\epsilon^2-8\epsilon+3)\Gamma(-2\epsilon)}
\Big(C_A(48\epsilon^3-230\epsilon^2+328\epsilon-138
\nonumber \\ && \nonumber -3d_1^2(4\epsilon^2-9\epsilon+5))
-4C_F(\epsilon-1)(16\epsilon^2-38\epsilon+21)
\\ && 
- 12n_f T_F(\epsilon-1)(1+d_2)\Big)-\frac{d_2+1}{2}
\frac{\beta_0}{\epsilon}\Bigg]+O(\alpha_s^2),
\label{eq:vm2s}\\
{\cal V}_{p}^{(soft)}(\alpha_s)
&=& \frac{\alpha_s}{4\pi}
\Bigg[\bigg(\frac{\mu^2}{{\bf q}^2}\bigg)^{\!\epsilon}
\frac{(-\epsilon)\Gamma(-\epsilon)^2\Gamma(\epsilon)
e^{\gamma_E\epsilon}\Big(C_A(56\epsilon^2-121\epsilon+57)
+ 12n_f T_F(\epsilon-1)\Big)}
{6(4\epsilon^2-8\epsilon+3)\Gamma(-2\epsilon)} 
\nonumber \\  
&& -\frac{\beta_0}{\epsilon}\Bigg]+O(\alpha_s^2),
\\
{\cal V}_{so}^{(soft)}(\alpha_s)
&=&\frac{\alpha_s}{4\pi}
\Bigg[\bigg(\frac{\mu^2}{{\bf q}^2}\bigg)^{\!\epsilon}
\frac{-\epsilon\Gamma(-\epsilon)^2\Gamma(\epsilon)e^{\gamma_E\epsilon}}
{6(4\epsilon^2-8\epsilon+3)\Gamma(-2\epsilon)}
\Big(C_A(d_3(8\epsilon^2-19\epsilon+11)
\nonumber \\
&&+2d_1(8\epsilon^2-15\epsilon+5))+ 4(2d_1+d_3)n_f
T_F(\epsilon-1)\Big)-\frac{2d_1+d_3}{3}\frac{\beta_0}{\epsilon}\Bigg]
\nonumber \\  &&+O(\alpha_s^2),
\\
{\cal V}_{hf}^{(soft)}(\alpha_s)
&=&\frac{\alpha_s}{4\pi}
\Bigg[\bigg(\frac{\mu^2}{{\bf q}^2}\bigg)^{\!\epsilon}
\frac{(-d_1^2)\epsilon(\epsilon-1)
\Gamma(-\epsilon)^2\Gamma(\epsilon)e^{\gamma_E\epsilon}
\Big(C_A(4\epsilon-5)+ 4n_f T_F\Big)}
{(8\epsilon^2-16\epsilon+6)\Gamma(-2\epsilon)}-
d_1^2\frac{\beta_0}{\epsilon}\Bigg]
\nonumber \\  &&+O(\alpha_s^2),
\\
{\cal V}_{s}^{(soft)}(\alpha_s)
&=&\frac{\alpha_s}{4\pi}
\bigg(\frac{\mu^2}{{\bf q}^2}\bigg)^{\!\epsilon}
\frac{d_1^2\epsilon\Gamma(-\epsilon)^2\Gamma(\epsilon)e^{\gamma_E\epsilon}
C_A}
{(8\epsilon-4)\Gamma(-2\epsilon)}+O(\alpha_s^2).
\label{eq:vm2slast}
\end{eqnarray}
Here the $d_i$ can be set to their tree-level value 1, 
since the entire expressions are already of order $\alpha_s$. 
An alternative method of calculating the soft
contribution is to extract the soft region directly from the 
QCD diagrams. In this way, there are fewer diagrams than in the 
NRQCD calculation (but more terms from the expansion). This calculation 
has also been performed and we checked that the results are exactly the
same. Note that with the second method we cannot get the result in
terms of the hard matching coefficients, but we obtain (\ref{eq:vm2s}) 
to (\ref{eq:vm2slast}) with $d_i=1$. \blue{The soft contributions 
to the potential were recently computed at the two-loop order 
\cite{Mishima:2024afk}, and contribute at N4LO, which is beyond 
the NNNLO accuracy that is currently achievable for the $t\bar{t}$ 
cross section.}

Some remarks are in order on the pole structure of the hard and soft 
$1/m^2$ potential coefficients. We first note that we did not 
yet add the relevant contribution from the counterterm (\ref{eq:counter}), 
so the expressions given are still divergent. The $1/\epsilon$ pole 
in the hard contribution is infrared, while the soft contribution 
contains both infrared and ultraviolet divergences. It is instructive 
to separate the two:
\begin{eqnarray}
\label{eq:vm2pole}
{\cal V}_{\delta}(\alpha_s)
&=& 1+ \frac{\alpha_s}{4\pi \epsilon} \,\Bigg\{
\bigg(\frac{\mu^2}{m^2}\bigg)^{\!\epsilon} 
\left[-\frac{13}{6} C_A-\frac{2}{3} C_F\right]_{\rm IR}
\nonumber\\
&& + \,\bigg(\frac{\mu^2}{{\bf q}^2}\bigg)^{\!\epsilon}
\left(
\left[\frac{13}{6} C_A+\frac{2}{3} C_F\right]_{\rm UV} +
\left[\frac{8}{3} C_A-\frac{16}{3} C_F\right]_{\rm IR} 
\right)
\Bigg\}+O(\alpha_s^2),
\\
{\cal V}_{p}(\alpha_s)
&=& 1+ \frac{\alpha_s}{4\pi \epsilon} 
\bigg(\frac{\mu^2}{{\bf q}^2}\bigg)^{\!\epsilon}
\left[\frac{8}{3}C_A \right]_{\rm IR} 
+O(\alpha_s^2),
\\
{\cal V}_{so}(\alpha_s)
&=& 1+ \frac{\alpha_s}{4\pi \epsilon} \left\{
\bigg(\frac{\mu^2}{m^2}\bigg)^{\!\epsilon} 
\left[\frac{4}{3} C_A\right]_{\rm IR}
+ \bigg(\frac{\mu^2}{{\bf q}^2}\bigg)^{\!\epsilon}
\left[-\frac{4}{3} C_A\right]_{\rm UV}
\right\}+O(\alpha_s^2),
\\
{\cal V}_{hf}(\alpha_s)
&=& 1+ \frac{\alpha_s}{4\pi \epsilon} \left\{
\bigg(\frac{\mu^2}{m^2}\bigg)^{\!\epsilon} 
\left[2 C_A\right]_{\rm IR}
+ \bigg(\frac{\mu^2}{{\bf q}^2}\bigg)^{\!\epsilon}
\left[-2 C_A\right]_{\rm UV} 
\right\}+O(\alpha_s^2),
\\
{\cal V}_{s}(\alpha_s)
&=& \frac{\alpha_s}{4\pi \epsilon} \left\{
\bigg(\frac{\mu^2}{m^2}\bigg)^{\!\epsilon} 
\left[-\frac{C_A}{2}\right]_{\rm IR}
+ \bigg(\frac{\mu^2}{{\bf q}^2}\bigg)^{\!\epsilon}
\left[\frac{C_A}{2} \right]_{\rm UV} 
\right\}+O(\alpha_s^2).
\label{eq:vm2polelast}
\end{eqnarray}
A term $\beta_0 \left(\left(\mu^2/{\bf q}^2\right)^{\!\epsilon}-1\right)$ 
that was included in (\ref{eq:vm2s}) 
to (\ref{eq:vm2slast}), which is related to the logarithms from charge 
renormalization, has been omitted from the expression in curly brackets
for all potentials except the 
last one, which has no tree-level term. 
We see that the IR poles from the hard region cancel the UV poles 
from the soft region, as it should be, since these singularities 
arise from hard-soft factorization. The remaining IR singularities 
in the soft contribution appear only in the spin-independent 
potentials ${\cal V}_{\delta}$, ${\cal V}_{p}$. They are precisely 
of the form of the remaining terms in the subtraction term 
(\ref{eq:counter}) and therefore cancel with UV divergences in the 
ultrasoft calculation. 
Again, the structure of divergences is as 
required by consistency, since the ultrasoft contribution is 
spin-independent at NNNLO.

\subsection{The spin-projected colour-singlet potential} 
Since the spin-dependent potentials appear first at NNLO,  
their double insertion is of higher order than NNNLO. Hence, when 
one computes the correlation function of the spin-triplet 
current (\ref{eq:GPNRQCD}), or the corresponding spin-singlet 
one, the spin-algebra can effectively be performed before the 
computation by working with spin-projected potentials. 
Given a potential with spin-dependence $a\otimes b$, where 
$a$ ($b$) refers to the spin-matrix on the quark (anti-quark) 
line, we replace
\begin{eqnarray}
&& \mbox{spin-triplet:} \qquad 
a\otimes b \rightarrow \frac{\mbox{tr}\,(\sigma^i a \sigma^i b)}
{\mbox{tr}\,(\sigma^i \sigma^i)} \,1\otimes 1 
=  \frac{\mbox{tr}\,(\sigma^i a \sigma^i b)}
{2 (d-1)} \,1\otimes 1
\\
&& \mbox{spin-singlet:} \qquad \!
a\otimes b \rightarrow \frac{\mbox{tr}\,(a b)}
{\mbox{tr}\,1} \,1\otimes 1 
=  \frac{\mbox{tr}\,(a b)}
{2} \,1\otimes 1
\end{eqnarray}
Note that the traces must be performed in $d-1$ space dimensions. 
Only the spin-triplet projection is relevant to the third-order 
top production cross section in $e^+ e^-$ collisions. For the 
three spin-dependent terms in the general potential 
(\ref{eq:potentialbeforespin}), the projections result in 
\begin{eqnarray}
&& [\sigma_i,\sigma_j]q_i p_j \otimes 1
 -1\otimes [\sigma_i,\sigma_j]q_i p_j\rightarrow 0,
\\ 
&& [\sigma_i,\sigma_j]q_j\otimes [\sigma_i,\sigma_k]q_k \rightarrow  
\frac{10-7d+d^2}{1-d}\, 4{\bf q}^2,
\\ 
&& [\sigma_i,\sigma_j]\otimes[\sigma_i,\sigma_j] \rightarrow
(-4)(10-7d+d^2)\,,
\end{eqnarray}
where we have omitted the trivial $1\otimes 1$ dependence as done earlier.

After the projection the four potentials ${\cal V}_\delta$, 
${\cal V}_{so}$, ${\cal V}_{hf}$, ${\cal V}_s$ in 
(\ref{eq:potentialbeforespin}) can be merged into 
a single expression ${\cal V}_{1/m^2}$, and we arrive at 
the spin-triplet, colour-singlet potential already presented 
in~\cite{Beneke:2007gj}:
\begin{eqnarray}
\label{eq:potentialafterspin} 
V({\bf p},{\bf p}^{\prime}) &=& 
  -\frac{4\pi \alpha_{s}C_F}{\bff{q}^2} \bigg[\,{\cal V}_C
  -{\cal V}_{1/m}\,\frac{\pi^2\,|\bf{q}|} {m}
  +{\cal V}_{1/m^2}\,\frac{\bff{q}^2}{m^2}
  +{\cal V}_{p}\,\frac{\bff{p}^2+\bff{p}^{\prime \,2}}{2m^2}\,
\bigg].
\end{eqnarray}
The Coulomb and $1/m$ potentials are as given earlier. The two 
$1/m^2$ terms read, up to the one-loop order: 
\begin{eqnarray}
{\cal V}_{p}^{\,(0)}&=&1\,,
\\
{\cal V}_{p}^{(1)} &=&
   \Bigg[\bigg( \frac{\mu^2}{\bf{q}^2} \bigg)^{\!\epsilon }-1
   \Bigg]\, \frac{1}{\epsilon}\,
   \bigg(\frac{8}{3}C_A+\beta_0\bigg)
 +\bigg( \frac{\mu^2}{\bf{q}^2} \bigg)^{\!\epsilon}\,
        v^{(1)}_p(\epsilon)\,,
\label{eq:vp1}
\\[0.2cm] 
{\cal V}_{1/m^2}^{\,(0)}&=& v_0(\epsilon)
=-\frac{4-\epsilon-2\,\epsilon^2}{6-4\epsilon}\,,
\label{eq:defv0}
\\ 
\label{eq:vc1}
{\cal V}_{1/m^2}^{(1)} &=&
   \Bigg[\bigg(\frac{\mu^2}{\bf{q}^2} \bigg)^{\!\epsilon }-1
   \Bigg]\,
   \frac{1}{\epsilon}\,
   \bigg(\,\frac{7}{3}C_F-\frac{11}{6}C_A\, + \beta_0\, v_{
     0}(\epsilon)\bigg)
\nonumber
\\
&&+\,\Bigg[\left(\frac{{\mu }^2}{m^2} \right)^{\!\epsilon}-1
   \Bigg]\, \frac{1}{\epsilon}\,
   \bigg(\,\frac{C_F}{3}+\frac{C_A}{2}\,\bigg)
  +\bigg(\frac{\mu^2}{\bf{q}^2} \bigg)^{\!\epsilon}\,
        v^{(1)}_{q}(\epsilon)
  +\bigg(\frac{{\mu }^2}{m^2} \bigg)^{\!\epsilon}\,
        v^{(1)}_{m}(\epsilon)\,.\qquad
\end{eqnarray}
The one-loop coefficients (expanded up to $O(\epsilon)$) are given by
\begin{eqnarray}
v^{(1)}_{p}(\epsilon) &=& \frac{31}{9} C_A -\frac{20}{9} \,T_F n_f +
\epsilon \left\{\bigg(\frac{188}{27} -\frac{19\pi^2}{36}\bigg)
C_A+\bigg(-\frac{112}{27} +\frac{\pi^2}{9}\bigg) T_F n_f
\right\}
\nonumber\\
&& + \,O(\epsilon^2) \,, 
\label{eq:vp1epsterms}\\[0.2cm]
v^{(1)}_{q}(\epsilon) &=& -\frac{C_F}{3} -\frac{11}{27} C_A
+\frac{40}{27} \,T_F n_f 
\label{eq:epsterms}\\ 
&& + \, \epsilon \left\{\bigg(-\frac{419}{81}+\frac{77\pi^2}{216}\bigg)C_A
+\bigg(2 -\frac{7\pi^2}{36}
\bigg)C_F+\bigg(\frac{274}{81} -\frac{2\pi^2}{27}\bigg) T_F n_f
\right\}+O(\epsilon^2) \,,
\nonumber \\[0.2cm]
v^{(1)}_{m}(\epsilon) &=& -\frac{C_F}{3} -\frac{29}{9} C_A
+\frac{4}{15}\,T_F + \epsilon \left\{\bigg(\frac{379}{54}
+\frac{\pi^2}{24}\bigg) C_A+\bigg(-10 +\frac{\pi^2}{36}\bigg) C_F
\right\}
\nonumber\\
&&+\,O(\epsilon^2)\,,
\label{eq:vm1epsterms}\end{eqnarray}
where now the $1/m^2$ pieces of the subtraction term 
(\ref{eq:counter}) have been added so that there are no $1/\epsilon$ poles. 
The four-dimensional expressions
$v_i^{(1)}(\epsilon=0)$ for $i=\{q,m,p\}$ agree with those
obtained from \cite{Kniehl:2002br,Wuester:2003}, and the 
$O(\epsilon)$ term agrees with \cite{Wuester:2003}.

\subsection{Matching of the NRQCD vector current}
\label{subsec:vectormatch}

Having determined the matching coefficients of the PNRQCD Lagrangian 
we now return to the question whether the NRQCD spin-triplet current 
is renormalized when it is matched to PNRQCD. In general, we may write, 
in analogy with (\ref{eq:QCDVectorCurrent}),
\begin{eqnarray}
\psi^{\dag}\sigma^i\chi_{|\rm NRQCD} &=& 
\tilde{c}_{v} \,\psi^{\dag}\sigma^i\chi_{|\rm PNRQCD} + 
\frac{\tilde{d}_{v1}}{6 m^2} \,\psi^\dag\sigma^i\,{\bf D^2}\chi_{|\rm PNRQCD}
+\ldots, \qquad \\[0.1cm]
\psi^\dag\sigma^i\,{\bf D^2}\chi_{|\rm NRQCD} &=& 
\tilde{d}_{v2} \,\psi^\dag\sigma^i\,{\bf D^2}\chi_{|\rm PNRQCD}
+\ldots.
\end{eqnarray}
Non-trivial ($\tilde{c}_v, \tilde{d}_{v2} \not=1$, 
$\tilde{d}_{v1} \not=0$) 
PNRQCD matching coefficients of the currents can 
arise from three sources: (1) Soft loops not accounted in the 
matching of the Lagrangian. This implies that the soft 
loop momentum must flow through the external current vertex. 
(2) Off-shell effects. Since the PNRQCD Lagrangian is 
matched on-shell, off-shell effects that are not reproduced 
by the Lagrangian interactions must be absorbed into a renormalization 
of the external currents. (3) ${\cal O}(\epsilon)$ terms of 
soft loops contributing to the matching of the Lagrangian that multiply 
$1/\epsilon$ poles of PNRQCD loops are local and must be 
absorbed into a renormalization 
of the external currents, when the \mbox{PNRQCD} matching coefficients 
(the potentials) are defined in four dimensions. As discussed before, 
we choose to work with $d$-dimensional potentials, hence these 
contributions are included in the PNRQCD calculation without a modification 
of the external current. We shall now prove that there is also 
no further renormalization of the currents from (1) and (2), that is, 
$\tilde{c}_v = 1$ to three loops, and $\tilde{d}_{v1} = 0$, 
$\tilde{d}_{v2} =1$ 
at one loop.\footnote{The arguments presented make it clear that 
this should remain true to any order in perturbation theory.}

\begin{figure}[t]
\begin{center}
\makebox[0cm]{\scalebox{0.9}{\rotatebox{0}{
\includegraphics[width=4cm]{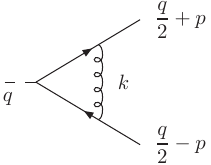}}}}
\end{center}
\vspace*{-0.3cm}
\caption{\label{fig:vertex1loop} One-loop vertex renormalization of 
the NRQCD current.}
\end{figure}

We first consider the issue (1) of soft renormalization of the NRQCD 
currents. The relevant vertex diagram at the one-loop order is  
shown in figure~\ref{fig:vertex1loop} with external momenta 
$q=(2 m+E,{\bm 0})$, $p_1=\frac{q}{2}+p = 
(m+E/2,{\bff p})$ and $p_2=\frac{q}{2}-p = 
(m+E/2,-{\bff p})$, and $p_1^2=p_2^2=m^2$. The NRQCD expression is
\begin{equation}
A_{\rm NRQCD} = (i g_s)^2 C_F \tilde\mu^{2\epsilon} 
\int \frac{d^d k}{(2\pi)^d}\,
\frac{-i}{k^2}\,
\frac{i}{\frac{E}{2}+k^0-\frac{({\bff p}+{\bff k})^2}{2 m}}
\frac{-i}{\frac{E}{2}-k^0-\frac{({\bff p}+{\bff k})^2}{2 m}}
\times \mbox{poly}({\bff p},{\bff k})\,,
\label{vertex1loopNRQCD}
\end{equation}
where the $+i\epsilon$ prescription for the propagators is 
left implicit. The unspecified polynomial factor arises from 
derivatives in the subleading NRQCD interactions or from the 
subleading external current. 

Power-counting for soft loop 
momentum $k^0\sim {\bff k}\sim m v$ shows that this integral 
can give rise to a ${\cal O}(\alpha_s)$ correction to 
$\tilde{c}_v, \tilde{d}_{v1}, \tilde{d}_{v2}$. However, in the 
soft region we pick the pole at $k^0=-|{\bff k}|+i\epsilon$ of 
the gluon propagator and expand the two quark propagators 
in $E$ and $({\bff p}+{\bff k})^2/(2 m)$. The resulting 
integral $\int d^{d-1}{\bff k}/|{\bff k}|^3\times  
\mbox{poly}({\bff p},{\bff k})$ is scaleless and vanishes in 
dimensional regularization. This holds to any order 
in the expansion in the soft region~\cite{Beneke:1997zp}, hence 
there is no soft one-loop correction to the matching of the 
PNRQCD currents.

\begin{figure}[t]
\begin{center}
\makebox[0cm]{\scalebox{0.8}{\rotatebox{0}{
\includegraphics[width=6cm]{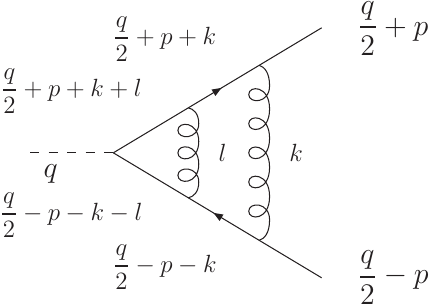}}}}
\end{center}
\vspace*{-0.2cm}
\caption{\label{fig:vertexplanar2loop} 
Planar 2-loop vertex renormalization of 
the NRQCD current.}
\end{figure}

Moving to the two-loop level, 
we consider as an example the planar vertex diagram in 
figure~\ref{fig:vertexplanar2loop}. The momentum regions of 
interest are s-s and s-p, where the first letter refers to the 
inner vertex subgraph and loop momentum $l$, the second to 
the box subgraph and loop momentum $k$. The p-s combination 
is not relevant, since the soft loop does not flow through 
the external vertex. Such contributions are included in the one-loop 
potentials. The inner vertex subdiagram in the s-s and s-p regions 
is an expression similar to (\ref{vertex1loopNRQCD}) with 
$k\to k+l$ in the quark propagator and $k^2\to l^2$ in the 
gluon propagator. Picking up the gluon propagator pole results 
in 
\begin{equation}
\int \frac{d^{d-1} \bff{l}}{(2\pi)^{d-1}}\, 
\frac{1}{|{\bff l}| \,(k^0-|{\bff l}|)^2} 
\end{equation}
for the inner integral. Now, in the s-p region $k$ is potential, and 
$k^0\sim mv^2 \ll |{\bff l}|$ must be expanded, in which case the 
integral is scaleless as before. If, as in the s-s region, $k$ is 
also soft, then the $k^0$-integration picks up the pole of 
the second gluon propagator at $k^0=-|{\bff k}|+i\epsilon$ and 
the entire two-loop integral is scaleless. This discussion evidently 
applies to all two-loop vertex diagrams. Since the vanishing of 
the integrals is due to the analytic structure of the propagator 
denominators, it generalizes to higher orders in the velocity expansion, 
which contains higher powers of the same propagators and numerator 
polynomials. Hence, we conclude that there are no s-p and s-s 
contributions to two-loop vertex diagrams in any order in the 
threshold expansion, in accordance with the results 
of~\cite{Beneke:1997zp}, 
and hence no soft renormalization of the currents at two loops. 

The structure of the analysis extends to higher loop orders. Either
one of the inner subgraphs is scaleless, because an outer loop 
momentum is potential. Or the entire diagram is soft and scaleless, 
because the external quark momenta are potential. 
We therefore conclude that there is no contribution to the matching of 
the NRQCD currents to PNRQCD from soft loops (item (1)).

We now turn to the discussion of off-shell effects, item (2), 
and start again with the one-loop diagram 
shown in figure~\ref{fig:vertex1loop}. Now, however, the loop 
momentum is potential, and we have to compare the NRQCD 
potential contribution contained in (\ref{vertex1loopNRQCD}) 
with the PNRQCD expression 
\begin{equation}
A_{\rm PNRQCD} = (i g_s)^2 C_F\tilde\mu^{2\epsilon} 
\int \frac{d^{d-1} {\bff k}}{(2\pi)^{d-1}}\,
\frac{1}{{\bff k}^2}\,
\frac{1}{E-\frac{({\bff p}+{\bff k})^2}{m}}
\times \mbox{poly}^\prime({\bff p},{\bff k})\,,
\label{vertex1loopPNRQCD}
\end{equation}
which arises from the single insertion of the tree-level 
PNRQCD potential and the ${\cal O}(\alpha_s)$ term of the Coulomb 
Green function. The factor $1/{\bff k}^2$ corresponds to the 
Coulomb potential insertion, while higher-order 
potentials as well as derivative factors from the external 
current are contained in the unspecified polynomial. 
The potential contribution in NRQCD is defined as the 
contribution from the quark-propagator pole at 
$k^0=E/2-({\bff p}+{\bff k})^2/(2 m)+i\epsilon$ in 
(\ref{vertex1loopNRQCD}).

The difference $\Delta A$ between $A_{\rm NRQCD}$ and $A_{\rm PNRQCD}$ 
contributes to the matching of the external current and 
arises as follows: when the tree-level PNRQCD potential 
is derived from the one-gluon exchange diagram, the external 
quark lines are assumed on-shell, which implies 
${\bff p}^2 = ({\bff p}+{\bff k})^2$ and $E_{p+k} - 
E_p = k^0=0$ with the momentum assignment as in 
figure~\ref{fig:vertex1loop}. However, no such restrictions are 
imposed on the loop momentum $k$ in the calculation of 
the NRQCD diagram, figure~\ref{fig:vertex1loop}. Thus the 
difference between  $A_{\rm NRQCD}$ and $A_{\rm PNRQCD}$ comes 
from the expansion of the gluon propagator in the potential 
region
\begin{equation}
\frac{1}{k^2} - \frac{1}{-{\bff k}^2} = 
- \frac{[k^0]^2}{{\bff k}^4} + {\cal O}(v^2),
\label{offshell1}
\end{equation}
and the difference of polynomial factors, which after a short 
computation can be determined 
to be 
\begin{equation}
\mbox{poly}({\bff p},{\bff k}) -  \mbox{poly}^\prime({\bff p},{\bff k})
= \frac{{\bff p}^2}{2m^2} - \frac{({\bff p}+{\bff k})^2}{2m^2} 
+ {\cal O}(v^4)\,.
\label{offshell2}
\end{equation}
Note that the leading contribution to $A_{\rm NRQCD}$ and 
$A_{\rm PNRQCD}$ is of order $\alpha_s/v$. The terms neglected in 
(\ref{offshell1}) and (\ref{offshell2}) are therefore of 
order $\alpha_s v^3$. These are fourth-order corrections to 
the cross section beyond the accuracy we aim at. In total, 
we arrive at 
\begin{eqnarray}
\Delta A &=& (i g_s)^2 C_F\tilde\mu^{2\epsilon} 
\!\int \frac{d^{d-1} {\bff k}}{(2\pi)^{d-1}}\,
\frac{1}{{\bff k}^2}\,
\frac{1}{E-\frac{({\bff p}+{\bff k})^2}{m}} \times 
\left\{\frac{{\bff p}^2}{2m^2} - \frac{({\bff p}+{\bff k})^2}{2m^2}
- \frac{[k^0]^2}{{\bff k}^2}\right\}_{\!|k^0 =
\frac{E}{2}-\frac{({\bff p}+{\bff k})^2}{2 m}}
\nonumber\\
&=&   (i g_s)^2 C_F\tilde\mu^{2\epsilon} 
\!\int \frac{d^{d-1} {\bff k}}{(2\pi)^{d-1}}\,
\frac{1}{{\bff k}^2}\,
\frac{1}{2 k^0} \times 
\left\{\frac{k^0}{m} - \frac{[k^0]^2}{{\bff k}^2}\right\}_{\!|k^0 =
\frac{E}{2}-\frac{({\bff p}+{\bff k})^2}{2 m}}\,.
\end{eqnarray}
To arrive at the second line we used that the on-shell condition implies 
$E=E_p+E_{-p}={\bff p}^2/m$. The two terms in curly brackets each cancel the 
heavy-quark propagator denominator $1/(2 k^0)$, after which the integral 
is scaleless and vanishes. This must be so, since a non-zero contribution 
to $\Delta A$ at this order would have scaled as $\alpha_s v$, but there 
is no ${\cal O}(v)$ production current.

\begin{figure}[t]
\begin{center}
\makebox[0cm]{\scalebox{0.9}{\rotatebox{0}{
\includegraphics[width=6cm]{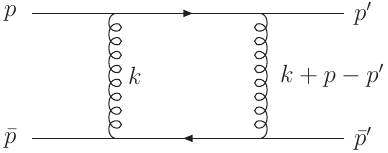}}}}
\end{center}
\vspace*{-0.2cm}
\caption{\label{fig:box1loop} 
One-loop box diagram, whose potential region is not completely 
reproduced by PNRQCD with potentials matched at tree-level.}
\end{figure}

That off-shell effects from the potential region {\em are} relevant in 
general can be seen from the calculation of the $1/m$-potential at 
${\cal O}(\alpha_s^2)$. The potential region of the one-loop box 
diagram shown in figure~\ref{fig:box1loop} is not completely reproduced 
by PNRQCD. The difference is a contribution to the $1/m$ potential, 
that is, a PNRQCD matching coefficient, which is crucial to obtain the 
gauge-invariant result given in~\cite{Beneke:1999qg} and (\ref{b1eps}). 
The difference between the box and the vertex diagram discussed above 
is that the box loop integral is not scaleless after the cancellation 
of the quark propagator by the off-shell terms, since there is a 
second gluon propagator $1/({\bff k}+{\bff p}-{\bff p^\prime})^2$.

Returning to the vertex diagrams, we now consider the planar two-loop 
diagram of figure~\ref{fig:vertexplanar2loop} in the p-p and p-s region, 
where the first letter refers again to the inner vertex subgraph. 
The corresponding PNRQCD diagrams are the two-loop vertex diagram 
with tree-level potentials and a one-loop vertex diagram with insertion 
of a one-loop potential, respectively. 
The off-shell terms of the NRQCD diagram 
in the p-p region are precisely the ones that contribute to the 
$1/m$ potential discussed in the previous paragraph; they are correctly 
reproduced by the PNRQCD diagram with insertion of the $1/m$ 
potential. The off-shell terms in the p-s region have a similar origin 
as in the one-loop vertex diagram. The soft box graph gives rise to 
a one-loop potential, but since the potentials are matched on-shell, 
the soft box graph is not completely reproduced when it appears as 
subgraph in a larger diagram. Since the leading p-s region is 
${\cal O}(\alpha_s^2/v)$, if all the off-shell corrections 
were of order $v^2$ relative to the leading term 
as in (\ref{offshell1}), (\ref{offshell2}), we could immediately dismiss 
them, since there is no ${\cal O}(\alpha_s^2 v)$ hard vertex correction
into which they could be absorbed. This is indeed true for the planar diagram 
but not in general.

\begin{figure}[t]
\begin{center}
\makebox[0cm]{\scalebox{0.8}{\rotatebox{0}{
\includegraphics[width=6cm]{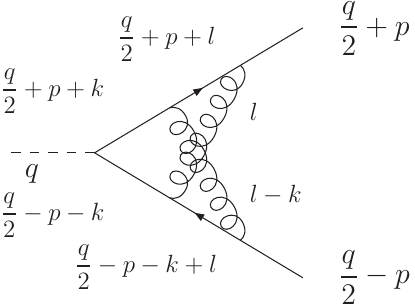}}}}
\end{center}
\vspace*{-0.2cm}
\caption{\label{fig:vertexnonplanar2loop} 
Non-planar 2-loop vertex renormalization of 
the NRQCD current.}
\end{figure}

As an example, we consider the non-planar NRQCD two-loop diagram 
shown in figure~\ref{fig:vertexnonplanar2loop}, which (neglecting 
constant factors) is given by the 
expression
\begin{eqnarray}
&& \int \frac{d^d k}{(2\pi)^d} \frac{d^d l}{(2\pi)^d}\,
\frac{i}{\frac{E}{2}+k^0-\frac{({\bff p}+{\bff k})^2}{2 m}}
\frac{-i}{\frac{E}{2}-k^0-\frac{({\bff p}+{\bff k})^2}{2 m}}
\times 
\nonumber\\
&& \hspace*{0cm} \times 
\frac{i}{\frac{E}{2}+l^0-\frac{({\bff p}+{\bff l})^2}{2 m}}
\frac{-i}{\frac{E}{2}-k^0+l^0-\frac{({\bff p}+{\bff k}-{\bff l})^2}{2 m}}
\frac{1}{l^2 \,(l-k)^2}\,.
\label{nonplanar2loopNRQCD}
\end{eqnarray}
In the p-s region\footnote{The p-p region is zero for the non-planar diagram.} 
the $l$-integral for the soft crossed-box subdiagram is exactly the same 
that appears in the computation of the one-loop Coulomb and $1/m$ potentials 
except for the additional $k^0$ in the last quark propagator that is 
absent when the quark lines of the inner vertex subgraph are on-shell 
(see discussion above). Since $k$ is potential, 
$k^0\sim m v^2$ must be expanded relative to $l^0\sim mv$, which 
results in a series of corrections beginning at ${\cal O}(v)$. If this 
off-shell correction were non-zero, it would result in a NNLO 
${\cal O}(\alpha_s^2)$ contribution to the coefficient function 
$\tilde c_v$. However, as $k$ is potential, the $k^0$-integral is 
the contribution from the pole of the quark propagator in the first 
line of (\ref{nonplanar2loopNRQCD}). Thus, in complete analogy with the 
discussion of the one-loop vertex diagram, the expansion in $k^0$ 
cancels the remaining quark propagator and renders the ${\bff k}$-integral 
scaleless. This cancellation is generic for all two-loop vertex diagrams 
in the p-s region.

The structure of the argument extends to the three-loop order. The possible 
off-shell terms are either already accounted in the matching of subleading 
potentials; or an inner vertex subgraph becomes scaleless due to 
a cancellation of the remaining quark-propagator of a potential loop.
Hence,  we conclude that there is no contribution to the matching of 
the NRQCD currents to PNRQCD from off-shell terms (item (2)), at least 
up to the NNNLO order.

\subsection{Equation of motion identities for current 
and potential insertions}
\label{sec:eqofmotion}

The integrals for PNRQCD diagrams with insertions of potentials or 
external currents such as (\ref{multipleinsertions}) can sometimes 
be simplified by the equation of motion for the PNRQCD 
quark-antiquark propagator. We will 
make use of this to reduce the number of independent insertions that 
need to be calculated (in part II of the paper), 
and provide the relevant identities here.

In the present context the equation of motion is the 
Lippmann-Schwinger equation (\ref{eq:lippmann-schwinger}) for the 
colour-singlet Coulomb Green function written in the form
\begin{eqnarray}
\frac{\bff{p}^2_2}{m} \,G_0(\bff{p}_1,\bff{p}_2;E) &=& 
E\,G_0(\bff{p}_1,\bff{p}_2;E)
+ (2\pi)^{d-1}\,\delta^{(d-1)}(\bff{p}_1 - \bff{p}_2)
\nonumber\\
&&+ \,4\pi \alpha_s C_F \,\tilde{\mu}^{2\epsilon}
\int\frac{d^{d-1}\bff{k}}{(2\pi)^{d-1}}\,
\frac{1}{(\bff{k}-\bff{p}_2)^2}\, G_0(\bff{p}_1,\bff{k};E) \,.
\end{eqnarray}
In the remainder of this subsection we drop the energy argument 
of the Green function, which is always $E$, and set $\tilde\mu=1$ 
to simplify the notation. 
\blue{The equation of motion identities are valid for complex $E$ 
in the domain of analyticity of the Green function.}

As our first example we consider the insertion of the subleading 
derivative current (\ref{eq:QCDVectorCurrent2}) into one of the vertices. 
The relevant integral is 
\begin{eqnarray}
&& \int \prod_{i=1}^2\frac{d^{d-1}{\bf
p}_i}{(2\pi)^{d-1}}\,G_0({\bf p}_1,{\bf p}_2)
\frac{\bff{p}^2_2}{m^2} 
= \frac{E}{m}\int \prod_{i=1}^2\frac{d^{d-1}{\bf
p}_i}{(2\pi)^{d-1}}\,G_0({\bf p}_1,{\bf p}_2) + 
\frac{1}{m}\int \frac{d^{d-1}{\bf
p}_1}{(2\pi)^{d-1}}
\nonumber\\
&& \hspace{0.5cm} + \, 
\frac{4\pi \alpha_s C_F}{m} \,\int\prod_{i=1}^2\frac{d^{d-1}{\bf
p}_i}{(2\pi)^{d-1}}\frac{d^{d-1}\bff{k}}{(2\pi)^{d-1}}\,
\frac{1}{(\bff{k}-\bff{p}_2)^2}\, G_0(\bff{p}_1,\bff{k})\,.
\end{eqnarray}
The second term on the right-hand side is a scaleless integral. The third 
one is seen to contain $\int d^{d-1}{{\bf p}_2}/\bff{p}_2^2 =0$ after 
shifting $\bff{p}_2 \to \bff{p}_2+\bff{k}$. Hence, the insertion of 
$\bff{p}^2_2/m^2$ can be replaced by the factor $E/m$. This holds true 
in expressions of the form
\begin{equation}
\int \Bigg[\prod_{i}\frac{d^{d-1}{\bf
p}_i}{(2\pi)^{d-1}}\Bigg] \,\frac{\bff{p}^2_1}{m^2}\, 
i{G_0}({\bf p}_1,{\bf p}_2)
i\delta{V}_1({\bf p}_2,{\bf p}_3) i{G_0}({\bf p}_3,{\bf p}_4)\, 
i\delta{V}_2({\bf p}_4,{\bf p}_5)  i{G_0}({\bf p}_5,{\bf p}_6)\, 
\ldots
\end{equation}
that contain multiple potential insertions.
This shows that the {\em unrenormalized, dimensionally regulated} 
matrix element of the 
subleading external current operator $\psi^\dagger \sigma_i \bff{D}^2\chi$ 
is $-m E$ times the unrenormalized matrix element of 
the leading-order current $\psi^\dagger \sigma_i \chi$, which has been 
used in (\ref{R1}).

As our second example we consider the insertion of the $\bff{p}^2/(m^2 
\bff{q}^2)$ potential. With $\bff{q}=\bff{p}_3-\bff{p}_2$ we find 
\begin{eqnarray}
&& \int \prod_{i=1}^4\frac{d^{d-1}{\bf p}_i}{(2\pi)^{d-1}}G_0({\bf p}_1,{\bf
p}_2) \frac{{\bf p}_2^2}{m^2 {\bf q}^2} G_0({\bf
p}_3,{\bf p}_4)
\nonumber \\
&& \hspace{0.5cm} = \,
\frac{E}{m} \int \prod_{i=1}^4\frac{d^{d-1}{\bf p}_i}{(2\pi)^{d-1}}G_0({\bf
p}_1,{\bf p}_2) \,\frac{1}{{\bf q}^2}\, G_0({\bf p}_3,{\bf p}_4) + 
\frac{1}{m} \int \prod_{i=1,3,4}\frac{d^{d-1}{\bf p}_i}{(2\pi)^{d-1}}
\,\frac{G_0({\bf p}_3,{\bf p}_4)}{(\bff{p}_1-{\bf p}_3)^2 }
\nonumber\\ 
&&  \hspace{1cm} + 
\frac{4\pi \alpha_s C_F}{m} \,\int\prod_{i=1}^4\frac{d^{d-1}{\bf
p}_i}{(2\pi)^{d-1}}\frac{d^{d-1}\bff{k}}{(2\pi)^{d-1}}\,
G_0(\bff{p}_1,\bff{k})\,\frac{1}{(\bff{k}-\bff{p}_2)^2}\,
\frac{1}{(\bff{p}_2-\bff{p}_3)^2}\,G_0(\bff{p}_3,\bff{p}_4)\nonumber\\ 
&&  \hspace{0.5cm}= \,\int \prod_{i=1}^4
\frac{d^{d-1}{\bf p}_i}{(2\pi)^{d-1}}G_0({\bf
p}_1,{\bf p}_2) \Bigg[\frac{E}{m} \frac{1}{{\bf q}^2} +\frac{\pi
\alpha_s C_F}{2m} \frac{\mu^{2\epsilon}k(0)}{[{\bf q}^2]^{\frac{1}{2}+\epsilon}}
\Bigg] G_0({\bf p}_3,{\bf p}_4)\,.
\end{eqnarray}
The final expression follows, since the second term on the left-hand side 
of the first equation contains the scaleless $\bff{p}_1$-integral, while 
in the third term the integration over $\bff{p}_2$ can be performed 
with the help of the integral 
\begin{eqnarray}
&& \tilde\mu^{2\epsilon} \int \frac{d^{d-1}\bff{k}}{(2\pi)^{d-1}}
\frac{1}{[\bff{k}^2]^{1+u} (\bff{q}-\bff{k})^2} 
\equiv \frac{\mu^{2\epsilon}}{[\bff{q}^2]^{\frac{1}{2}+u+\epsilon}}\,
\frac{k(u)}{8}
\end{eqnarray}
with
\begin{equation}
k(u)=\frac{e^{\gamma_E\epsilon}\, \Gamma(\frac{1}{2}+u+\epsilon)
\Gamma(\frac{1}{2}-u-\epsilon)\Gamma(\frac{1}{2}-\epsilon)}
{\pi^{\frac{3}{2}}\,\Gamma(1+u)\Gamma(1-u-2\epsilon)}\,,
\label{eq:def-k(u)}
\end{equation}
whose general form for $u\not =0$ will be needed below. This shows that 
the insertion of the  $\bff{p}^2/(m^2 
\bff{q}^2)$ potential can be eliminated in favour of the insertion of the 
Coulomb and the $d$-dimensional $1/(m|\bff{q}|)$ potential.

In the general case, after applying the spin projection, only six 
different types of insertions are needed for the NNNLO calculation:
\begin{eqnarray}
\label{eq:insertiontypes}
\frac{1}{{\bf q}^2} \bigg(\frac{\mu^2}{{\bf
q}^2}\bigg)^{a\epsilon}\!\!,\,\,\, \frac{1}{|{\bf
q}|}\bigg(\frac{\mu^2}{{\bf q}^2}\bigg)^{a\epsilon}\!\!,\,\,\,
\bigg(\frac{\mu^2}{{\bf q}^2}\bigg)^{a\epsilon}\!\!,\,\,\,
\frac{\bff{p}^2+\bff{p}^{\prime \,2}}{2{\bf
q}^2}\bigg(\frac{\mu^2}{{\bf q}^2}\bigg)^{a\epsilon}\!\!,\,\,\,
\bff{p}^4\delta^{(d-1)}({\bf q}),\,\,\, \bff{p}^2\delta^{(d-1)}({\bf q})\,.
\qquad
\end{eqnarray}
The first four come from the potentials in (\ref{eq:potentialafterspin}), 
the fifth is the kinetic energy correction
and the last might be used for the conversion of the pole scheme to 
threshold mass scheme as discussed in 
paper~II.\footnote{As a matter of fact, the implementation will be done 
in a different way, and the result is given here only for 
completeness.} The identities given below show that the last three
types of insertions 
can be reduced by using the equation of motion to the first three
and the delta-function potential $\delta^{(d-1)}({\bf q})$ without factors 
of $\bff{p}^2$. At third order, both, single insertions and
double insertions with an additional Coulomb potential insertion 
have to be considered. For the single insertions, the 
equation of motion relations read:
\begin{eqnarray}
\label{eq:eqofmotionpot} \delta^{(d-1)}({\bf q})\frac{{\bf
p}_2^2}{m}:&&\int \prod_{i=1}^4\frac{d^{d-1}{\bf
p}_i}{(2\pi)^{d-1}}G_0({\bf p}_1,{\bf p}_2)
(2\pi)^{d-1}\delta^{(d-1)}({\bf q})\frac{{\bf p}_2^2}{m} G_0({\bf
p}_3,{\bf p}_4)\nonumber
\\ \nonumber &=&E\int
\prod_{i=1}^3\frac{d^{d-1}{\bf p}_i}{(2\pi)^{d-1}}G_0({\bf p}_1,{\bf
p}_2) G_0({\bf p}_2,{\bf p}_3) +\int \prod_{i=1}^2\frac{d^{d-1}{\bf
p}_i}{(2\pi)^{d-1}}G_0({\bf p}_1,{\bf p}_2)
\\  &+&4\pi C_F \alpha_s\int
\prod_{i=1}^4\frac{d^{d-1}{\bf p}_i}{(2\pi)^{d-1}}G_0({\bf p}_1,{\bf
p}_2) \frac{1}{{\bf q}^2} G_0({\bf p}_3,{\bf p}_4),
\\\delta^{(d-1)}({\bf q})\frac{{\bf p}_2^4}{m^3}:&&\int
\prod_{i=1}^4\frac{d^{d-1}{\bf p}_i}{(2\pi)^{d-1}}G_0({\bf p}_1,{\bf
p}_2) (2\pi)^{d-1}\delta^{(d-1)}({\bf q})\frac{{\bf p}_2^4}{m^3}
G_0({\bf p}_3,{\bf p}_4) \nonumber
\\ \nonumber &=&\frac{E^2}{m}\int
\prod_{i=1}^3\frac{d^{d-1}{\bf p}_i}{(2\pi)^{d-1}}G_0({\bf p}_1,{\bf
p}_2) G_0({\bf p}_2,{\bf p}_3)+\frac{2E}{m}\int
\prod_{i=1}^2\frac{d^{d-1}{\bf p}_i}{(2\pi)^{d-1}}G_0({\bf p}_1,{\bf
p}_2)\nonumber
\\  \nonumber &+&\frac{8\pi C_F
\alpha_s E}{m} \int\prod_{i=1}^4\frac{d^{d-1}{\bf
p}_i}{(2\pi)^{d-1}}G_0({\bf p}_1,{\bf p}_2) \frac{1}{{\bf
q}^2}G_0({\bf p}_3,{\bf p}_4)
\\  &+&\frac{(4\pi C_F \alpha_s)^2}{m}\int
\prod_{i=1}^4\frac{d^{d-1}{\bf p}_i}{(2\pi)^{d-1}}G_0({\bf p}_1,{\bf
p}_2) \frac{\mu^{2\epsilon}}{[{\bf q}^2]^{\frac{1}{2}+\epsilon}} \frac{k(0)}{8}
G_0({\bf p}_3,{\bf p}_4),
\\
\frac{{\bf p}_2^2+{\bf p}_3^2}{2m^2 {\bf q}^2}:&&\int
\prod_{i=1}^4\frac{d^{d-1}{\bf p}_i}{(2\pi)^{d-1}}G_0({\bf p}_1,{\bf
p}_2) \frac{{\bf p}_2^2+{\bf p}_3^2}{2m^2 {\bf q}^2} G_0({\bf
p}_3,{\bf p}_4)
\label{eompoverq} \\
&=&\int \prod_{i=1}^4\frac{d^{d-1}{\bf p}_i}{(2\pi)^{d-1}}G_0({\bf
p}_1,{\bf p}_2) \Bigg[\frac{E}{m} \frac{1}{{\bf q}^2} +\frac{\pi
C_F\alpha_s}{2m} \frac{\mu^{2\epsilon} k(0)}{[{\bf q}^2]^{\frac{1}{2}+\epsilon}}
\Bigg] G_0({\bf p}_3,{\bf p}_4).\nonumber
\end{eqnarray}
The  delta-function potential $\delta^{(d-1)}({\bf q})$ appears implicitly in 
the first two relations in integrands of the form $G_0({\bf p}_1,{\bf
p}_2) G_0({\bf p}_2,{\bf p}_3)$, where the delta-function has been 
eliminated to set two arguments equal.
The equation of motion identities  
for the double insertions with a Coulomb potential (with
${\bf q}_1={\bf p}_3-{\bf p}_2$ and ${\bf q}_2={\bf p}_5-{\bf p}_4$)
read:
\begin{eqnarray}
\delta^{(d-1)}({\bf q_1})\frac{{\bf p}_2^2}{m}:&&\int
\prod_{i=1}^6\frac{d^{d-1}{\bf p}_i}{(2\pi)^{d-1}}G_0({\bf p}_1,{\bf
p}_2) (2\pi)^{d-1}\delta^{(d-1)}({\bf q}_1)\frac{{\bf p}_2^2}{m}
 \frac{G_0({\bf p}_3,{\bf p}_4)G_0({\bf
p}_5,{\bf p}_6)}{[{\bf q}_2^2]^{1+u}}\nonumber
\\ \nonumber  &=&E\int
\prod_{i=1,3-6}\frac{d^{d-1}{\bf p}_i}{(2\pi)^{d-1}} G_0({\bf
p}_1,{\bf p}_3)
\frac{G_0({\bf p}_3,{\bf p}_4)G_0({\bf p}_5,{\bf p}_6)}{[{\bf
q}_2^2]^{1+u}}
\\ \nonumber &+&\int
\prod_{i=1}^4\frac{d^{d-1}{\bf p}_i}{(2\pi)^{d-1}} \frac{G_0({\bf
p}_1,{\bf p}_2)G_0({\bf p}_3,{\bf p}_4)}{[{\bf q}_1^2]^{1+u}}
\\  &+&4\pi C_F \alpha_s\int
\prod_{i=1}^6\frac{d^{d-1}{\bf p}_i}{(2\pi)^{d-1}}G_0({\bf
p}_1,{\bf p}_2)\frac{1}{{\bf q}_1^2} \frac{G_0({\bf p}_3,{\bf p}_4)G_0({\bf p}_5,{\bf
p}_6)}{[{\bf q}_2^2]^{1+u}},
\\
\delta^{(d-1)}({\bf q}_1)\frac{{\bf p}_2^4}{m^3}:&&\int
\prod_{i=1}^6\frac{d^{d-1}{\bf p}_i}{(2\pi)^{d-1}}G_0({\bf p}_1,{\bf
p}_2) (2\pi)^{d-1}\delta^{(d-1)}({\bf q}_1)\frac{{\bf p}_2^4}{m^3}
\frac{G_0({\bf p}_3,{\bf p}_4)G_0({\bf p}_5,{\bf p}_6)}{[{\bf
q}_2^2]^{1+u}}\nonumber
\\ \nonumber  &=&\frac{E^2}{m}\int
\prod_{i=1,3-6}\frac{d^{d-1}{\bf p}_i}{(2\pi)^{d-1}} G_0({\bf
p}_1,{\bf p}_3)
\frac{G_0({\bf p}_3,{\bf p}_4)G_0({\bf p}_5,{\bf p}_6)}{[{\bf
q}_2^2]^{1+u}}
\\ \nonumber&+&\int
\prod_{i=1}^4\frac{d^{d-1}{\bf p}_i}{(2\pi)^{d-1}}G_0({\bf p}_1,{\bf
p}_2) \left[\frac{2E}{m}\frac{1}{[{\bf q}_1^2]^{1+u}}+ 
\frac{\pi C_F \alpha_s}{2 m}\frac{\mu^{2\epsilon}k(u)}{[{\bf
q}_1^2]^{\frac{1}{2}+u+\epsilon}}\right] G_0({\bf p}_3,{\bf p}_4)
\\ \nonumber&+&4\pi C_F \alpha_s\int
\prod_{i=1}^6\frac{d^{d-1}{\bf p}_i}{(2\pi)^{d-1}}G_0({\bf p}_1,{\bf
p}_2) \left[\frac{2E}{m}\frac{1}{{\bf q}_1^2}+ \frac{\pi C_F
\alpha_s}{2m}\frac{\mu^{2\epsilon}k(0)}{[{\bf q}_1^2]^{\frac{1}{2}+\epsilon}}\right]
\\ &&\times\frac{G_0({\bf p}_3,{\bf p}_4)G_0({\bf p}_5,{\bf p}_6)}{[{\bf
q}_2^2]^{1+u}},
\label{eq:eomkindouble}\\
\frac{{\bf p}_2^2+{\bf p}_3^2}{2 m^2 {\bf q}_1^2}:&&\int
\prod_{i=1}^6\frac{d^{d-1}{\bf p}_i}{(2\pi)^{d-1}}G_0({\bf p}_1,{\bf
p}_2) \frac{{\bf p}_2^2+{\bf p}_3^2}{2 m^2 {\bf q}_1^2}
\frac{G_0({\bf p}_3,{\bf p}_4)G_0({\bf p}_5,{\bf p}_6)}{[{\bf
q}_2^2]^{1+u}}\nonumber
\\
&&\hspace*{-1cm} =\,\int \prod_{i=1}^6\frac{d^{d-1}{\bf p}_i}{(2\pi)^{d-1}}G_0({\bf
p}_1,{\bf p}_2) \left[\frac{E}{m} \frac{1}{{\bf q}_1^2} +\frac{\pi
C_F\alpha_s}{2m} \frac{\mu^{2\epsilon}k(0)}{[{\bf
q}_1^2]^{\frac{1}{2}+\epsilon}}\right]
\frac{G_0({\bf p}_3,{\bf p}_4)G_0({\bf p}_5,{\bf p}_6)}{[{\bf
q}_2^2]^{1+u}}\nonumber
\\  
&&\hspace*{-1cm}  +\, \frac{\mu^{2\epsilon}k(u)}{16m}\int\prod_{i=1}^4\frac{d^{d-1}{\bf
p}}{(2\pi)^{d-1}}
 \frac{G_0({\bf p}_1,{\bf p}_2)G_0({\bf p}_3,{\bf p}_4)}{[{\bf q}_1^2]^{\frac{1}{2}+u+\epsilon}}
\,.
\end{eqnarray}
We note that we have kept the insertion of the Coulomb potential in the 
more general form $1/[\bff{q}_2^2]^{1+u}$, since in the double insertions 
we may also need the $d$-dimensional Coulomb potential, 
implying $u=\epsilon$.

\subsection{Ultrasoft interaction}
\label{sec:ultrasoft}

At third order there is for the first time 
a contribution from an ultrasoft loop momentum 
region. The ultrasoft correction to the heavy-quark correlation 
function has been computed separately in \cite{Beneke:2008cr} and 
will be incorporated in the results shown in paper~II. Here we 
provide a short overview of the ultrasoft
calculation and discuss some issues of the factorization,
which are important to understand the splitting of various divergent
parts.

The relevant ultrasoft interaction 
terms in the PNRQCD Lagrangian (\ref{eq:pnrqcd}) are 
given by
\begin{equation}
g_s\psi^\dagger(x) \big[A_0(t,\bff{0})
-\bff{x}\cdot\bff{E}(t,\bff{0})\big]\psi(x) + 
g_s\chi^\dagger(x)\big[A_0(t,\bff{0})
-\bff{x}\cdot \bff{E}(t,\bff{0})\big]\chi(x).
\label{usterms}
\end{equation}
The derivation of the chromoelectric dipole 
interaction from the multipole expansion of the 
NRQCD Lagrangian can be found in \cite{Beneke:1999zr}. The 
interaction with $A_0(t,\bff{0})$ can be removed by a field redefinition 
involving a time-like Wilson line. This modifies the external current 
that creates the heavy-quark pair, as discussed 
in \cite{Beneke:2010da}. In the present case of colour-singlet production 
in $e^+ e^-$ collisions the Wilson lines cancel, and the $A_0(t,\bff{0})$ 
terms in (\ref{usterms}) can be dropped. With $\bff{x} \sim 1/v$, 
and $g_s  \bff{E} \sim v^{9/2}$ for ultrasoft gluon fields, it follows that 
the chromoelectric dipole interaction is suppressed by $v^{3/2}$ 
relative to the kinetic term in the action. Two ultrasoft 
interaction vertices are required to form a loop, from which it 
follows that the leading ultrasoft contribution arises first 
at the third order.
 
The ultrasoft correction can be expressed in the form
\begin{eqnarray}
\delta^{us} G(E)&=& ig_s^2 C_F\int d^3{\bf r} \,d^3{\bf r}'\int
\frac{d^4{k}}{(2\pi)^4}\, \Bigg[\frac{k_0^2\,{\bf r}\cdot {\bf r}'- ({\bf
r}\cdot {\bf
      k})({\bf r}'\cdot {\bf k})}{k^2+i\varepsilon}
\nonumber\\
&& \times\,G^{(1)}_0(\bff{0},{\bf r};E)G^{(8)}_0({\bf r},{\bf r}';E-k_0)
G^{(1)}_0({\bf r}',\bff{0};E)\Bigg],\, 
\label{eq:gfcorr}
\end{eqnarray}
with the understanding that one picks up only the pole 
at $k^0=|\bff{k}|-i\epsilon$ in the gluon propagator. 
Here $G^{(1)}_0$ is the colour-singlet and $G^{(8)}_0$ the colour-octet 
Coulomb Green function (\ref{eq:lippmann-schwinger}). 
However, as explained in~\cite{Beneke:2007pj}, this expression 
cannot be used in practice, because it is ultraviolet (UV) divergent.
The regularization and subtraction of divergences must be 
done consistently with the calculation of potential insertions 
and hard matching coefficients, which have been done in dimensional 
regularization. In order to apply dimensional regularization to 
the ultrasoft contribution, (\ref{eq:gfcorr}) is transformed 
to momentum space. It also turns out to be convenient to 
revert the derivation of the PNRQCD ultrasoft interaction 
(\ref{usterms}) and to instead use the NRQCD vertices. The reason 
for this is that the derivation of (\ref{usterms}) uses the 
PNRQCD equation of motion, which reshuffles the loop expansion, 
and employs four-dimensional identities \cite{Beneke:1999zr}. 
The correspondence between UV divergences in the ultrasoft calculation
and IR divergences in the potential and hard matching 
calculations is more directly seen at the level of NRQCD diagrams, 
and the correct evaluation of the finite terms requires the 
consistent use of dimensional regularization in every loop order.

The UV divergences arise from the integral over the three-momentum 
$\bff{k}$ of the ultrasoft gluon, and from the subsequent potential 
loop integrations. The former divergence is related to the
factorization of the ultrasoft scale from the other scales, 
and cancels when all pieces of the calculation are added. The
UV-divergent part of the ultrasoft integral has the form
of a single insertion of a third-order potential and of a one-loop
correction to the coefficient $d_v$ of the 
derivative current in (\ref{eq:dv1loop}). We therefore define the ultrasoft 
correction by adding counterterms that cancel these ultrasoft 
subdivergences. With these 
subtractions, the ultrasoft correction reads~\cite{Beneke:2007pj} 
\begin{eqnarray}
\delta^{us} G(E)&=& \big[\tilde\mu^{2\epsilon}\big]^2\int
\frac{d^{d-1}\bffmath{\ell}}{(2\pi)^{d-1}} \frac{d^{d-1}\bffmath{\ell
^\prime}}{(2\pi)^{d-1}} \bigg\{\delta d_v^{\rm div}\,(-1)
\frac{\bffmath{\ell}^2+\bffmath{\ell}^{\prime\,2}} {6 m^2}\,
G^{(1)}_0(\bffmath{\ell},\bffmath{\ell}^\prime;E)
\nonumber\\[0.2cm]
&& \hspace*{-1.5cm} 
+\,\big[\tilde\mu^{2\epsilon}\big]^2 \int \frac{d^{d-1}{\bf
p}}{(2\pi)^{d-1}} \frac{d^{d-1}{\bf p^\prime}}{(2\pi)^{d-1}}
\,G^{(1)}_0(\bffmath{\ell},{\bf p};E) \,i\Big[\delta U -\delta 
V_{c.t.}\Big] \,iG^{(1)}_0({\bf p^\prime},\bffmath{\ell^\prime};E)
\bigg\}\,.\qquad 
\label{eq:defUS}
\end{eqnarray}
Here $\delta V_{c.t.}$ represents the potential 
subtraction (\ref{eq:counter}), and 
$\delta U$ is the ultrasoft insertion (containing the octet Green
function).\footnote{Note that we define $\delta U$ with an 
opposite sign compared to~\cite{Beneke:2007pj}.} 
The first line of (\ref{eq:defUS}) is related to 
the renormalization of the $1/m^2$ suppressed vector current 
$j_{1/m^2}^i = \psi^\dagger \sigma^i \bff{D^2} \chi$. If 
$[j_{1/m^2}]_{\rm ren} = Z_{1/m^2} [j_{1/m^2}]_{\rm bare}$, 
the one-loop counterterm $Z_{1/m^2}-1$ equals the infrared divergence 
$\delta d_v^{\rm div}$ that was subtracted to obtain the finite expression 
(\ref{eq:dv1loop}). The explicit expression is 
\begin{equation}
\delta d_v^{\rm div}  = Z_{1/m^2}-1 = - \frac{\alpha_s}{4\pi}\,
\frac{16 C_F}{\epsilon}\,.
\end{equation}
The remaining divergences in (\ref{eq:defUS}) are associated with 
the three-loop hard matching coefficient $c_3$. Note that 
this divergence structure implies that the NRQCD current mixes 
with the $1/m^2$ suppressed current through ultrasoft interactions. 
Schematically, 
\begin{equation}
T\left(\psi^\dagger \sigma^i\chi, 
\left[\int d^4 x\,{\cal L}_{\rm us}\right]^{2}\,\right)_{| \rm us} 
= \mbox{const}\times\frac{\alpha_s}{\epsilon} \,
\frac{1}{6 m^2}\,\psi^\dagger \sigma^i \bff{D^2}\chi
+\ldots \,,
\label{eq:usdiv}
\end{equation}
where the ellipses denote the remaining divergences related to 
time-ordered products with potentials and the leading-power 
current itself.

The subtracted expression is then simplified and reduced to a number 
of integrations that can mostly be done only numerically. The code that 
computes the ultrasoft correction was developed in conjunction 
with \cite{Beneke:2008cr}, and is implemented in the code 
\texttt{QQbar\_threshold}~\cite{Beneke:2016kkb} for the 
third-order cross section.

\section{Master formula for the third-order 
cross section}
\label{sec:master}

We have now collected all prerequisites to write down the expansion of 
the non-relativistic correlation function 
\begin{equation}
G(E) =\frac{i}{2 N_c (d-1)} \int d^{d} x\, e^{iEx^0}\,
\langle 0| \,T(\,
[\chi^{\dag}\sigma^i\psi](x)\,
[\psi^{\dag}\sigma^i\chi](0))
|0\rangle_{| \rm PNRQCD}
\end{equation}
(see (\ref{eq:G})) to third order in non-relativistic (PNRQCD) 
perturbation theory. Adopting the operator notation from 
section~\ref{sec:qm}, the expansion is given by
\begin{equation}
\label{greenexpand}
G(E) = G_0(E) + \delta_1 G(E) + \delta_2 G(E) + \delta_3 G(E) 
+ \ldots
\end{equation}
with $G_0(E) = \langle \bff{0}| \hat G_{0}(E) |\bff{0}\rangle = 
G_0(0,0;E)$ as given in (\ref{eq:G00MSbar}), and 
\begin{eqnarray}
\delta_1 G(E) &=& \langle \bff{0}| \hat G_0(E)i\delta V_1 
i\hat G_0(E)|\bff{0}\rangle,\\[0.2cm]
\delta_2 G(E) &=& \langle \bff{0}| \hat G_0(E)i \delta V_1 
i\hat G_0(E) i\delta V_1 i\hat G_0(E)|\bff{0}\rangle + 
 \langle \bff{0}| \hat G_0(E)i\delta V_2 
i\hat G_0(E)|\bff{0}\rangle,\\[0.2cm]
\delta_3 G(E) &=& \langle \bff{0}| \hat G_0(E)i \delta V_1 
i\hat G_0(E) i \delta V_1 
i\hat G_0(E) i\delta V_1 i\hat G_0(E)|\bff{0}\rangle 
\nonumber\\ && +\, 
2 \langle \bff{0}| \hat G_0(E)i \delta V_1 
i\hat G_0(E) i\delta V_2 i\hat G_0(E)|\bff{0}\rangle + 
 \langle \bff{0}| \hat G_0(E)i\delta V_3 
i\hat G_0(E)|\bff{0}\rangle \nonumber\\ 
&&+ \, \delta^{us}G(E)\,.
\label{masterthirdorder}
\end{eqnarray} 
In momentum space these expressions are of the form of single 
and multiple insertions of potentials as defined in 
(\ref{multipleinsertions}) plus the ultrasoft correction. 
$\delta V_n$ denotes a potential correction of order $n$.
The first-order potential consists only of the one-loop 
correction to the Coulomb potential:
\begin{equation}
\delta V_1 = -\frac{4\pi \alpha_{s}C_F}{\bff{q}^2}\,\hat {\cal V}_C^{(1)}\,.
\end{equation}
At second order, we have the two-loop Coulomb potential, the one-loop 
$1/(m|\bf{q}|)$ potential and the $v^2$-suppressed potentials at
tree-level. Together with the kinetic energy correction, we obtain
\begin{eqnarray}
\delta V_2 &=& 
  -\frac{4\pi \alpha_{s}C_F}{\bff{q}^2} \bigg[\,\hat {\cal V}_C^{(2)}
  -\hat {\cal V}_{1/m}^{(1)}\,\frac{\pi^2\,|\bf{q}|} {m}
  +\hat {\cal V}_{1/m^2}^{(0)}\,\frac{\bff{q}^2}{m^2}
  +\hat {\cal V}_{p}^{(0)}\,\frac{\bff{p}^2+\bff{p}^{\prime \,2}}{2m^2}\,
\bigg] 
\nonumber\\[-0.1cm] 
&& - \,2\times \frac{\bff{p}^4}{8 m^3}\,(2\pi)^{d-1}\delta^{(d-1)}(\bff{q}).
\end{eqnarray}
The notation for the potentials is defined in
(\ref{eq:potentialafterspin}), where also the explicit expressions 
are given. Note that different from (\ref{eq:calVexp}) the 
$n$-loop potential  
$\hat {\cal V}_X^{(n)}$ includes the coupling constant factor, 
i.e. $\hat {\cal V}_X^{(n)} = \left(\frac{\alpha_s}{4\pi}\right)^n\,
{\cal V}_X^{(n)}$. There are no new potentials appearing 
at third order, hence
\begin{eqnarray}
\delta V_3 &=& 
  -\frac{4\pi \alpha_{s}C_F}{\bff{q}^2} \bigg[\,\hat {\cal V}_C^{(3)}
  -\hat {\cal V}_{1/m}^{(2)}\,\frac{\pi^2\,|\bf{q}|} {m}
  +\hat {\cal V}_{1/m^2}^{(1)}\,\frac{\bff{q}^2}{m^2}
  +\hat {\cal V}_{p}^{(1)}\,\frac{\bff{p}^2+\bff{p}^{\prime \,2}}{2m^2}\,
\bigg].
\end{eqnarray}
Note that there is no kinetic energy term in $\delta V_3$, because 
the kinetic energy term in the Lagrangian is not renormalized.
These potentials appear as single insertions in $\delta_n G(E)$. 
In addition, $\delta_n G(E)$ receives contributions from multiple 
insertions of the lower-order potentials. 

To complete the 
perturbative expansion of the third-order cross section, the 
expansion (\ref{greenexpand}) 
of the Green function is inserted into (\ref{R1}), 
\blue{
\begin{equation}
R = 12\pi e_t^2\,\mbox{Im}
\left[\frac{N_{c}}{2m^{2}}\left(c_v 
\left[c_v-\frac{E}{m}\,\left(c_v+\frac{\mathcal{E}}{E}\frac{d_v}{3}
\right)\right] G(\mathcal{E}) + \ldots\right) \right],
\end{equation}
here written in the form of (\ref{eq:pitoNRQCD}), which distinguishes 
between complex energy $\mathcal{E}$ and its real part $E$, which 
will be used in part II. }
The coefficient functions $c_v$ and $d_v$ are likewise expanded, 
and product terms of order higher than three in non-relativistic
perturbation theory are dropped. Note 
that $E/m\sim v^2$ counts as second order in this expansion.

This work concerns the third-order correction $\delta_3 G(E)$. 
The triple insertion of the first-order Coulomb potential is 
algebraically complicated but has no UV or IR divergences and can 
therefore be computed numerically as done in~\cite{Beneke:2005hg}. 
The ultrasoft correction was obtained in~\cite{Beneke:2008cr}. 
In part II of the present work we will give the details of the 
calculation of the remaining single and double insertions in the 
second line of (\ref{masterthirdorder}) in dimensional regularization 
as is necessary for a consistent combination with the matching 
calculations performed and summarized in this part I. A more precise 
master formula for the third-order correction to the Green function 
that accounts for the pole structure of the $d$-dimensional 
potentials will be given in part II. \blue{In that part, we further discuss the 
implementation of a mass renormalization scheme that avoids large 
corrections present in the pole scheme, the treatment of finite-width 
effects relevant to the top threshold, and provide a detailed 
numerical analysis of the QCD third-order correction.}

\noindent
\subsubsection*{Acknowledgement}
We thank A.~Maier, J.~Piclum and M.~Steinhauser for comments on the text.
This work has been supported by the DFG
Sonder\-forschungs\-bereich/Transregio~9 ``Computer\-gest\"utzte
Theoretische Teil\-chen\-physik'', the DFG Graduiertenkolleg
``Elementar\-teil\-chen\-physik an der TeV-Skala'', the DFG 
cluster of excellence ``Origin and Structure of the Universe'' and 
by JSPS KAKENHI Grant Number JP22K03602.

\addcontentsline{toc}{section}{\numberline{}References}
\bibliographystyle{JHEP-2}
\bibliography{paper}

\end{document}